\documentclass[12pt]{scrartcl}

\usepackage{
adjustbox,
amsmath,
amssymb,
amsthm,
booktabs,
cite,
enumerate,
float,
graphicx,
mathtools,
multicol,
multirow,
pdflscape,
tabularx,
tikz,
tikz-cd,
xcolor,
xspace
}

\usepackage[pagebackref,hidelinks]{hyperref}
\usepackage[all]{xy}
\usepackage[width=1.09\textwidth]{caption}
\usepackage[LGR,T1]{fontenc}
\usepackage[utf8]{inputenc}
\usepackage[english]{babel}

\usepackage{cleveref}

\usetikzlibrary{decorations.markings, calc}
\tikzstyle{stuff_fill_green}=[circle,draw,fill=green!]
\tikzstyle{stuff_fill_red}=[circle,draw,fill=red!40]
\tikzstyle{stuff_fill_blue}=[circle,draw,fill=cyan!70]
\tikzstyle{stuff_fill_connect}=[circle,draw,fill=orange!70]

\newcommand{\oref}{\Cref}

\def\KB{{\overline{K}_{B_3}}}
\def\bbP{{\mathbb{P}}}

\newcommand{\one}[0]{\ensuremath{\mathbf{1} }\xspace}
\newcommand{\two}[0]{\ensuremath{\mathbf{2} }\xspace}

\newcommand{\three}[0]{\ensuremath{\mathbf{3} }\xspace}

\newcommand{\xrightarrowdbl}[2][]{%
  \xrightarrow[#1]{#2}\mathrel{\mkern-14mu}\rightarrow
}

\newtheoremstyle{break}  
  {\topsep}   
  {\topsep}   
  {}  
  {0pt}       
  {\bfseries} 
  {:}         
  {\newline}  
  {}          

\theoremstyle{break}
\newtheorem{defi}{Definition}
\newtheorem{lemma}[defi]{Lemma}
\newtheorem{corollary}[defi]{Corollary}
\newtheorem{prop}[defi]{Proposition}

\makeatletter
\let\@addpunct\@gobble
\makeatother
\newenvironment{myproof}[1][\bfseries \proofname]{%
  \begin{proof}[#1]$ $\par\nobreak\ignorespaces
}{%
  \end{proof}
}

\makeatletter
\DeclareOldFontCommand{\sc}{\normalfont\scshape}{\@nomath\sc}
\makeatother

\newcommand{\textgreek}[1]{\begingroup\fontencoding{LGR}\selectfont#1\endgroup}

\numberwithin{equation}{section}

\hypersetup{
    pdftitle={Root Bundles and Towards Exact Matter Spectra of F-theory MSSMs},
    pdfauthor={Martin Bies, Mirjam Cvetic, Ron Donagi, Muyang Liu, Marielle Ong},
    pdfsubject={Root bundles, Limit roots, Line bundles, Global sections, F-theory, MSSM, Standard Model, vector-like spectra}
}

\allowdisplaybreaks
\begin{document}


\vspace*{-2cm}
\begin{flushright}
{\texttt{UPR-1309-T}}\\
\end{flushright}

\vspace*{0.8cm}
\begin{center}
{\LARGE Root Bundles and\\[0.5em]Towards Exact Matter Spectra of F-theory MSSMs}


\vspace*{1.8cm}
{Martin Bies$^{1,2}$, Mirjam Cveti{\v c}$^{2,1,3}$, Ron Donagi$^{1,2}$, Muyang Liu${^2}$, Marielle Ong$^{1}$}

\vspace*{1cm}

{\textit{$^1$Department of Mathematics, University of Pennsylvania,  \\
 Philadelphia, PA 19104-6396, USA}}

\bigskip
{\textit{$^2$Department of Physics and Astronomy, University of Pennsylvania,  \\
 Philadelphia, PA 19104-6396, USA}}

\bigskip
{\textit{$^3$Center for Applied Mathematics and Theoretical Physics, University of Maribor, \\
 Maribor, Slovenia}}

\vspace*{0.8cm}
\end{center}
%
\noindent

Motivated by the appearance of fractional powers of line bundles in studies of vector-like spectra in 4d F-theory compactifications, we analyze the structure and origin of these bundles. 
Fractional powers of line bundles are also known as root bundles and can be thought of as generalizations of spin bundles. We explain how these root bundles are linked to inequivalent F-theory gauge potentials of a \mbox{$G_4$-flux}.

While this observation is interesting in its own right, it is \mbox{particularly} valuable for F-theory Standard Model constructions. In aiming for MSSMs, it is desired to argue for the absence of vector-like exotics. We work out the root bundle constraints on all matter curves in the largest class of \mbox{currently-known} F-theory Standard Model constructions without chiral exotics and gauge coupling unification. On each matter curve, we conduct a systematic ``bottom''-analysis of all solutions to the root bundle constraints and all spin bundles. Thereby, we derive a lower bound for the number of combinations of root bundles and spin bundles whose cohomologies satisfy the physical demand of absence of vector-like pairs.

On a technical level, this systematic study is achieved by a well-known diagrammatic description of root bundles on nodal curves. We extend this description by a counting procedure, which determines the cohomologies of so-called limit root bundles on full blow-ups of nodal curves. By use of deformation theory, these results constrain the vector-like spectra on the smooth matter curves in the actual F-theory geometry.


\newpage
\tableofcontents
\newpage



\section{Introduction}

String theory elegantly couples gauge dynamics to gravity. This makes string theory a leading candidate for a unified theory of quantum gravity. As such, it must account for all aspects of our physical reality, especially the low energy particle physics that we observe. As a first order approximation, one desires an explicit demonstration in which one can actually obtain the particle spectrum of the Standard Model from string theory.

In the past decades, enormous efforts have been undertaken to achieve this goal. Many of these models concentrated on perturbative corners of string theory, such as the $E_8 \times E_8$ heterotic string \cite{Candelas:1985en,Greene:1986ar,Braun:2005ux,Bouchard:2005ag,Bouchard:2006dn,Anderson:2009mh,Anderson:2011ns,Anderson:2012yf} or intersecting branes models in type II \cite{Berkooz:1996km,Aldazabal:2000dg,Aldazabal:2000cn,Ibanez:2001nd,Blumenhagen:2001te,Cvetic:2001tj,Cvetic:2001nr} (see also \cite{Blumenhagen:2005mu} and references therein). These perturbative models were among the first compactifications from which the Standard Model gauge sector emerged with its chiral or, in the case of \cite{Bouchard:2005ag,Bouchard:2006dn}, even the vector-like spectrum. Unfortunately, these constructions are limited due to their perturbative nature in the string coupling, and they typically suffer from chiral and vector-like exotic matter. Among these perturbative models, the first globally consistent MSSM constructions are \cite{Bouchard:2005ag,Bouchard:2006dn} (see \cite{Gomez:2005ii,Bouchard:2008bg} for more details on the subtle global conditions for slope-stability of vector bundles).

The non-perturbative effects in string theory are elegantly described by F-theory \cite{Vafa:1996xn,oai:arXiv.org:hep-th/9602114,oai:arXiv.org:hep-th/9603161}. As a non-perturbative extension of type IIB string theory, the framework of F-theory describes the gauge dynamics on 7-branes \emph{including} their back-reactions (to all orders in the string coupling) onto the compactification geometry $B_n$. These back-reactions are encoded in the geometry of an elliptically fibered Calabi-Yau space $\pi \colon Y_{n + 1} \twoheadrightarrow B_n$. By studying this space $Y_{n + 1}$ with well-established tools of algebraic geometry, one can then ensure the global consistency conditions of the physics in $10-2n$ non-compact dimensions.

An important characteristic of 4d $\mathcal{N} = 1$ F theory compactifications (i.e., $n = 3$), which must match the particle physics that we observe, is the chiral fermionic spectrum. In F-theory, this spectrum is uniquely fixed by a background gauge flux, which in turn is most conveniently specified by the internal $C_3$ profile in the dual M-theory geometry. The chiral spectrum then only depends on the flux $G_4 = dC_3 \in H^{(2,2)} (Y_4)$. By now, there exists an extensive toolbox for creating and enumerating the so-called primary vertical subspace of $G_4$ configurations \cite{oai:arXiv.org:1111.1232,Krause:2012yh,Braun:2013nqa,Cvetic:2013uta,Cvetic:2015txa,Lin:2015qsa,Lin:2016vus}. The application of these tools led to the construction of globally consistent chiral F-theory models \cite{Krause:2011xj,Cvetic:2015txa,Lin:2016vus,Cvetic:2018ryq}, which recently culminated in the largest class of explicit string vacua that realize the Standard Model gauge group with their exact chiral spectrum and gauge coupling unification \cite{Cvetic:2019gnh}.

However, these methods are insufficient to determine the exact vector-like spectrum of the chiral zero modes (i.e., not just the difference between chiral and anti-chiral fields). This is because the zero modes depend not only on the flux $G_4$, but also on the flat directions of the potential $C_3$. The complete information is encoded in the \mbox{so-called} \emph{Deligne cohomology}. In\cite{Bies:2014sra,Bies:2017fam,Bies:2018uzw}, methods for determining the exact vector-like spectra were put forward. This approach exploits the fact that (a subset of) the Deligne cohomology can be parameterized by Chow classes. By use of this parameterization, one can extract line bundles $L_{\mathbf{R}}$ that are defined on curves $C_\mathbf{R}$ in $B_3$. In the dual IIB picture, this can be interpreted as localization of gauge flux on matter curves, which lifts some vector-like pairs on these curves. Explicitly, the zero modes are counted by the sheaf cohomologies of $L_{\mathbf{R}}$ and we have $h^{0}(C_{\mathbf{R}}, \, {L}_{\mathbf{R}})$ massless chiral and $h^1 (C_{\mathbf{R}}, \, {L}_{\mathbf{R}})$ massless anti-chiral superfields in representation $\mathbf{R}$ on $C_{\mathbf{R}}$.

Although this procedure works in theory for any compactification, technical limitations arise in practical applications. Intuitively, one may think of the technical difficulties as reflections of the delicate complex structure dependence of the line bundles cohomologies. Even state-of-the-art algorithms such as \cite{ToricVarietiesProject,M2,sagemath} (see also \cite{Bies:2017fam,Bies:2018uzw}) on supercomputers specifically designed for such computations (such as \texttt{Plesken} at \textit{Siegen University}), can oftentimes not perform the necessary operations in realistic compactification geometries. For instance, the models studied in \cite{Bies:2014sra,Bies:2017fam,Bies:2018uzw} focused on computationally simple geometries as a result. While this led to a proof of principle, these models have unrealistically large numbers of chiral fermions. Therefore --- even though it is expected --- it remains an open question whether or not F-theory compactifications can actually realize effective theories that resemble the matter spectra of the Standard Model.

Recently, \cite{Bies:2020gvf} the complex structure dependence of line bundle cohomologies was \linebreak investigated . This analysis was inspired from the F-theory GUT models discussed in \cite{Bies:2018uzw} and focused on simple geometries, in which the algorithms in \cite{ToricVarietiesProject} could \mbox{generate} a large data set \cite{Database}. This data was analyzed by use of data science techniques, in particular decision trees. A theoretical understanding of this data was achieved by \mbox{supplementing} the data science results by Brill-Noether theory \cite{Brill1874} (see \cite{Eisenbud1996} for a modern exposition and \cite{Watari:2016lft} for an earlier application of Brill-Noether theory in F-theory). These insights led to a quantitative study of jumps of charged matter vector pairs as a function of the complex structure parameters of the matter curves.

\paragraph{Results}

In globally consistent F-theory constructions with the exact chiral spectra of the Standard Model and gauge coupling unification \cite{Cvetic:2019gnh}, the vector-like spectra on the low-genus matter curves are encoded in cohomologies of a line bundle, which are identified with a fractional power of the canonical bundle. On high-genus curves, these fractional powers of the canonical bundle are further modified by contributions from Yukawa points.

In order to make sense of these fractional powers, we study the $G_4$-flux in more detail. The models in \cite{Cvetic:2019gnh} consider a background $G_4$-flux, which not only leads to the exact chiral spectra but also satisfies global consistency conditions, such as the D3-tadpole cancelation and masslessness of the $U(1)$-gauge boson. We lift this very background $G_4$-flux to a gauge potential in the \emph{Deligne cohomology} to identify the line bundles ${L}_{\mathbf{R}}$. This process requires an understanding of the intermediate Jacobian $J^2( Y_4 )$, which labels inequivalent gauge backgrounds. A naive analysis, which does not properly take the intermediate Jacobian into account, leads to the fractional line bundle powers mentioned above. In past works \cite{Bies:2014sra,Bies:2017fam,Bies:2018uzw}, such scenarios were avoided as it is not immediately clear how to interpret these fractional expressions. However, since these expressions appear ubiquitously in compact models with realistic chiral indices, this work analyzes the origin and meaning of these bundles in detail.

The objects we are therefore interested in are fractional powers of line bundles, also known as root bundles. They may be thought of as generalizations of spin bundles. Similar to spin bundles, root bundles are far from unique. The mathematics of root bundles indicates that we should think of the different root bundles as being induced from \mbox{inequivalent} gauge potentials for a given $G_4$-flux. While \cite{Bies:2014sra,Bies:2017fam,Bies:2018uzw} has already anticipated that inequivalent gauge potentials for a given $G_4$-flux lead to different vector-like spectra, the root bundle interpretation allows one to test this expectation.

In general, not all root bundles on the matter curves are induced from F-theory gauge potentials in the Deligne cohomology $H^4_D( \widehat{Y}_4, \mathbb{Z}(2) )$. This mirrors the expectation that only some of the spin bundles on the matter curves are consistent with the F-theory geometry $\widehat{Y}_4$. This raises the interesting and important question of identifying which roots and spin bundles are induced top-down. While this work does not answer this question, we hope that it initiates and facilitates this analysis by providing a systematic approach to all root bundles and spin bundles on the matter curves. In particular, we identify pairs of root bundles and spin bundles such that their tensor product is a line bundle whose cohomologies satisfy the physical demand of the presence/absence of vector-like pairs.

On a technical level, this requires a sufficient understanding of root bundles and their cohomologies on a matter curve $C_{\mathbf{R}}$. We gain this control from a deformation $C_{\mathbf{R}} \to C^\bullet_{\mathbf{R}}$ into a nodal curve. On the latter, root bundles are described in a diagrammatic way by so-called \emph{limit roots} \cite{2004math4078C}. We extend these ideas to a counting procedure for the global sections of limit roots, which we use to infer the cohomologies of root bundles on $C_{\mathbf{R}}$. This approach is demonstrated in the largest class of currently-known constructions of globally consistent F-theory Standard Models without chiral exotics and gauge coupling unification \cite{Cvetic:2019gnh}. In one particular such geometry, we derive a lower bound to the number of pairs of root bundles and spin bundles whose tensor product is a line bundle without vector-like exotics.

\paragraph{Outline}

In \oref{sec:ZeroModesInQSM}, we recall zero mode counting in F-theory and the \mbox{appearance} of fractional line bundle powers. We explain that these fractional powers of line \mbox{bundles}, also known as root bundles, relate to inequivalent gauge potentials in F-theory. These ideas are subsequently applied to the largest currently-known family of globally \linebreak consistent F-theory Standard Model constructions without chiral exotics and gauge \mbox{coupling} unification \cite{Cvetic:2019gnh}. We explain how the background $G_4$-flux, which satisfies global consistency conditions such as the cancellation of the D3-tadpole and the \mbox{masslessness} of the $U(1)$-gauge boson, induces root bundles on the matter curves. Details of this derivation are summarized in \cref{sec:LineBundleComputations}. This derivation heavily relies on a detailed \mbox{understanding} of the elliptically fibered 4-fold F-theory geometry $\widehat{Y}_4$, including \linebreak intersection numbers in the fiber over the Yukawa points. We supplement the \mbox{earlier} works \cite{Klevers:2014bqa,Cvetic:2015txa,Cvetic:2019gnh} with a complete list of all fiber intersection numbers in \cref{sec:FibreStructure}.

In \oref{sec:LimitRoots} we first summarize well-known results about root bundles before we describe the limit root constructions, which were originally introduced in \cite{2004math4078C}. We extend the limit root constructions by a counting procedure for the global sections of limit roots on full blow-ups of nodal curves. In fortunate instances, this even provides a means to explore \emph{Brill-Noether theory} of \emph{limit roots}, which we demonstrate in an example inspired from \cite{Farkas_2017}.

Finally, we apply these ideas to globally consistent constructions of F-theory Standard Models without chiral exotics and gauge coupling unification \cite{Cvetic:2019gnh} in \oref{sec:ApplicationToMSSMs}. In an explicit base space geometry, we deform the matter curves to nodal curves, construct limit roots on these nodal curves, identify the number of global sections of these limit roots and finally use deformation theory to relate these counts to the cohomologies of root bundles in the actual F-theory geometry. Thereby, we explicitly prove the existence of root bundle solutions without vector-like pairs. Technical details of the specific base geometry and the limit root constructions are summarized in \cref{sec:ExampleGeometry}.

\section{Root bundles in F-theory} \label{sec:ZeroModesInQSM}

\subsection{The appearance of root bundles} \label{subsec:RootBundleAppearance}

\paragraph{Zero mode counting in F-theory}

We consider an F-theory compactification to four dimensions given by a singular, elliptically fibered 4-fold $\pi \colon Y_4 \twoheadrightarrow {B}_3$. We assume that this fibration has a section $s \cong {B}_3$ and admits a smooth, flat, crepant resolution $\widehat{\pi} \colon \widehat{Y_4} \twoheadrightarrow {B}_3$. In such a compactification, the flux $G_4 \in H^{(2,2)}( \widehat{Y}_4 )$ is subject to the quantization condition \cite{Witten:1996md}
\begin{align}
G_4 + \frac{1}{2} c_2( T_{\widehat{Y}_4} ) \in H^{(2,2)}_{\mathbb{Z}} ( \widehat{Y}_4 ) = H^{(2,2)} ( \widehat{Y}_4 ) \cap H^{4} ( \widehat{Y}_4, \mathbb{Z} ) \, . \label{equ:QuantizationCondition}
\end{align}
For simplicity, we focus on compactifications with even $c_2( T_{\widehat{Y}_4} )$, which holds true for F-theory compactifications on an elliptically fibered smooth Calabi-Yau 4-fold with globally defined Weierstrass model \cite{Collinucci:2010gz}.\footnote{For zero mode counting of half-integer quantized ${G}_4$-fluxes, see e.g. \cite{Bies:2014sra}.} Under the simplifying assumption that $c_2( T_{\widehat{Y}_4} )$ is even, \oref{equ:QuantizationCondition} requires that $G_4 \in H^{(2,2)}_{\mathbb{Z}} ( \widehat{Y}_4 )$. We will show an example of this and the following root bundle analysis in the largest \mbox{currently-known} class of F-theory Standard Model constructions with gauge coupling unification and no chiral exotics \cite{Cvetic:2019gnh} in \oref{{subsec:RootBundlesInF-theoryMSSM}}.

Over the codimension-2 matter curves $C_{\mathbf{R}} \subseteq {B}_3$, the reducible fibers of $\widehat{Y}_4$ contain a chain of $\mathbb{P}^1$s. A state with weight $\mathbf{w}$ in the representation $\mathbf{R}$ corresponds to a linear combination of these $\mathbb{P}^1$'s. By fibering this linear combination over the matter curve $C_{\mathbf{R}}$, one obtains the matter surface $S_{\mathbf{R}}$.\footnote{In general, a $G_4$-flux can induce different chiral indices and vector-like spectra on the different weight states. In such instances, it makes sense to keep track of $\mathbf{w}$ and write $S^{\mathbf{w}}_{\mathbf{R}}$. However, in anticipation of \cite{Cvetic:2019gnh}, we focus on gauge invariant $G_4$-fluxes, which induce the same chiral index and vector-like spectra for all weight states.} The chiral index of the massless matter localized on the matter curve $C_{\mathbf{R}} \subseteq {B}_3$ is then specified by \cite{Donagi:2008ca,Hayashi:2008ba,Donagi:2009ra,Heckman:2010bq,Marsano:2011hv,Grimm:2011fx,Braun:2011zm} \cite{Krause:2011xj,Krause:2012yh,Intriligator:2012ue} as
\begin{align}
\chi \left( \mathbf{R} \right) = \int \limits_{S_{\mathbf{R}}}{G_4} \, . \label{equ:ChiralIndex}
\end{align}
The vector-like spectrum induced by a $G_4$-flux has been analyzed in \cite{Bies:2014sra, Bies:2017fam, Bies:2018uzw}. We employ the short exact sequence
\begin{align}
0 \to J^2( \widehat{Y}_4 ) \xhookrightarrow{\iota} H^4_D ( \widehat{Y}_4, \mathbb{Z}(2) ) \xrightarrowdbl{\widehat{c}} H^{(2,2)}_{\mathbb{Z}} ( \widehat{Y}_4 ) \to 0 \, , \label{equ:DefiDeligne}
\end{align}
where there exists a surjection $\widehat{c}$ that maps the gauge potential $A \in H^4_D ( \widehat{Y}_4, \mathbb{Z}(2) )$ as an element of the Deligne cohomology group to its $G_4$-flux. The Deligne cohomology classes encode the full gauge background data. Therefore, they parallel the internal \mbox{3-form} potentials $C_3$ in the dual M-theory picture in which $G_4 = d C_3$. As long as $C_3^\prime - C_3$ is a closed 3-form, $C_3^\prime$ has the same field strength $G_4$ as $C_3$. In F-theory, such closed 3-form potentials are encoded by the intermediate Jacobian. Put differently, two inequivalent ${A}^\prime, A \in H^4_D( \widehat{Y}_4, \mathbb{Z}(2) )$ with $\widehat{c}( A^\prime ) = \widehat{c}( A ) = G_4$ differ by $A^\prime - A = \iota( B )$, where $B \in J^2( \widehat{Y}_4 )$ is an element of the intermediate Jacobian corresponding to a closed M-theory 3-form potential.\footnote{Equivalently, different gauge potentials in $H^4_D( \widehat{Y}_4, \mathbb{Z}(2) )$ differ by their Wilson lines \cite{Greiner:2015mdm, Greiner:2017ery}.}

The Deligne cohomology group $H^4_D ( \widehat{Y}_4, \mathbb{Z}(2) )$ is rather hard to handle in explicit \mbox{computations}. However, we can parametrize (at least a subset of) $H^4_D ( \widehat{Y}_4, \mathbb{Z}(2) )$ by the Chow group $\mathrm{CH}^2( \widehat{Y}_4, \mathbb{Z} )$. This is summarized in the commutative diagram\footnote{For more details, see \cite{Bies:2014sra} and references therein.}
\begin{equation}
\begin{tikzcd}
0 \arrow{r}{} & \mathrm{CH}^2_{\text{hom}} ( \widehat{Y}_4, \mathbb{Z} ) \arrow{d}{} \arrow{r}{} & \mathrm{CH}^2( \widehat{Y}_4, \mathbb{Z} ) \arrow{d}{\widehat{\gamma}} \arrow{r}{\gamma} & H^{(2,2)}_{\text{alg}} ( \widehat{Y}_4 ) \arrow{d}{} \arrow{r}{} & 0 \\
0 \arrow{r}{} & J^2( \widehat{Y}_4 ) \arrow{r}{} & H^4_D ( \widehat{Y}_4, \mathbb{Z}(2) ) \arrow{r}{\widehat{c}} & H^{(2,2)}_{\mathbb{Z}} ( \widehat{Y}_4 ) \arrow{r}{} & 0
\end{tikzcd} \label{equ:CommutativeDiagramLifts}
\end{equation}
Unless stated differently, the symbol $\mathcal{A}$ is reserved for an element $\mathcal{A} \in \mathrm{CH}^2( \widehat{Y}_4, \mathbb{Z} )$, by which we specify an F-theory gauge potential $A = \widehat{\gamma} (\mathcal{A})$.

In order to count the zero modes in representation $\mathbf{R}$ in the presence of such a gauge potential $\widehat{\gamma} ( \mathcal{A} ) \in H^4_D( \widehat{Y}_4, \mathbb{Z}(2) )$, we consider the matter surface $S_{\mathbf{R}}$ with
\begin{align}
\iota_{S_{\mathbf{R}}} \colon S_{\mathbf{R}} \hookrightarrow \widehat{Y}_4 \, , \qquad \pi_{S_{\mathbf{R}}} \colon S_{\mathbf{R}} \twoheadrightarrow C_{\mathbf{R}} \, .
\end{align}
The cylinder map, which sends $\mathcal{A} \in \mathrm{CH}^2( \widehat{Y}_4, \mathbb{Z} )$ to a class $D_R( \mathcal{A} ) \in \mathrm{Pic} \left( C_{\mathbf{R}} \right) $, is the \mbox{restriction} to $S_R$ followed by integration over the fibers to $C_R$:
\begin{align}
D_{\mathbf{R}} \left( \mathcal{A} \right) = \pi_{S_{\mathbf{R}}\ast} \left( \iota_{S_{\mathbf{R}}}^\ast \left( \mathcal{A} \right) \right) \in \mathrm{Pic} \left( C_{\mathbf{R}} \right) \, . \label{equ:DivisorFromChow}
\end{align}
The matter spectrum is then determined in terms of sheaf cohomology groups:
\begin{align}
\begin{split} \label{actual}
h^0 \left( C_{\mathbf{R}}, L_{\mathbf{R}} \left( \mathcal{A} \right) \right) & \quad \leftrightarrow \quad \text{chiral zero modes} \, , \\
h^1 \left( C_{\mathbf{R}}, L_{\mathbf{R}} \left( \mathcal{A} \right) \right) & \quad \leftrightarrow \quad \text{anti-chiral zero modes} \, ,
\end{split}
\end{align}
where 
\begin{align}
L_{\mathbf{R}} \left( \mathcal{A} \right) = \mathcal{O}_{C_{\mathbf{R}}} \left( D_{\mathbf{R}} \left( \mathcal{A} \right) \right) \otimes_{\mathcal{O}_{C_{\mathbf{R}}}} \mathcal{O}_{C_{\mathbf{R}}}^{\text{spin}} \, ,
\end{align}
with $\mathcal{O}_{C_{\mathbf{R}}}^{\text{spin}}$ an appropriate spin bundle on $C_{\mathbf{R}}$. This is a refinement of \oref{equ:ChiralIndex}, since Riemann-Roch gives
\begin{equation}
  \resizebox{0.89\textwidth}{!}{%
$\begin{aligned}
\chi({\mathbf{R}}) = h^0 \left( C_{\mathbf{R}}, L_{\mathbf{R}} \left( \mathcal{A} \right) \right) - h^1 \left( C_{\mathbf{R}}, L_{\mathbf{R}} \left( \mathcal{A} \right) \right) = \chi \left( L_{\mathbf{R}}( \mathcal{A}) \right) = deg \left( D_{\mathbf{R}}( \mathcal{A} ) \right) = \int_{S_{\mathbf{R}}}{G_4} .
\end{aligned}$%
}
\end{equation}

\paragraph{Roots of F-theory gauge potentials}

For an F-theory model, we need an F-theory gauge potential, i.e. a class in the Deligne cohomology group $H^4_D ( \widehat{Y}_4, \mathbb{Z}(2) ) $. This will be specified as $ \widehat{\gamma} (\mathcal{A})$ for some ``potential" $\mathcal{A} \in \mathrm{CH}^2( \widehat{Y}_4, \mathbb{Z} )$. We find that the geometry determines a class $\widehat{\gamma}(\mathcal{A}^\prime) \in H^4_D( \widehat{Y}_4, \mathbb{Z}(2) )$ and an integer $\xi \in \mathbb{Z}_{> 0}$ such that $\mathcal{A}$ is subject to the two constraints:
\begin{align}
\gamma( \mathcal{A} ) = G_4 \, , \qquad \xi \cdot \widehat{\gamma}( \mathcal{A} ) \sim \widehat{\gamma} ( \mathcal{A}^\prime ) \, . \label{equ:Conditions}
\end{align}
The condition $\gamma( \mathcal{A} ) = G_4$ immediately follows from \cref{equ:CommutativeDiagramLifts} and it means that $\widehat{\gamma}( \mathcal{A} )$ is an F-theory gauge potential for the given $G_4$-flux. We will illustrate with several examples below that the absence of chiral exotics in the F-theory Standard Models boils down to the second constraint. It is important to notice that the gauge potential $\mathcal{A}$ specified by the two conditions in \cref{equ:Conditions} is in general not unique. It is difficult to say much about solutions in the Chow group itself, but going to the bottom row in \eqref{equ:CommutativeDiagramLifts}, we see that the collection of all $\xi$-th roots of ${\widehat{\gamma}} ( \mathcal{A}^\prime )$ (if non empty) is a coset of the group of all $\xi$-th roots of $0$. In particular, the number of solutions is $\xi^{2 \cdot \mathrm{dim}_{\mathbb{C}} \left( J^2( \widehat{Y}_4 ) \right)}$.

All these solutions lead to the same chiral spectrum  \eqref{equ:ChiralIndex}, since they all have the same degree when restricted to the curves $C_{\mathbf{R}}$, hence the same index. However, they could differ in their actual spectrum \eqref{actual}. This extra flexibility is the key tool that we intend to use to produce a desirable spectrum such as the MSSM.

\paragraph{Roots on the matter curves}

In theory, we could simply analyze the algebraic cycles $\mathcal{A}$ which satisfy \cref{equ:Conditions}. However, as we will demonstrate momentarily, we can explicitly construct $\mathcal{A}^\prime \in \mathrm{CH}^2( \widehat{Y}_4, \mathbb{Z} )$. Therefore, we have a sufficient level of arithmetic control over $\widehat{\gamma}(\mathcal{A}^\prime)$ and it is natural to ask how the induced divisors $D_{\mathbf{R}}( \mathcal{A}^\prime )$ and $D_{\mathbf{R}}( \mathcal{A} )$ are related. The map $D_{\mathbf{R}} \colon \mathrm{CH}^2 ( \widehat{Y}_4, \mathbb{Z} ) \to \mathrm{CH}^{1}( C_{\mathbf{R}}, \mathbb{Z} ) \cong \mathrm{Pic} ( C_{\mathbf{R}} )$, as defined in \cref{equ:DivisorFromChow}, factors through $\widehat{\gamma}$ and a group homomorphism
\begin{align}
H^4_D ( \widehat{Y}_4, \mathbb{Z}(2) ) \to \mathrm{Pic} ( C_{\mathbf{R}} ) \, .
\end{align}
Thus, it follows that
\begin{align}
\xi \cdot D_{\mathbf{R}} \left( \mathcal{A} \right) \sim D_{\mathbf{R}} \left( \mathcal{A}^\prime \right) \in \mathrm{Pic}( C_{\mathbf{R}} ) \, . \label{equ:Refined}
\end{align}
This means that the F-theory gauge potential $A = \widehat{\gamma}( \mathcal{A} )$ induces \emph{a} divisor $D_{\mathbf{R}}( \mathcal{A})$, whose $\xi$-th multiple is linearly equivalent to the divisor $D_{\mathbf{R}} \left( \mathcal{A}^\prime \right)$ that is induced by the F-theory gauge potential $A^\prime = \widehat{\gamma}( \mathcal{A}^\prime ) \in H^4_D( \widehat{Y}_4, \mathbb{Z}(2) )$. Such a divisor $D_{\mathbf{R}} \left( \mathcal{A} \right)$ is termed a $\xi$-th root of $D_{\mathbf{R}} \left( \mathcal{A}^\prime \right)$.

In general, $\xi$-th roots of $D_{\mathbf{R}} \left( \mathcal{A}^\prime \right)$ do not exist. When they do, they are not unique. This is particularly well known for the case $\xi = 2$ and $D_{\mathbf{R}} \left( \mathcal{A}^\prime \right) = K_{{\mathbf{R}}}$, where the 2nd roots of the canonical bundle are the spin structures on $C_{\mathbf{R}}$. If $C_{\mathbf{R}}$ is a curve of genus $g$, then it admits $2^{2g}$ spin structures (see e.g. \cite{atiyah1971riemann, mumford1971theta}). This easily extends to $\xi > 2$ and $D_{\mathbf{R}} \left( \mathcal{A}^\prime \right) \neq K_{{\mathbf{R}}}$. While we will provide more details on root bundles in \oref{subsec:MthRootsOfDivisors}, it suffices to state here that $\xi$-th roots of $D_{\mathbf{R}} \left( \mathcal{A}^\prime \right)$ do exist if and only if $\xi$ divides $\mathrm{deg} \left( D_{\mathbf{R}} \left( \mathcal{A}^\prime \right) \right)$. So on a genus $g$ curve, there exist $\xi^{2g}$ $\xi$-th roots of $D_{\mathbf{R}} \left( \mathcal{A}^\prime \right)$.

In general, it cannot be expected that $\xi$-th roots of $D_{\mathbf{R}} \left( \mathcal{A}^\prime \right) \in \mathrm{Pic} ( C_{\mathbf{R}})$ and $\xi$-th roots of $A^\prime = \widehat{\gamma}( \mathcal{A}^\prime ) \in H^4_D( \widehat{Y}_4, \mathbb{Z}(2) )$ are one-to-one. Rather, only a subset of the $\xi$-th roots of $D_{\mathbf{R}} \left( \mathcal{A}^\prime \right)$ will be realized from F-theory gauge potentials in $H^4_D( \widehat{Y}_4, \mathbb{Z}(2) )$. In this sense, the root bundle constraint in \cref{equ:Refined} is necessary but not sufficient to conclude that the divisor $D_{\mathbf{R}} \left( \mathcal{A} \right)$ stems from an F-theory gauge potential.

It is an interesting question to investigate which roots of $D_{\mathbf{R}} \left( \mathcal{A}^\prime \right) \in \mathrm{Pic} ( C_{\mathbf{R}})$ are induced from F-theory gauge potentials. While this work does not answer this question, we hope to initiate and facilitate this study by providing a systematic approach to all $\xi$-th roots of $D_{\mathbf{R}} \left( \mathcal{A}^\prime \right)$. In particular, we will provide a counting procedure, which allows one to infer the cohomologies of some of these root bundles. This allows one to search for roots which satisfy the physical demand of the presence/absence of vector-like pairs.

Before we show an example of these notions in the F-theory Standard Models \cite{Cvetic:2019gnh}, let us briefly comment on spin bundles $\mathcal{O}_{C_{\mathbf{R}}}^{\text{spin}}$. Recall that Freed-Witten anomaly \mbox{cancelation} requires $\text{spin}^c$-structures on \mbox{$D7$-branes} in perturbative IIB-compactifications \cite{Freed:1999vc}. As \mbox{explained} in \cite{Beasley:2008dc}, this extends to the demand of $\text{spin}^c$-structures on gauge surfaces \mbox{$S \subset {B}_3$} in F-theory compactifications. Then, a choice of $\text{spin}^c$-structure on $N_{C_{\mathbf{R}}|S}$ induces a unique $\text{spin}^c$-structure on $C_{\mathbf{R}}$ \cite{Lawson:1998yr}. Therefore, the question of which $\text{spin}^c$-structures are realized from the F-theory geometry $\widehat{Y}_4$ arises. While this is a fascinating question, we will not answer it in this work. Rather, we will systematically study all $\xi$-th roots of $D_{\mathbf{R}} \left( \mathcal{A}^\prime \right)$ and all spin bundles on $C_{\mathbf{R}}$. Our goal is to identify combinations of root and spin bundles such that their tensor product is a line bundle whose cohomologies satisfy the physical demand of presence/absence of vector-like pairs.

\subsection{Root bundles in F-theory Standard Models} \label{subsec:RootBundlesInF-theoryMSSM}

We will now exhibit an example of the root bundle analysis in the largest class of currently-known globally consistent F-theory Standard Model constructions that support gauge coupling unification and avoid chiral exotics \cite{Cvetic:2019gnh}. Earlier geometric details can be found in the works \cite{Klevers:2014bqa,Cvetic:2015txa}. For convenience, we briefly summarize the geometry before we discuss the $G_4$-flux and its lifts.

The analysis of the induced line bundles, i.e., evaluating \oref{equ:DivisorFromChow}, is both tedious and lengthy. It makes use of the intersection numbers in the fibers over the matter curves and Yukawa points. As an extension of the past works on this class of F-theory geometries, we list exhaustive details of the fiber geometry in \oref{sec:FibreStructure}. The necessary intersection computations are detailed in \oref{sec:LineBundleComputations}. The latter includes a section on topological intersection numbers of non-complete intersections, which we determine rigorously from the Euler characteristic of the structure sheaf of the intersection variety.

\subsubsection{The resolved elliptic fibration \texorpdfstring{$\mathbf{\widehat{Y}_4}$}{Y4}} \label{subsec:Y4}

\paragraph{4-fold geometry}

For a base 3-fold ${B}_3$, the resolved elliptic fibration $\widehat{Y}_4$ is a \linebreak \mbox{hypersurface} in the space $X_5 = {B}_{3} \times \bbP_{F_{11}}$. The fiber ambient space $\bbP_{F_{11}}$ is the toric surface with the following toric diagram. In the accompanying table, we indicate its $\mathbb{Z}^6$-graded Cox ring:
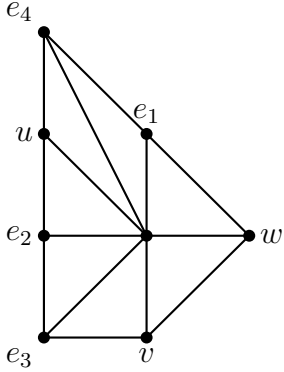
\begin{equation}
  \hspace{2em}
  \begin{minipage}{.4\textwidth}
    \begin{tikzpicture}[scale=0.9,baseline=(current  bounding  box.center)]

      \def\w{1.5};
      
      \draw[thick] (-\w,2*\w)--(\w,0)--(0,-\w)--(-\w,-\w)--(-\w,2*\w);
      
      \draw[thick] (0,0) -- (\w,0);
      \draw[thick] (0,0) -- (0,\w);
      \draw[thick] (0,0) -- (0,-\w);
      \draw[thick] (0,0) -- (-\w,2*\w);
      \draw[thick] (0,0) -- (-\w,\w);
      \draw[thick] (0,0) -- (-\w,0);
      \draw[thick] (0,0) -- (-\w,-\w);
      
      \fill (0,0) circle[radius=2.5pt];
      \fill (\w,0) circle[radius=2.5pt];
      \fill (0,\w) circle[radius=2.5pt];
      \fill (-\w,2*\w) circle[radius=2.5pt];
      \fill (-\w,\w) circle[radius=2.5pt];
      \fill (-\w,0) circle[radius=2.5pt];
      \fill (-\w,-\w) circle[radius=2.5pt];
      \fill (0,-\w) circle[radius=2.5pt];
      
      \node at (\w,0) [right] {$w$};
      \node at (0,\w) [above] {$e_1$};
      \node at (-\w,2*\w) [above left] {$e_4$};
      \node at (-\w,\w) [left] {$u$};
      \node at (-\w,0) [left] {$e_2$};
      \node at (-\w,-\w) [below left] {$e_3$};
      \node at (0,-\w) [below] {$v$};
      
      \end{tikzpicture}
  \end{minipage}%
  \hspace{-2em}\begin{minipage}{.5\textwidth}
  {\footnotesize
  \begin{tabular}{c|ccc|cccc}
    \toprule
    & u & v & w & $e_1$ & $e_2$ & $e_3$ & $e_4$ \\
    \midrule
    H & 1 & 1 & 1 \\
    \midrule
    $E_1$ & -1 &    & -1 & 1 \\
    $E_2$ & -1 & -1 &    &    & 1 \\
    $E_3$ &    & -1 &    &    & -1 & 1 \\
    $E_4$ & -1 &    &    & -1 &    &   & 1 \\
    \bottomrule
    \end{tabular}}
  \end{minipage}
  \end{equation}
Equivalently, the Stanley-Reisner ideal of $\bbP_{F_{11}}$ is given by
\begin{align}
\begin{split}\label{eq:SRF11}
I_{\text{SR}} \left( \bbP_{F_{11}} \right) = \langle & e_4 w, e_4 e_2, e_4 e_3, e_4 v, e_1 u, e_1 e_2, e_1 e_3, e_1 v, w u, w e_2, w e_3, v e_2, u v, e_3 u \rangle \, .
\end{split}
\end{align}
Consider sections $s_i \in H^0( {B}_3, \overline{K}_{{B}_3} )$. Then, in the space $X_5$, the resolved 4-fold $\widehat{Y}_4$ is the hypersurface $V( p_{F_{11}} )$ with
\begin{align}
\begin{split} \label{eq:pF11}
p_{F_{11}} &= s_1 e_1^2 e_2^2 e_3 e_4^4 u^3 + s_2 e_1 e_2^2 e_3^2 e_4^2 u^2 v + s_3 e_2^2 e_3^3 u v^2 + s_5 e_1^2 e_2 e_4^3 u^2 w \\
           &\qquad \qquad \qquad \qquad \qquad \qquad \qquad \qquad + s_6 e_1 e_2 e_3 e_4 u v w + s_9 e_1 v w^2 \, .
\end{split}
\end{align}
It is instructive to note that
\begin{align}
\left\{ e_1^2 e_2^2 e_3 e_4^4 u^3, e_1 e_2^2 e_3^2 e_4^2 u^2 v, e_2^2 e_3^3 u v^2, e_1^2 e_2 e_4^3 u^2 w, e_1 e_2 e_3 e_4 u v w, e_1 v w^2 \right\}
\end{align}
is a basis of $H^0 \left( \mathbb{P}_{F_{11}}, \overline{K}_{\mathbb{P}_{F_{11}}} \right)$. Since $X_5 = {B}_{3} \times \bbP_{F_{11}}$ and $s_i \in H^0( {B}_3, \overline{K}_{{B}_3} )$, it follows from the Künneth-formula that $p_{F_{11}}$ is a section of $\overline{K}_{X_5}$. Consequently, $\widehat{Y}_4$ is a smooth elliptically fibered Calabi-Yau 4-fold.

\paragraph{Gauge group, matter curves and Yukawa points}

Over $V( s_3 ) = \{ s_3 = 0 \} \subset B_3$, the fibration $\widehat{Y}_4$ admits an $SU(2)$ gauge enhancement. Similarly, there is an $SU(3)$ \mbox{enhancement} over $V( s_9 )$. The fibration $\widehat{\pi} \colon \widehat{Y}_4 \twoheadrightarrow {B}_3$ admits two independent sections $s_0 = V( v )$ and $s_1 = V( e_4 )$. We call $s_0 = V( v )$ the zero section and employ the Shioda map to associate a $U(1)$-gauge symmetry to $s_1$. Consequently, $\widehat{Y}_4$ admits an \mbox{$SU(3) \times SU(2) \times U(1)$} gauge symmetry with zero section $s_0 = V( v )$.

We label the matter curves by the representations of $SU(3) \times SU(2) \times U(1)$ in which the zero modes, localized on these curves, transform:
\begin{align}
C_{(\mathbf{3},\mathbf{2})_{1/6}} &= V( s_3, s_9 ) \, , & C_{(\mathbf{1},\mathbf{2})_{-1/2}} &= V \left( s_3, s_2 s_5^2 + s_1 ( s_1 s_9 - s_5 s_6 ) \right) \, , \\
C_{(\overline{\mathbf{3}},\mathbf{1})_{-2/3}} &= V( s_5, s_9 ) \, , & C_{(\overline{\mathbf{3}},\mathbf{1})_{1/3}} &= V \left( s_9, s_3 s_5^2 + s_6 ( s_1 s_6 - s_2 s_5 ) \right) \, , \\
C_{(\mathbf{1},\mathbf{1})_{1}} &= V( s_1, s_5 ) \, . \label{equ:MatterCurves}
\end{align}
These curves intersect in the Yukawa loci
\begin{align}
Y_1 = V( s_3, s_5, s_9 ) \, , & &Y_2 = V( s_3, s_9, s_2 s_5 - s_1 s_6 ) \, , & & Y_3 = V( s_3, s_6, s_9 ) \, , \\
Y_4 = V( s_1, s_3, s_5 ) \, , & &Y_5 = V( s_9, s_5, s_6^2 ) \, , & & Y_6 = V( s_1, s_5, s_9 ) \, .
\end{align}
We represent the intersections among the matter curves including the physically relevant self-intersections as follows:
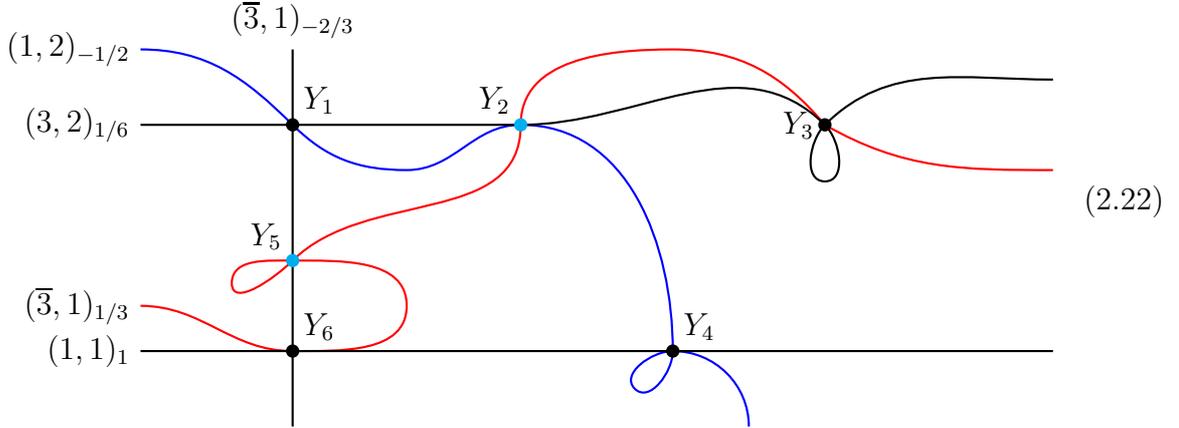
\begin{equation}
\begin{tikzpicture}[baseline=(current  bounding  box.center)]

\def\w{10};
\def\h{3};
\def\o{1};

\draw[thick,blue] (-2*\o,\o) to [out=0,in=135] (0,0)
                      to [out=-45,in=-180] (0.15*\w,-0.2*\h)
                      to [out = 0, in = -180] (0.3*\w,0)
                      to [out = 0, in = 90] (0.5*\w,-\h)
                      to [out = -90, in = -45] (0.45*\w,-\h-0.5*\o)
                      to [out = 135, in = 180] (0.5*\w,-\h)
                      to [out = 0, in = 90] (0.6*\w,-\h-\o);

\draw[thick,red] (-2*\o,-0.8*\h) to [out=0,in=-180] (0,-\h)
                      to [out=0,in=-90] (0.15*\w,-0.8*\h)
                      to [out = 90, in = 0] (0,-0.6*\h)
                      to [out = 180, in = 90] (-0.08*\w,-0.7*\h)
                      to [out = -90, in = -135] (0,-0.6*\h)
                      to [out = 45, in = -90] (0.3*\w,0)
                      to [out = 90, in = -180] (0.5*\w,\o)
                      to [out = 0, in = 130] (0.7*\w,0)
                      to [out = -30, in = -180] (\w,-0.2*\h);

\draw[thick,black] (-2*\o,0) to [out = 0, in = 180] (0.3*\w,0)
                             to [out=0,in=135] (0.7*\w,0)
                             to [out=-45,in=0] (0.7*\w,-0.25*\h)
                             to [out = 180, in = -135] (0.7*\w,0)
                             to [out = 45, in = 180] (\w,0.2*\h);

\draw[thick] (0,\o) -- (0,-\h-\o);
\draw[thick] (-2*\o,-\h) -- (\w,-\h);

\fill (0,0) circle[radius=2.5pt];
\fill[cyan] (0.3*\w,0) circle[radius=2.5pt];
\fill (0.7*\w,0) circle[radius=2.5pt];
\fill[cyan] (0,-0.6*\h) circle[radius=2.5pt];
\fill (0,-\h) circle[radius=2.5pt];
\fill (0.5*\w,-\h) circle[radius=2.5pt];

\node at (0,0) [above right] {$Y_1$};
\node at (0.3*\w,0) [above left] {$Y_2$};
\node at (0.7*\w,0) [left] {$Y_3$};
\node at (0.5*\w,-\h) [above right] {$Y_4$};
\node at (0,-0.6*\h) [above left] {$Y_5$};
\node at (0,-\h) [above right] {$Y_6$};

\node at (-2*\o,\o) [left] {$(1,2)_{-1/2}$};
\node at (-2*\o,0) [left] {$(3,2)_{1/6}$};
\node at (-2*\o,-\h) [left] {$(1,1)_{1}$};
\node at (-2*\o,-0.8*\h) [left] {$(\overline{3},1)_{1/3}$};
\node at (0,\o) [above] {$(\overline{3},1)_{-2/3}$};

\end{tikzpicture} \label{equ:IntersectionsOfCurves}
\end{equation}
The topological intersection number is $\overline{K}_{B_3}^3$ at $Y_1$, $Y_3$, $Y_4$, $Y_6$ and $2 \cdot \overline{K}_{B_3}^3$ at $Y_2$, $Y_5$.

\subsubsection{\texorpdfstring{$\mathbf{G_4}$}{G4}-flux} \label{subsubsec:G4-flux}

Let us identify the root bundles whose sections count the localized zero modes in the presence of the (candidate) $G_4$-flux introduced in \cite{Cvetic:2019gnh}. This flux is a \emph{base dependent} linear combination of the $U(1)$-flux $\omega \wedge \sigma$, where $\sigma$ is the Shioda $(1,1)$-form associated to the divisor $s_1 = V( e_4 )$, and of the matter surface flux $G_4^{(\three, \two)_{1/6}}$ on the curve $C_{(\three, \two)_{1/6}}$:
\begin{align}
G_4( a, \omega ) = a \cdot G_4^{(\three, \two)_{1/6}} + \omega \wedge \sigma \in H^{(2,2)}_{\mathrm{alg}}( \widehat{Y}_4 ) \, .
\end{align}
The parameters $a \in \mathbb{Q}$ and $\omega \in \pi^\ast \left( H^{(1,1)}( {B}_3 ) \right)$ are subjected to flux quantization, $D_3$-tadpole cancelation, masslessness of the $U(1)$-gauge boson, and exactly three chiral families on all matter curves. These conditions are solved by
\begin{align}
\omega = \frac{3}{\overline{K}_{B_3}^3} \cdot \overline{K}_{{B}_3} \, , \qquad a = \frac{15}{\overline{K}_{B_3}^3} \, .
\end{align}
Explicitly, the resulting flux candidate can be expressed as (see \oref{subsec:FluxDetails} for details)
\begin{equation}
\resizebox{0.9\textwidth}{!}{%
$\begin{aligned}
    G_4 = \frac{-3}{\KB^3} \cdot & \bigg( 5 [ e_1 ] \wedge [ e_4 ] \\
                                 & \qquad + \left. \widehat{\pi}^\ast \left( \KB \right) \wedge \left( - 3 [ e_1 ] - 2 [ e_2 ] - 6 [ e_4 ] + \widehat{\pi}^\ast \left( \KB \right) - 4 \left[ u \right] + \left[ v \right] \right) \bigg)  \right|_{\widehat{Y}_4} \, .
\end{aligned}$%
} \label{equ:G4-FieldStrength}
\end{equation}
In this expression, $[e_1] = \gamma(V(e_1)) \in H^{(1, 1)}_{\mathrm{alg}}(X_5)$ is the image of the divisor $V( e_1 ) \subseteq X_5$ under the cycle map $\gamma$. Also, we use the projection map $\widehat{\pi} \colon \widehat{Y}_4 \twoheadrightarrow {B}_3$. This $G_4$-flux candidate cancels the D3-tadpole and ensures the masslessness of a $U(1)$-gauge boson.

We must verify that the $G_4$-flux candidate in \cref{equ:G4-FieldStrength} satisfies the flux quantization $G_4 + \frac{1}{2} c_2 ( T_{\widehat{Y}_4} ) \in H^{(2,2)}_{\mathbb{Z}}( \widehat{Y}_4 )$\cite{Witten:1996md}. As a necessary check, in \cite{Cvetic:2019gnh}, the integrals of $G_4 + \frac{1}{2} c_2( T_{\widehat{Y}_4} )$ over all matter surfaces $S_{\mathbf{R}}$ and complete intersections of toric divisors were worked out. By employing the results in \cite{Collinucci:2010gz, He:2017gam}, these were found to be integral. A sufficient check for \cref{equ:G4-FieldStrength} to be properly quantized is computationally very demanding and currently beyond our arithmetic abilities. Therefore, the authors of \cite{Cvetic:2019gnh} proceeded under the assumption that this candidate $G_4$-flux is properly quantized. We will also follow this line of thought.

Furthermore, we slightly extend this result. Namely, we integrate only $c_2( T_{\widehat{Y}_4} )$ over the matter surfaces and over the complete intersections of toric divisors. By the reduction technique in \cite{Lin:2016vus}, we can relate these integrals to intersection numbers in the base $\mathcal{B}_3$. An explicit computation reveals that the only quantities which are not manifestly even are
\begin{align}\label{eq:NecessaryConditionForIntegralC2}
	 \int_{B_3} {c_2(B)\wedge \KB} \, , \qquad \int_{B_3} {\alpha \wedge (c_2(B_3) + \KB^2)} \, \text{ for all } \, \alpha \in H^{1,1}(B_3,\mathbb{Z})\, ,
\end{align}
where $c_2(B_3)$ is the second Chern class of $B_3$. For smooth 3-folds $B_3$ that appear as a base of a smooth elliptic Calabi--Yau 4-fold, it is known \cite{Collinucci:2010gz} that $c_2(B_3) + \KB^2$ is an \emph{even} class. Furthermore, \cite{He:2017gam} states that $\int_{B_3}c_2(B_3)\cdot\KB=24$ is \emph{even} as well. It thus follows that $c_2( T_{\widehat{Y}_4} )$ passes the necessary conditions for being even. Likewise, we can integrate the $G_4$-flux candidate \cref{equ:G4-FieldStrength} over the matter surfaces and over the complete intersections of toric divisors. All of those are found to be integral. Since a sufficient check is currently beyond our arithmetic abilities, we proceed under the assumption that $c_2( T_{\widehat{Y}_4} )$ is even and that the $G_4$-flux candidate \cref{equ:G4-FieldStrength} is integral.

It should be mentioned that the $G_4$-flux (candidate) \cref{equ:G4-FieldStrength} was chosen so that the F-theory Standard Model vacua are stable, that is the D3-tadpole can be canceled. This requires
\begin{align}
n_{D3} = \frac{\chi \left( T_{\widehat{Y}_4} \right)}{24} - \frac{1}{2} \int \limits_{\widehat{Y}_4}{G_4 \wedge G_4} \stackrel{!}{\in} \mathbb{Z}_{\geq 0} \, . \label{equ:Tadpole}
\end{align}
Moreover, the masslessness condition for the $U(1)$-gauge boson was enforced:
\begin{align}
\begin{split}
    \forall \eta \in H^{1,1}(B_3) : \quad \int_{Y_4} G_4 \wedge \sigma \wedge \pi^*\eta \stackrel{!}{=} 0 \, . \label{equ:Massless}
\end{split}
\end{align}
Here, $\sigma$ is the $(1,1)$-form that relates to the so-called \emph{Shioda-divisor} associated with the $U(1)$ \cite{Park:2011ji,Morrison:2012ei}.

\subsubsection{Zero modes and root bundles} \label{subsubsec:BundleExpressions}

We now discuss the zero modes in the presence of the flux in \oref{equ:G4-FieldStrength}. As \mbox{explained} in \oref{subsec:RootBundleAppearance}, we thus look for a lift to $H^4_D( \widehat{Y}_4, \mathbb{Z}(2) )$ in the diagram \linebreak \oref{equ:CommutativeDiagramLifts}. For computational simplicity, we aim to parametrize such a lift as $A = \widehat{\gamma}( \mathcal{A} )$ with $\mathcal{A} \in\mathrm{CH}^2( \widehat{Y}_4, \mathbb{Z} )$. To describe a candidate, we recall that the cycle map $\gamma \colon \mathrm{CH}^2( \widehat{Y}_4, \mathbb{Z} ) \to H^{2,2}_{\text{alg}}( \widehat{Y}_4 )$ is a ring homomorphism in which the intersection product in $\mathrm{CH}^*(X_5, \mathbb{Z})$ is compatible with the cup product in $H^{*}(X_5, \mathbb{C})$. By De Rham's theorem and the Hodge decomposition, it follows that
\begin{align}
H^{2k}(X_5, \mathbb{C}) \cong H^{2k}_{DR}(X_5,\mathbb{C}) = \bigoplus_{p + q = 2k} H^{p, q}(X_5) \, .
\end{align}
The cup product in $H^{*}(X_5, \mathbb{C})$ respects the grading, and restricts to the wedge product of $(p, q)$-forms. For any two divisors $V(r)$ and $V(s)$ on $X_5$, it therefore follows that \mbox{$\gamma(V(r, s)) = \gamma(V(r) \cdot V(s)) = [r] \wedge [s].$}  This also shows that $[r] \wedge [s] \in H^{2, 2}_{\mathrm{alg}}(\widehat{Y}_4)$ is in the image of $\gamma$. With this in mind, it is natural to consider
\begin{align}
\begin{split}
  \mathcal{A}^\prime = -3 \cdot & \bigg( 5 V( e_1, e_4 ) - 3 V( e_1, t_1 ) - 2 V( e_2, t_2 ) - 6 V( e_4, t_3 ) \\
                                    & \qquad \qquad \left. + V( t_4, t_5 ) - 4 V( t_6, u ) + V( t_6, v ) \bigg) \right|_{\widehat{Y}_4} \in \mathrm{CH}^2( \widehat{Y}_4, \mathbb{Z} ) \, ,
\end{split} \label{equ:G4-Potential}
\end{align}
where $t_i \in H^0( X_5, \alpha^\ast( \KB ) )$ and $\alpha \colon X_5 = {B}_3 \times \mathbb{P}_{F_{11}} \twoheadrightarrow {B}_3$. Note that $\gamma( \mathcal{A}^\prime ) = \KB^3 \cdot G_4$. Therefore, the gauge potential $A^\prime = \widehat{\gamma}( \mathcal{A}^\prime )$ would induce chiral exotics unless we ``divide" it by $\xi = \KB^3$. Hence, we are led to consider gauge potentials $A = \widehat{\gamma}( \mathcal{A} ) \in H^4_D( \widehat{Y}_4, \mathbb{Z}(2) )$ with
\begin{align}
\gamma( \mathcal{A} ) = G_4 \, , \qquad \xi \cdot \widehat{\gamma}( \mathcal{A} ) \sim \widehat{\gamma}( \mathcal{A}^\prime ) \, .
\end{align}
Hence, we can infer that the line bundles induced from $A = \widehat{\gamma}(\mathcal{A})$ are $\overline{K}_{B_3}^3$-th roots of the ones induced from $A^\prime = \widehat{\gamma}( \mathcal{A}^\prime )$. We can explicitly compute the latter from \cref{equ:G4-Potential} and \cref{equ:DivisorFromChow}. As an example, let us consider the curve $C_{(\three, \two)_{1/6}}$. For this curve, we find (details in \oref{sec:LineBundleComputations})
\begin{align}
D_{(\three, \two)_{1/6}} \left( \mathcal{A}^\prime \right) = 3 \cdot V( t,s_3,s_9) = 3 \cdot \left. \KB \right|_{C_{(\three, \two)_{1/6}}} \, , \qquad t \in H^0( {B}_3, \KB ) \, ,
\end{align}
where the last equality follows from the adjunction formula. From this, we conclude that $D_{(\three, \two)_{1/6}} \left( \mathcal{A} \right)$ satisfies
\begin{align}
\xi \cdot D_{(\three, \two)_{1/6}} \left( \mathcal{A} \right) = \KB^3 \cdot D_{(\three, \two)_{1/6}} \left( \mathcal{A} \right) &\sim D_{(\three, \two)_{1/6}} \left( \mathcal{A}^\prime \right) = 3 \cdot \left. \KB \right|_{C_{(\three, \two)_{1/6}}} \, . \label{equ:FirstRootEquation}
\end{align}
The zero modes in the representation $(\three, \two)_{1/6}$ are counted by the tensor product of the line bundle associated with $D_{(\three, \two)_{1/6}} \left( \mathcal{A} \right)$ and a spin bundle. 
Let us emphasize again that we wish to provide a systematic study of all $\xi$-th roots of $D_{(\three, \two)_{1/6}} \left( \mathcal{A}^\prime \right)$ and all the spin bundles on $C_{(\three, \two)_{1/6}}$. To this end, recall the defining property of spin bundles on $C_{(\three, \two)_{1/6}}$:
\begin{align}
2 \cdot D^{\text{spin}}_{(\three, \two)_{1/6}} \sim {K}_{(\three, \two)_{1/6}} \sim \left. \KB \right|_{C_{(\three, \two)_{1/6}}} \, . \label{equ:SpinQuarkDoublet}
\end{align}
Consequently, we notice
\begin{equation}
  \resizebox{0.9\textwidth}{!}{%
$\begin{aligned}
2 \KB^3 \cdot \left( D_{(\three, \two)_{1/6}} \left( \mathcal{A} \right) + D^{\text{spin}}_{(\three, \two)_{1/6}} \right) &\sim 2 \cdot \left( \KB^3 \cdot D_{(\three, \two)_{1/6}} \left( \mathcal{A} \right) \right) + \KB^3 \cdot \left( 2 \cdot D^{\text{spin}}_{(\three, \two)_{1/6}} \right) \\
&\sim \left( 6 + \KB^3 \right) \cdot \left. \KB \right|_{C_{(\three, \two)_{1/6}}} \, . \label{equ:LineBundleQuarkDoublet}
\end{aligned}$%
}
\end{equation}
Expressed in line bundles, we thus conclude that\footnote{Inspired by \textgreek{Ρίζα}, which is greek for root, $P$ refers to root bundles throughout this article.}
\begin{align}
\begin{split}
P_{(\three, \two)_{1/6}}^{\otimes 2 \KB^3} \sim \left( \left. \KB \right|_{C_{(\three, \two)_{1/6}}} \right)^{\otimes \left( 6 + \KB^3 \right)} \sim K_{(\three, \two)_{1/6}}^{\otimes \left( 6 + \KB^3 \right)} \, .
\end{split}
\end{align}
By repeating this computation for the other matter curves, one finds the following root bundle constraints:
\begin{align}
\begin{tabular}{cp{0.2em}l}
\toprule
curve & & root bundle constraint \\
\midrule
$C_{(\mathbf{3},\mathbf{2})_{1/6}} = V( s_3, s_9 )$ & & $P_{(\three,\two)_{1/6}}^{\otimes 2 \KB^3} = K_{{(\three,\two)_{1/6}}}^{\otimes \left( 6 + \KB^3 \right)}$ \\
$C_{(\mathbf{1},\mathbf{2})_{-1/2}}= V \left( s_3, P_H \right)$ & & $P_{(\mathbf{1},\mathbf{2})_{-1/2}}^{\otimes 2 \KB^3} = K_{{(\mathbf{1},\mathbf{2})_{-1/2}}}^{\otimes \left( 4 + \KB^3 \right)} \otimes \mathcal{O}_{C_{(\mathbf{1},\mathbf{2})_{-1/2}}} \left( - 30 \cdot Y_1 \right)$ \\
$C_{(\overline{\mathbf{3}},\mathbf{1})_{-2/3}} = V( s_5, s_9 )$ & & $P_{(\overline{\mathbf{3}},\mathbf{1})_{-2/3}}^{\otimes 2 \KB^3} = K_{{(\overline{\mathbf{3}},\mathbf{1})_{-2/3}}}^{\otimes \left( 6 + \KB^3 \right)}$ \\
$C_{(\overline{\mathbf{3}},\mathbf{1})_{1/3}} = V \left( s_9, P_R \right)$ & & $P_{(\overline{\mathbf{3}},\mathbf{1})_{1/3}}^{\otimes 2 \KB^3} = K_{{(\overline{\mathbf{3}},\mathbf{1})_{1/3}}}^{\otimes \left( 4 + \KB^3 \right)} \otimes \mathcal{O}_{C_{(\overline{\mathbf{3}},\mathbf{1})_{1/3}}} \left( - 30 \cdot Y_3 \right)$ \\
$C_{(\mathbf{1},\mathbf{1})_{1}} = V( s_1, s_5 )$ & & $P_{(\mathbf{1},\mathbf{1})_{1}}^{\otimes 2 \KB^3} = K_{{(\mathbf{1},\mathbf{1})_{1}}}^{\otimes \left( 6 + \KB^3 \right)}$ \\
\bottomrule
\end{tabular} \label{equ:GeneralRootBundleExpressions}
\end{align}
In this table, we use $P_H = s_2 s_5^2 + s_1 ( s_1 s_9 - s_5 s_6 )$ and $P_R = s_3 s_5^2 + s_6 ( s_1 s_6 - s_2 s_5 )$. Note that the line bundles on the Higgs curve $C_{(\mathbf{1},\mathbf{2})_{-1/2}}$ and the curve $C_{(\overline{\mathbf{3}},\mathbf{1})_{1/3}}$ depend on the Yukawa points $Y_1 = V( s_3, s_5, s_9 )$ and $Y_3 = V( s_3, s_6, s_9 )$ (see \oref{sec:LineBundleComputations} for details). It must also be noted that for two divisors $D$ and $E$, 
\begin{align}
D \sim E \qquad \Rightarrow \qquad n \cdot D \sim n \cdot E \, .
\end{align}
The converse is not true. This is why we do not cancel common factors. Finally, let us point out that the toric base spaces for these F-theory Standard Model constructions must satisfy $\KB^3 \in \{ 6, 10, 18, 30 \}$ \cite{Cvetic:2019gnh}. We provide an explicit list of the root bundles constraints for these values of $\KB^3$ in \oref{subsec:LineBundlesKbarCubeDifferent}.

\section{Root bundles from limit roots} \label{sec:LimitRoots}

In the previous section, we explained that root bundles feature prominently in \mbox{vector-like} spectra in F-theory. For the largest currently-known class of globally consistent \mbox{F-theory} Standard Model constructions without chiral exotics and gauge coupling unification \cite{Cvetic:2019gnh}, we have worked out these bundle expressions explicitly and summarized them in \mbox{\oref{subsec:LineBundlesKbarCubeDifferent}}. In aiming for MSSM constructions, i.e., vacua without vector-like \mbox{exotics}, the cohomologies of root bundles beg to be investigated. Therefore, our goal is to \mbox{construct} roots whose number of global sections is exactly the amount required by the physical considerations.

Before we exhibit an example of this in \oref{sec:ApplicationToMSSMs}, we first summarize well-known facts about root bundles in general. In particular, we outline an argument for the existence of such bundles on smooth, irreducible curves. From this argument, it can be extrapolated that explicit constructions of root bundles on smooth, irreducible curves -- not to mention an explicit count of their sections -- are very challenging at best. Fortunately, we can employ deformation theory to simplify the task. Namely, it is possible to relate root bundles on smooth, irreducible curves to so-called limit roots on nodal curves. This follows from the detailed study in \cite{2004math4078C}. For convenience to the reader, we summarize the essential steps in these limit root constructions before we extend these ideas. Namely, we provide a simple counting procedure for the global sections of many limit root bundles. Our analysis in \oref{sec:ApplicationToMSSMs} will employ exactly this counting strategy in order to gain insights into the vector-like spectra of F-theory Standard Models.

\subsection{Root bundles} \label{subsec:MthRootsOfDivisors}

Let us look at root bundles on a smooth, complete Riemann surface (or curve) $C$ of genus $g$. We focus on a line bundle $L \in \mathrm{Pic}(C)$ and an integer $n$ with $n \geq 2$, and $n|\mathrm{deg}(L)$. We first recall the following definition.

\begin{defi}[$n$-th root bundle]
An \emph{$n$-th root bundle of $L$} is a line bundle $P$ such that $P^n \cong L$. Collectively, we denote the $n$-th roots of $P$ by $\mathrm{Roots} ( n, L )$
\end{defi}

Equivalently, in the language of divisors, an $n$-th root $D$ of a divisor $\widetilde{D}$ is a \mbox{divisor} such that $nD \sim \widetilde{D}$. The first important result about root bundles concerns their \mbox{existence}. While this seems to be a well-known fact, we were surprised to notice that well-established references, such as \cite{10.2307/j.ctt1b9x2g3, freitag2011complex, Griffiths:433962}, do not give an explicit proof. Not only does the proof nicely illustrates the challenge in constructing root bundles on high genus curves, it also allows us to easily understand why there are $n^{2g}$ root bundles and why their differences are torsion divisors. For all these reasons, let us give a proof for the existence of root bundles on smooth, complete Riemann surfaces.

\begin{prop} \label{prop:ExistenceOfRoots}
Let $n \in \mathbb{Z}$ with $n \geq 2$. For every $L \in \mathrm{Pic}(C)$, there exists an $n$-th root bundle $P$ of $L$ if and only if $n|\mathrm{deg}(L)$.
\end{prop}

\begin{myproof} For the forward direction, if there exists an $n$-th root bundle $P$ such that $P^n \cong L$, then $n\deg(P) = \deg(L)$ and $n |\deg(L)$. Conversely, suppose that $n| \mathrm{deg}(L)$. Recall that $J(C)$ is a complex torus of the form $V/\Lambda$, where $V$ is a vector space of dimension $g$, and $\Lambda$ is a discrete subgroup of $V$ of rank $2g$. Denote the $n$-fold tensor product of line bundles by $[n]: P \mapsto nP$.

First, we will describe some properties of the map $[n]: J(C) \rightarrow J(C)$. Observe that its kernel is given by
\begin{align}
\ker([n]) = \{P \in  V/\Lambda: nP + \Lambda = \Lambda\} =  ((1/n)\Lambda)/\Lambda \cong \Lambda/n\Lambda \cong (\mathbb{Z}/n\mathbb{Z})^{2g} \, ,
\end{align}
because $\Lambda$ is a discrete subgroup of rank $2g$. Hence, $\ker([n]) = [n]^{-1}(0)$ is finite, and has dimension 0. Since $J(C)$ is a complete variety, its image $[n](J(C))$ is a closed subvariety. For any $a \in [n](J(C))$, the translation map
\begin{align}
t_a: [n]^{-1}(0) \rightarrow [n]^{-1}(a), \quad x \mapsto x + a \, , \label{equ:Iso}
\end{align}
is an isomorphism. It follows from the dimension formula that
\begin{align}
0 = \dim [n]^{-1}(0) = \dim [n]^{-1}(a) \geq \dim(J(C)) - \dim [n](J(C)) \geq 0 \, .
\end{align}
Hence, $[n]: J(C) \rightarrow J(C)$ is surjective.

Now, consider the following commutative diagram, where $\deg$ is the degree map, and $(\times n)$ is the multiplication of integers by $n$.
\begin{align}
\begin{tikzpicture}[node distance=1.8cm, auto, baseline=(current  bounding  box.center)]
\node (A) {$0$};
\node (B) [right of = A] {$J(C)$};
\node (C) [right of = B] {$\mathrm{Pic}(C)$};
\node (D) [right of = C] {$\mathbb{Z}$};
\node (E) [right of = D] {$0$};
\node (F) [below of = A] {$0$};
\node (G) [right of = F] {$J(C)$};
\node (H) [right of = G] {$\mathrm{Pic}(C)$};
\node (I) [right of = H] {$\mathbb{Z}$};
\node (J) [right of = I] {$0$};
\draw[->] (A) to node {} (B);
\draw[->] (B) to node {} (C);
\draw[->] (C) to node {$\deg$} (D);
\draw[->] (D) to node {} (E);
\draw[->] (F) to node {} (G);
\draw[->] (G) to node {} (H);
\draw[->] (H) to node {$\deg$} (I);
\draw[->] (I) to node {} (J);
\draw[->] (B) to node {$[n]$} (G);
\draw[->] (C) to node {$[n]$} (H);
\draw[->] (D) to node {$\times n$} (I);
\end{tikzpicture}
\end{align}
Applying the Snake Lemma yields an exact sequence of the cokernels of the vertical maps, i.e.
\begin{align}
0 = J(C)/[n](J(C)) \rightarrow  \mathrm{Pic}(C)/[n](\mathrm{Pic}(C)) \xrightarrow{\cong} \mathbb{Z}/n\mathbb{Z} \rightarrow 0 \, ,
\end{align}
where the isomorphism is provided by the degree map. Since $n|\deg(L),$ we have that $L \in [n](\mathrm{Pic}(C))$. So, there exists $P \in \mathrm{Pic}(C)$ such that $P^n \cong L,$ i.e. an $n$-th root bundle.
\end{myproof}

There are two important lessons that we can learn from this proposition. First, there are $n^{2g}$ $n$-th root bundles $P$ of $L$. This is because \oref{equ:Iso} is an isomorphism, and so,
\begin{align}
t_{\deg(L)}: (\mathbb{Z}/n\mathbb{Z})^{2g} \cong \ker([n]) = [n]^{-1}(0)  \rightarrow [n]^{-1}(\deg(L)) = \mathrm{Roots}(n, L) \, ,
\end{align}
is an isomorphism as well. If $L = K_C$, then the $n$-th roots are called \emph{$n$-spin bundles} \cite{natanzon2016higher}.

The second lesson concerns the difference between two root bundles. Any two $n$-th root bundles differ by an $n$-torsion line bundle, i.e. a line bundle $M \in J(C)$ such that $M^n \cong \mathcal{O}_C$.

The Jacobian and the linear equivalence of divisors is well-understood for elliptic curves $E$ (see \cite{10.2307/j.ctt1b9x2g3, freitag2011complex, Griffiths:433962} for background). This allows us to exhibit examples of the above notions fairly explicitly. First, recall that $E \cong J( E ) \cong \mathbb{C} / \Lambda$ and $\Lambda = \mathbb{Z} \oplus \mathbb{Z} \cdot \tau$, where $\tau \in \mathbb{C}$ is the complex structure modulus of the elliptic curve. We denote $D \in \mathrm{Div}( E )$ by
\begin{align}
D = \sum_{i = 1}^{n}{n_i \cdot \left( p_i \right)} \, , \qquad n_i \in \mathbb{Z}, \quad p_i \in E \, ,
\end{align}
i.e. we place the points $p_i \in E$ in round brackets for notational clarity. Note that $\mathrm{deg}( D ) = \sum_{i = 1}^{n}{n_i}$ and that a zero degree divisor satisfies
\begin{align}
D \sim 0 \quad \Leftrightarrow \quad \left( \sum_{i = 1}^{n}{n_i \cdot p_i} \right) \in \Lambda \, .
\end{align}
From this, we can work out the divisor classes of the 2-torsion divisors in $J( E )$:
\begin{align}
D_1 &= \left[ 0 \right] \, , & D_2 &= \left[ -1 \cdot ( 0 ) + 1 \cdot \left( \frac{1}{2} \right) \right] \, , \\
D_3 &= \left[ -1 \cdot ( 0 ) + 1 \cdot \left( \frac{\tau}{2} \right) \right] \, , & D_4 &= \left[ -1 \cdot ( 0 ) + 1 \cdot \left( \frac{1+\tau}{2} \right) \right] \, .
\end{align}
It follows $\mathrm{ker} ( [ 2 ] ) \cong \left( \mathbb{Z} / 2 \mathbb{Z} \right)^2$, which we can intuitively collect in the following picture:
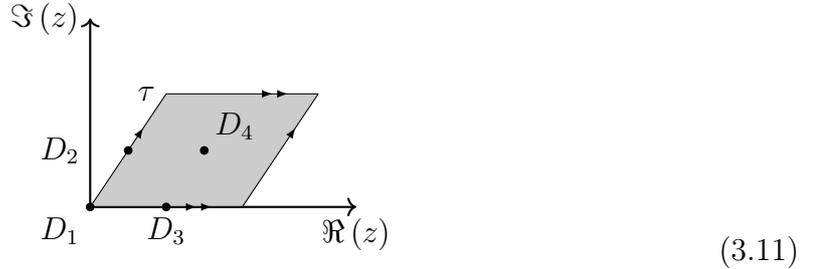
\begin{equation}
\begin{tikzpicture}[scale=1]

\draw[->, thick] (0,0) -- (0,2.5);
\node [left] at (0, 2.5) {$\Im \left( z \right)$};
\draw[->,thick] (0,0)--(3.5,0);
\node [below] at (3.5, 0) {$\Re \left( z \right)$};

\fill[gray!80,opacity=0.5] (0,0)--(1,1.5)--(3,1.5)--(2,0)--(0,0);

\draw[postaction={decorate},decoration={markings,mark=at position .7 with {\arrow{latex}}}](0,0) -- (1,1.5);
\draw[postaction={decorate},decoration={markings,mark=at position .7 with {\arrow{latex}}}](2,0) -- (3,1.5);
\draw[postaction={decorate},decoration={markings,mark=at position .7 with {\arrow{latex}},mark=at position .8 with {\arrow{latex}}}](0,0) -- (2,0);
\draw[postaction={decorate},decoration={markings,mark=at position .7 with {\arrow{latex}},mark=at position .8 with {\arrow{latex}}}](1,1.5) -- (3,1.5);

\node [below left] at (0, 0) {$D_1$};
\draw [fill] (0,0) circle [radius = 0.05];

\node [left] at (0,0.75) {$D_2$};
\draw [fill] (0.5,0.75) circle [radius = 0.05];

\node [below] at (1,0) {$D_3$};
\draw [fill] (1,0) circle [radius = 0.05];

\node [above right] at (1.5,0.75) {$D_4$};
\draw [fill] (1.5,0.75) circle [radius = 0.05];

\node [left] at (1, 1.5) {$\tau$};

\end{tikzpicture}
\end{equation}
Note that $\{ D_1, D_2, D_3, D_4 \}$ are exactly the four spin structures on $E$.

In making contact with our physics applications, we should next investigate the sheaf cohomologies of root bundles. Generally speaking, this is a very challenging task. \mbox{However}, on elliptic curves, the situation simplifies and we can achieve a complete \mbox{classification}. Recall that any line bundle $L \in \mathrm{Pic}(E)$ with $\mathrm{deg} ( L ) \neq 0$ is in the Kodaira stable regime, i.e. we can infer its cohomologies from its degree. For $\mathrm{deg}( L ) = 0$ we have
\begin{align}
h^0(E, L) = 1 \, \quad \Leftrightarrow \quad L \cong \mathcal{O}_E \, .
\end{align}
Therefore, it merely remains to study the cohomologies of roots $P$ of a line bundle $L \in \mathrm{Pic}(E)$ with $\mathrm{deg} ( L ) = 0$. This is achieved by the following proposition.

\begin{prop} \label{prop:RootsElliptic}
Let $L \in \mathrm{Pic}(E)$ with $\deg(L) = 0$ and consider an integer $n$ with $n \geq 2$. Then,
\begin{enumerate}[(i)]
    \item $L \cong \mathcal{O}_E$: Exactly one $n$-th root $P$ of $L$ has $h^0(E, P) = 1$, and the remaining $n$-th roots $Q$ have $h^0(E, Q) = 0$.
    \item $L \not \cong \mathcal{O}_E$: All $n$-th roots $P$ of $L$ have $h^0(E, P) = 0$.
\end{enumerate}
\end{prop}

\begin{myproof}
\begin{enumerate}[(i)]
    \item Since $\mathcal{O}_E$ is an $n$-th root of itself, there is one $n$-th root $P = \mathcal{O}_E$ with $h^0(E, P) = 1$. Any other $n$-th root $Q$ differs from $P$ by a non-trivial $n$-torsion line bundle. As such, $Q$ is non-trivial and has $h^0(E, Q) = 0$.
    \item $L$ is non-trivial. Hence, all $n$-th roots $P$ are non-trivial and have $h^0(E, P) = 0$. \qedhere
\end{enumerate}
\end{myproof}

For example, among the $9$ 3rd roots $P$ of a line bundle $L \in \mathrm{Pic}( E )$ with $\mathrm{deg} ( L ) = 0$, at least $8$ have $h^0( E, P ) = 0$. We will make use of this simple result in \oref{subsec:RootBundlesF-theoryMSSMs}.

\subsection{Deformation theory and global sections} \label{subsec:Approximation}

For applications in F-theory, we wish to generalize \oref{prop:RootsElliptic} to matter curves $C_{\mathbf{R}}$ with $g > 1$. Unfortunately on such curves, it is very hard to tell if a divisor is linearly equivalent to zero. This is due to the current lack of practical understanding of the Abel-Jacobi map $ \text{Div}_0(C_{\mathbf{R}}) \to J( C_{\mathbf{R}} )$, whose kernel is exactly given by the (classes of) trivial divisors. This in turn makes it very challenging to identify $n$-torsion bundles, which forms a measure-$0$ subset of the Jacobian $J( C_{\mathbf{R}} )$. Consequently, it becomes almost impossible to explicitly identify a single $n$-th root bundle $P$ of a line bundle $L$ on $C_{\mathbf{R}}$.

To overcome this hurdle, we wonder if it is possible to simplify the matter curves $C_{\mathbf{R}}$. However, recall that the geometry of the matter curves is dictated by that of the elliptic fibration $\widehat{\pi} \colon \widehat{Y}_4 \twoheadrightarrow {B}_3$. Therefore, even though special, non-generic elliptic fibrations $\widehat{Y}_4$ may contain matter curves $C_{\mathbf{R}}$ with simple geometries, it can be expected that such fibrations lead to physically unwanted gauge enhancements.

Therefore, in order to remain on physically solid grounds, we stick to the geometry of the matter curves $C_{\mathbf{R}}$ as enforced by the generic fibration $\widehat{Y}_4$. In this situation, there is still a way to improve our situation. Namely, suppose that $\varphi \colon C^{\text{simple}}_{\mathbf{R}} \to C_{\mathbf{R}}$ is a deformation of a curve $C^{\text{simple}}_{\mathbf{R}}$, whose simple geometry allows easy access to root bundles and their cohomologies, into the actual physical matter curve $C_{\mathbf{R}} \subset \widehat{Y}_4$. Then, we can wonder if the root bundles $P^{\text{simple}}_{\mathbf{R}}$ on $C^{\text{simple}}_{\mathbf{R}}$ approximate the roots $P_{\mathbf{R}}$ on $C_{\mathbf{R}}$.

In general, this sort of question leads to a deep discussion of deformation theory (see e.g. \cite{hartshorne2009deformation, greuel2007introduction} for a modern exposition). In this work, we will not attempt to give a complete answer. Rather, we make a special choice for $C^{\text{simple}}_{\mathbf{R}}$. Inspired by \cite{2004math4078C}, we focus on curves $C_{\mathbf{R}}^{\text{simple}}$ with singularities, which locally look like $\left\{ x \cdot y = 0 \right\}$, i.e. are nodes. On such nodal curves $C^\bullet_{\mathbf{R}}$, roots $P^\bullet_{\mathbf{R}}$ admit a description in terms of weighted diagrams \cite{2004math4078C}. Even more, there are exactly as many roots $P_{\mathbf{R}}^\bullet$ as there are roots $P_{\mathbf{R}}$ and we can, at least in theory, identify them with each other by tracing them along the deformation $C^\bullet_{\mathbf{R}} \to C_{\mathbf{R}}$.

That said, the next question is in what sense we can use the roots $P_{\mathbf{R}}^\bullet$ on $C_{\mathbf{R}}^\bullet$ to approximate the cohomologies of the roots $P_{\mathbf{R}}$ on the physical matter curve $C_{\mathbf{R}}$. To this end, we first recall that refined section counting mechanisms exist for line bundles on singular curves \cite{Bies:2020gvf}. In exactly this spirit, we are able to extend the ideas from \cite{2004math4078C}. In \oref{subsec:LimitRoots} we will argue that it is often possible to count the number of global sections of roots $P_{\mathbf{R}}^\bullet$ on a nodal curve $C_{\mathbf{R}}^\bullet$ from simple combinatorics.

It now remains to relate the cohomologies $h^i( C^\bullet_{\mathbf{R}}, P^\bullet_{\mathbf{R}} )$ to $h^i( C_{\mathbf{R}}, P_{\mathbf{R}} )$. Since, the chiral index is fixed from topology, it suffices to study how $h^0( C^\bullet_{\mathbf{R}}, P^\bullet_{\mathbf{R}} )$ relates to $h^0( C_{\mathbf{R}}, P_{\mathbf{R}} )$. Since $C^\bullet_{\mathbf{R}}$ is singular (and therefore non-generic) and $C_{\mathbf{R}}$ expected to be smooth, a tendency is known. This tendency goes by the name upper semi-continuity. It means that the number of global sections of $P_{\mathbf{R}}$ must not increase when traced along $C^\bullet_{\mathbf{R}} \to C_{\mathbf{R}}$ to the root $P_{\mathbf{R}}$, i.e.
\begin{align}
h^0 \left( C_{\mathbf{R}}, P_{\mathbf{R}} \right) \leq h^0 \left( C^\bullet_{\mathbf{R}}, P^\bullet_{\mathbf{R}} \right) \, .
\end{align}
It is a very interesting but also very challenging question to distinguish the roots $P^\bullet_{\mathbf{R}}$ that lose sections along $C^\bullet_{\mathbf{R}} \to C_{\mathbf{R}}$ from the roots with a constant number of sections. While we hope to return to this question in the future, the physics applications in \oref{sec:ApplicationToMSSMs} focus on a subset of roots, which do not lose sections. Namely, if $\chi \left( P^\bullet_{\mathbf{R}} \right) \geq 0$ and
\begin{align}
h^0 \left( C^\bullet_{\mathbf{R}}, P^\bullet_{\mathbf{R}} \right) = \chi \left( P^\bullet_{\mathbf{R}} \right) \, ,
\end{align}
then $P^\bullet_{\mathbf{R}}$ cannot lose sections, since its numbers of sections is already minimal. In particular, it then holds $h^0 \left( C_{\mathbf{R}}, P_{\mathbf{R}} \right) = h^0 \left( C^\bullet_{\mathbf{R}}, P^\bullet_{\mathbf{R}} \right)$.

\subsection{Limit roots} \label{subsec:LimitRoots}

In \oref{prop:ExistenceOfRoots} we saw that $n$-th roots $P$ of a line bundle $L$ on a smooth curve $C$ exist if $\deg(L)$ is divisible by $n$. This is not the case for reducible, nodal curves $C^\bullet$. Indeed, a root $P^\bullet$ of a line bundle $L^\bullet$ on such curves should restrict to a root on the irreducible components. However, even if $n$ divides the degree of $L^\bullet$, it may not divide $\deg(L|_Z)$ for some irreducible component $Z$ of $C^\bullet$. This is elegantly circumvented by passing to limit $n$-th root bundles $P^\circ$ on (partial) blow-ups $C^\circ$ of $C^\bullet$, as originally introduced in \cite{2004math4078C}. Just as every nodal curve $C^\bullet$ can be described through its dual graph, these limit $n$-th root bundles $P^\circ$ are determined by weighted graphs. This combinatorial data can be exploited to make the task of section-counting more tractable. For convenience to the reader, let us outline the important steps in these constructions before we explain the section counting for limit roots. For more material on limit roots, we refer the reader to \cite{Jarvis2001, 1998math......9138J} in which the pushforwards of these limits roots along the blow-up map, and their moduli are extensively studied.

\subsubsection{Nodal curves and blow-ups}
A point is a node if it has a neighborhood where the curve locally looks like $\{ xy = 0 \}$ in $\mathbb{C}^2$. A \emph{nodal curve} is a complete algebraic curve such that every point is either smooth, or a node. Let $C^\bullet$ be a connected (possibly reducible) nodal curve of arithmetic genus $g$. We associate to $C^\bullet$ a \emph{dual graph} $\Pi_{C^\bullet}$ in which
\begin{enumerate}[(i)]
\item every vertex corresponds to an irreducible component $C^\bullet_i$ of $C^\bullet$,
\item every half-edge emanating from a vertex $C^\bullet_i$ is a node on $C^\bullet_i$.
\end{enumerate}
If a node lies on both $C^\bullet_i$ and $C^\bullet_j$, then the half-edges exiting from $C^\bullet_i$ and $C^\bullet_j$ join together to form an edge. For example, consider the \emph{Holiday lights} -- a nodal curve $H^\bullet$ with 11 components given by:\footnote{In all base spaces ${B}_3$ of the globally consistent F-theory Standard Model constructions discussed in \oref{subsec:RootBundlesInF-theoryMSSM} and originally introduced in \cite{Cvetic:2019gnh}, the matter curves $C_{\mathbf{R}}$ are contained in K3-surfaces. Motivated from \cite{Farkas_2017}, it stands to wonder if the matter curves $C_{\mathbf{R}}$ admit a deformation into such a \emph{Holiday lights}. Even more, \emph{Holiday lights} allow easy access to Brill-Noether theory of limit roots as we will see momentarily. As such, they are very favorable nodal curves for our study. We hope to return to this question in the future.}
\begin{itemize}
 \item a rational curve $\Gamma$ with genus $g(\Gamma) = 0$,
 \item 10 elliptic curves $E_1, \dots, E_{10}$ with genus $g(E_i) = 1$.
\end{itemize}
Each elliptic curve intersects no other curve except $\Gamma$, hence the name \emph{Holiday lights}. Its dual graph can be visualized as follows:
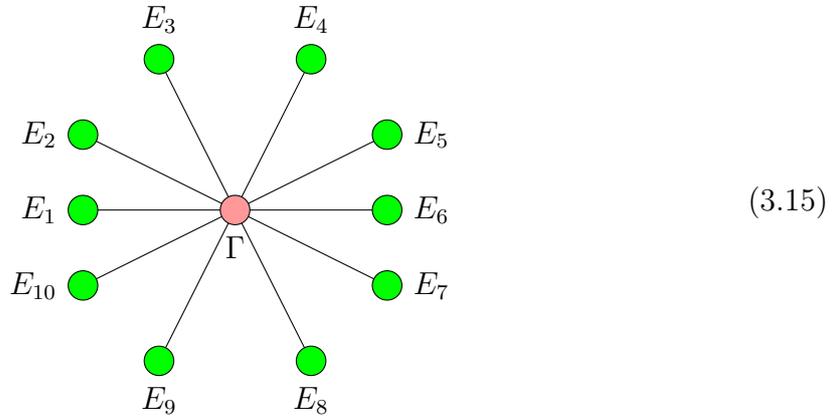
\begin{equation}
\begin{tikzpicture}[baseline=(current  bounding  box.center)]

\def\s{2};
\draw (-\s,0) -- (\s,0);
\draw (-\s,0.5*\s) -- (\s,-0.5*\s);
\draw (-\s,-0.5*\s) -- (\s,0.5*\s);
\draw (-0.5*\s,\s) -- (0.5*\s,-\s);
\draw (0.5*\s,\s) -- (-0.5*\s,-\s);

\node at (0,0) [stuff_fill_red, label=below:$\Gamma$] {};
\node at (-\s,0)  [stuff_fill_green, label=left:$E_1$] {};
\node at (-\s,0.5*\s) [stuff_fill_green, label=left: $E_2$] {};
\node at (-0.5*\s,\s) [stuff_fill_green, label=above:$E_3$] {};
\node at (0.5*\s,\s) [stuff_fill_green, label=above:$E_4$] {};
\node at (\s,0.5*\s) [stuff_fill_green, label=right:$E_5$] {};
\node at (\s,0) [stuff_fill_green, label=right:$E_6$] {};
\node at (\s,-0.5*\s) [stuff_fill_green, label=right:$E_7$] {};
\node at (0.5*\s,-\s) [stuff_fill_green, label=below:$E_8$] {};
\node at (-0.5*\s,-\s) [stuff_fill_green, label=below:$E_9$] {};
\node at (-\s,-0.5*\s)[stuff_fill_green, label=left:$E_{10}$] {};
\end{tikzpicture}
\end{equation}
Each elliptic curve $E_i$ is represented by a green vertex, while the rational curve $\Gamma$ is represented by the pink vertex.

If $\pi: C^\circ \rightarrow C^\bullet$ is a blow-up of $C^\bullet$, then for every node $n_i \in C^\bullet$, we denote the exceptional components by
\begin{align}
\pi^{-1}(n_i) = \mathcal{E}_i \cong \mathbb{P}^1 \, .
\end{align}
Set $C^N = \overline{C^\circ \setminus\cup_i \mathcal{E}_i}$. Then, $\pi|_{C^N}: C^N \rightarrow C^\bullet$ is the normalization of $C^\bullet$. For every node $n_i$, the points in $(\pi|_{C^N})^{-1}(n_i) = \mathcal{E}_i \cap C^N = \{p_i, q_i\}$ are called the \emph{exceptional nodes}.

From this point forward, we will consider blow-ups of $C^\bullet$ on the full set of nodes unless stated otherwise. We will often refer to this setup as a \emph{full blow-up}. More general statements exist for partial blow-ups, the details of which are fully treated in \cite{2004math4078C}.

\subsubsection{Limit $n$-th roots and weighted graphs}

Let $n$ be a positive integer, and $L^\bullet$ be a line bundle on $C^\bullet$ so that $n| \deg(L^\bullet)$. Denote the full set of nodes by $\Delta_{C^\bullet}$.
\begin{defi}
A \emph{limit $n$-th root} of $L^\bullet$ associated to $\Delta_{C^\bullet}$ is a triple $(C^\circ, P^\circ, \alpha)$ consisting of: 
\begin{itemize}
\item the (full) blow-up $\pi: C^\circ \rightarrow C^\bullet$,
\item a line bundle $P^\circ$ on $C^\circ$,
\item a homomorphism $\alpha: (P^\circ)^n \rightarrow \pi^*(L^\bullet)$,
\end{itemize}
satisfying the following properties:
\begin{enumerate}[(i)]
\item $\deg(P^\circ|_{\mathcal{E}_i}) = 1$ for every exceptional component $\mathcal{E}_i$,
\item $\alpha$ is an isomorphism at all points of $C^\circ$ outside of the exceptional components.
\item for every exceptional component $\mathcal{E}_i$ of $C^\circ$, the orders of vanishing of $\alpha$ at the exceptional nodes $p_i$ and $q_i$ add up to $n$. 
\end{enumerate}
\end{defi}
We can also define limit $n$-th roots associated to a subset $\Delta \subseteq \Delta_{C^\bullet}$ in which the full blow-up is replaced by the partial blow-up at $\Delta$, see \cite{2004math4078C} for details.

Limit $n$-th roots over $C^\bullet$ carry some combinatorial data, in the form of weighted graphs, that takes into account the combinatorial aspects of the nodal curve $C^\bullet$. \linebreak Conversely, these weighted graphs allow one to construct and recover limit $n$-th roots. Although the correspondence between limit $n$-th roots and these weighted graphs are not one-to-one, it allows for a convenient parametrization of limit roots. First, let us \mbox{introduce} the weighted graphs in question. Let $\widetilde{\Delta}_{C^\bullet}$ be the exceptional nodes \linebreak corresponding to $\Delta_{C^\bullet}$.
\begin{defi}\label{Defi:weightedgraph} A \emph{weighted graph associated to a limit $n$-th root $(C^\circ, P^\circ, \alpha)$ of $L^\bullet$}
is the dual graph $\Pi_{C^\bullet}$ endowed with weights assigned by the weight function
\begin{align}
w: \widetilde{\Delta}_{C^\bullet} \rightarrow \{1, \dots, n-1\} \, ,
\end{align}
where $w(p_i) = u_i$ and $w(q_i) = v_i$ are the orders of the vanishing of $\alpha$ at $p_i$ and $q_i$ respectively. 
\end{defi}
Such weighted graphs naturally satisfy two conditions:
\begin{enumerate}[(A)]
\item $w(p_i) + w(q_i) = u_i + v_i = n,$
\item For every irreducible component $C^\bullet_i$ of $C^\bullet$, the sum of all weights assigned to the vertex corresponding to $C^\bullet_i$ is congruent to $\deg_{C^\bullet_i} L^\bullet \pmod n$.
\end{enumerate}
We illustrate an example of a weighted graph by returning to the \emph{Holiday lights} $H^\bullet$. We wish to find the limit 3rd roots of $K_{H^\bullet}^2$. If $C^\bullet_i$ is a component of $H^\bullet$, then set $k_i = \# C^\bullet_i \cap (\overline{H^\bullet \setminus C^\bullet_i})$. Therefore, \mbox{$\deg(K_{H^\bullet}|_{C^\bullet_i}) = 2g(C^\bullet_i) -2 + k_i$}, and the multi-degree of $K_{H^\bullet}$ is
\begin{align}
(\deg(K_{H^\bullet}|_\Gamma), \deg(K_{H^\bullet}|_{E_1}), .., \deg(K_{H^\bullet}|_{E_{10})}) = (8, 1, \dots , 1) \, ,
\end{align}
which has total degree is $2g(H^\bullet)-2 = 18$. So, the multi-degree of $K_{H^\bullet}^2$ is $(16, 2, \dots , 2).$ A weighted graph associated to the limit 3rd roots of $K_{H^\bullet}^2$, as well as the multi-degrees of $K_{H^\bullet}^2$, is given below. The labels inside the vertices are the multi-degrees of $K_{H^\bullet}^2$, while the labels outside the vertices are the weights.
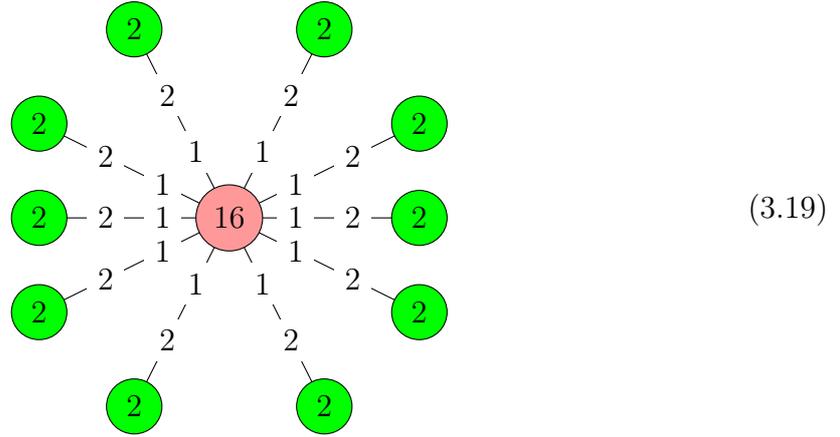
\begin{equation} \label{christmas_weighted_graph}
\begin{tikzpicture}[baseline=(current  bounding  box.center)]

\def\s{2.5};

\path[-, every node/.append style={fill=white}] (-\s,0) edge node[pos=0.35] {2} node[pos=0.65] {1} (0,0);
\path[-, every node/.append style={fill=white}] (\s,0) edge node[pos=0.35] {2} node[pos=0.65] {1} (0,0);
\path[-, every node/.append style={fill=white}] (-\s,0.5*\s) edge node[pos=0.35] {2} node[pos=0.65] {1} (0,0);
\path[-, every node/.append style={fill=white}] (\s,-0.5*\s) edge node[pos=0.35] {2} node[pos=0.65] {1} (0,0);
\path[-, every node/.append style={fill=white}] (-\s,-0.5*\s) edge node[pos=0.35] {2} node[pos=0.65] {1} (0,0);
\path[-, every node/.append style={fill=white}] (\s,0.5*\s) edge node[pos=0.35] {2} node[pos=0.65] {1} (0,0);
\path[-, every node/.append style={fill=white}] (-0.5*\s,\s) edge node[pos=0.35] {2} node[pos=0.65] {1} (0,0);
\path[-, every node/.append style={fill=white}] (0.5*\s,-\s) edge node[pos=0.35] {2} node[pos=0.65] {1} (0,0);
\path[-, every node/.append style={fill=white}] (0.5*\s,\s) edge node[pos=0.35] {2} node[pos=0.65] {1} (0,0);
\path[-, every node/.append style={fill=white}] (-0.5*\s,-\s) edge node[pos=0.35] {2} node[pos=0.65] {1} (0,0);

\node at (0,0) [stuff_fill_red] {$16$};
\node at (-\s,0)  [stuff_fill_green] {$2$};
\node at (-\s,0.5*\s) [stuff_fill_green] {$2$};
\node at (-0.5*\s,\s) [stuff_fill_green] {$2$};
\node at (0.5*\s,\s) [stuff_fill_green] {$2$};
\node at (\s,0.5*\s) [stuff_fill_green] {$2$};
\node at (\s,0) [stuff_fill_green] {$2$};
\node at (\s,-0.5*\s) [stuff_fill_green] {$2$};
\node at (0.5*\s,-\s) [stuff_fill_green] {$2$};
\node at (-0.5*\s,-\s) [stuff_fill_green] {$2$};
\node at (-\s,-0.5*\s)[stuff_fill_green] {$2$};

\end{tikzpicture}
\end{equation}
Given a weighted graph satisfying conditions A and B, we have a recipe for constructing limit $n$-th roots of $L^\bullet$.
\begin{prop} \label{Prop:limitrootconstruction}
Every weighted graph, whose underlying graph is $\Pi_{C^\bullet}$, and whose weight function $w: \widetilde{\Delta}_{C^\bullet} \rightarrow \{1, \dots, n-1\}$ satisfies conditions A and B, encodes a limit $n$-th root $(C^\circ, P^\circ, \alpha)$ of $L$. Moreover, this weighted graph coincides with the weighted graph associated to $(C^\circ, P^\circ, \alpha)$ of $L^\bullet$.
\end{prop}

\begin{myproof}
Suppose we have a weighted graph satisfying the hypothesis of the proposition, and let $\pi|_{C^N}: C^N \rightarrow C^\bullet$ be the normalization of $C^\bullet$. Thanks to condition B, the line bundle
\begin{align}
(\pi|_{C^N})^*(L^\bullet)\left(-\sum_{p_i, q_i \in \widetilde{\Delta}} (u_ip_i +v_iq_i)\right)
\end{align}
has on each irreducible component of $C^N$ degree divisible by $n$. Thus, on each irreducible component of $C^N$ it admits an $n$-th root. The collection formed from an $n$-th root on each irreducible component is a line bundle $P^N \in \mathrm{Pic}(C^N)$. Let $\pi: C^\circ \rightarrow C^\bullet$ be the full blow-up. Over each exceptional component, glue a degree one line bundle to $P^N$ to obtain a line bundle $P^\circ \in \mathrm{Pic}(C^\circ)$. Finally, define $\alpha: (P^\circ)^n \rightarrow \pi^*(L^\bullet)$ to be zero on the exceptional components, and
\begin{align}
\alpha|_{C^N}: (P^N)^n = (\pi|_{C^N})^*(L^\bullet)\left(-\sum_{p_i, q_i \in \widetilde{\Delta}} (u_ip_i + v_iq_i)\right) \hookrightarrow (\pi|_{C^N})^*(L^\bullet)
\end{align}
on $C^N$. Then, $(C^\circ, P^\circ, \alpha)$ is the desired limit $n$-th root of $L^\bullet$ associated to $\Delta_{C^\bullet}.$
\end{myproof}
The same statements follow when $\Delta_{C^\bullet}$ is replaced by a subset $\Delta$. In this case, the limit roots associated to $\Delta$ give rise to weighted subgraphs satisfying conditions A and B. Using the same procedure from \oref{Prop:limitrootconstruction}, we can construct a limit root from a weighted subgraph.

Every nodal curve $C^\bullet$ and line bundle $L^\bullet$ on $C^\bullet$ has a total of $n^{b_1(\Pi_{C^\bullet})}$ weighted subgraphs satisfying conditions A and B, where
\begin{align}
b_1(\Pi_{C^\bullet}) = \# \text{edges} + \#\text{connected components} - \# \text{vertices} \, ,
\end{align}
is the first Betti number of $\Pi_{C^\bullet}.$ We emphasize that this counts all of the weighted subgraphs, whose edge sets coincide with subsets of $\Delta_{C^\bullet}$. Curves, whose dual graphs are trees, will have zero $b_1$, and thus, will only have one weighted graph. These curves are said to be of compact type. This is certainly the case for the \emph{Holiday lights} $H^\bullet$. Here, $b_1(\Pi_{H^\bullet}) = 0$ and the weighted graph depicted in eq. \eqref{christmas_weighted_graph} is the only possible weighted graph for the 3rd limit roots of $K_{H^\bullet}^2$.

The correspondence between limit $n$-th roots and weighted graphs satisfying \mbox{conditions} $A$ and $B$ is not one-to-one. Indeed, the construction detailed in \oref{Prop:limitrootconstruction} involves a choice of a root $P^N$ of $(\pi|_{C^N})^*(L^\bullet)(-\sum (u_ip_i + v_i q_i))$. A careful count reveals that there are $n^{2g}$ limit $n$-th roots \cite{2004math4078C}.

We will apply the limit root construction in \oref{Prop:limitrootconstruction} to describe the limit 3rd roots of $K_{H^\bullet}^2$ on the \emph{Holiday lights} $H^\bullet$. We proceed as follows:
\begin{enumerate}
 \item Blow-up all nodal singularities, and denote the exceptional component at the $i$-th node by $\mathcal{E}_i \cong \mathbb{P}^1$. This $\mathbb{P}^1$ touches $E_i$ at the exceptional node $p_i$ and $\Gamma$ at $q_i$.
 \item Let $H^N$ be the (full) normalization of $H^\bullet$, and consider the bundle 
\begin{align}
 (\pi|_{H^N})^\ast ( K_{H^\bullet}^2 ) \left( - 2 \sum_{i = 1}^{10}p_i - \sum_{i = 1}^{10}q_i \right) \, ,
\end{align}
which has multi-degree $\left( 16 - 10, 2 - 2, \dots, 2 - 2 \right) = ( 6, 0, \dots, 0 )$. This bundle \mbox{admits} 3rd roots on $H^N$, namely $3^{2g(E_i)} = 9$ roots on each elliptic curve $E_i$ and $3^{2g(\Gamma)} = 1$ root on $\Gamma$. Hence, there are $9^{10} = 3^{20}$ roots, and each has multi-degree $\left( 2, 0, \dots, 0 \right)$.
 \item Pick a 3rd root $P^N$, and glue to it a degree one bundle over every $\mathcal{E}_i$. The resulting \emph{limit 3rd root $P^\circ$ of $K_{H^\bullet}^2$} has multi-degree $(2, 0, \dots, 0, 1, \dots, 1)$ over $H^\circ$, where \begin{align} \deg(P^\circ|_{C^\bullet_i}) = \deg(P^N|_{C^\bullet_i}), \quad  \deg(P^\circ|_{\mathcal{E}_i}) = 1.
 \end{align} These limit roots can be represented as follows:

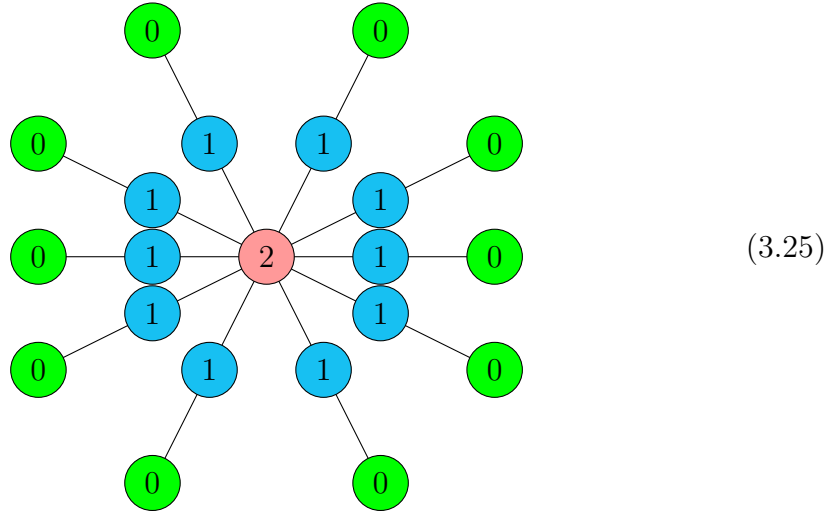
\begin{equation} 
\begin{tikzpicture}[baseline=(current  bounding  box.center)]

\def\s{1.5};
\draw (-2*\s,0) -- (2*\s,0);
\draw (-2*\s,\s) -- (2*\s,-\s);
\draw (-2*\s,-\s) -- (2*\s,\s);
\draw (-\s,2*\s) -- (\s,-2*\s);
\draw (\s,2*\s) -- (-\s,-2*\s);

\node at (0,0) [stuff_fill_red] {$2$};
\node at (-2*\s,0)  [stuff_fill_green] {$0$};
\node at (-2*\s,\s)[stuff_fill_green] {$0$};
\node at (-\s,2*\s) [stuff_fill_green] {$0$};
\node at (\s,2*\s) [stuff_fill_green] {$0$};
\node at  (2*\s,\s) [stuff_fill_green] {$0$};
\node at (2*\s,0)[stuff_fill_green] {$0$};
\node at (2*\s,-\s) [stuff_fill_green] {$0$};
\node at (\s,-2*\s) [stuff_fill_green] {$0$};
\node at (-\s,-2*\s) [stuff_fill_green] {$0$};
\node at (-2*\s,-\s)[stuff_fill_green] {$0$};
\node at (-\s,0) [stuff_fill_blue] {$1$};
\node at (-\s,0.5*\s) [stuff_fill_blue] {$1$};
\node at (-0.5*\s,\s) [stuff_fill_blue] {$1$};
\node at (0.5*\s,\s) [stuff_fill_blue] {$1$};
\node at (\s,0.5*\s) [stuff_fill_blue] {$1$};
\node at (\s,0) [stuff_fill_blue] {$1$};
\node at (\s,-0.5*\s) [stuff_fill_blue] {$1$};
\node at (0.5*\s,-\s) [stuff_fill_blue] {$1$};
\node at (-0.5*\s,-\s) [stuff_fill_blue] {$1$};
\node at (-\s,-0.5*\s) [stuff_fill_blue] {$1$};
\end{tikzpicture}
\end{equation}
As before, the green vertices represent the elliptic curves $E_i$, and the pink vertex represents $\Gamma$. The blue vertices represent the exceptional component $\mathcal{E}_i$, which intersects $E_i$ and $\Gamma$. The multi-degrees of the limit root $P^\circ$ is written inside the vertices. In particular, $P^\circ$ restricts to a degree 1 line bundle over each exceptional component.
\end{enumerate}

\subsection{Global sections of full blow-up limit roots} \label{subsec:SectionTechnology}
Of ample importance for our analysis is the number of global sections of the limit roots. These arise from gluing sections on the irreducible components of the nodal curve across exceptional divisors, which is addressed in the next lemma. 

\begin{lemma}
Let $p_1, p_2$ be two distinct points on $\mathbb{P}^1$, and $a_1, a_2 \in \mathbb{C}$. For every $p_3 \in \mathbb{P}^1 \setminus \{p_1, p_2\}$, there exists a unique section $s \in H^0(\mathbb{P}^1, \mathcal{O}_{\mathbb{P}^1}(p_3))$ such that $s(p_1) = a_1$ and $s(p_2) = a_2.$
\end{lemma}

\begin{myproof}
Endow $\mathbb{P}^1$ with its standard open cover  $\{U_0, U_1\}$. Let $z \in U_0$ and $w \in U_1$ be local coordinates so that $w = \frac{1}{z}$ in $U_0 \cap U_1.$ Since $\mathrm{ PGL }(2)$ acts transitively on $\mathbb{P}^1$, we may assume that
\begin{align} p_1 = 0, \quad p_2 = 1, \quad p_3 = \infty \, , \end{align}
without loss of generality. The desired section $s$ is given by
\begin{align}
s|_{U_0}(z) = (a_2-a_1)z + a_1 \, , \quad s|_{U_1} = (a_2 - a_1)+ a_1w \, .
\end{align}
It remains to show uniqueness. Recall that every section $t \in \Gamma(\mathbb{P}^1, \mathcal{O}_{\mathbb{P}^1}(p_3))$ is given by two analytic functions $t_0 = t|_{U_0} \in \Gamma(U_0,\mathcal{O}_{\mathbb{P}^1}(p_3))$ and $t_1 = t|_{U_1} \in \Gamma(U_1,\mathcal{O}_{\mathbb{P}^1}(p_3))$ such that over $U_0 \cap U_1$,
\begin{align}
t_0(z) = zt_1(w) = zt_1(1/z) \, .
\end{align}
If $t_0(z) = \sum_{k \geq 0} \alpha_k z^k$ and $t_1(w) = \sum_{k \geq 0} \beta_k w^k$, the above implies that 
\begin{align}
  \sum_{k \geq 0} \alpha_k z^k = z\left(\sum_{k \geq 0} \beta_k w^k \right) = z\left(\sum_{k \geq 0} \beta_k z^{-k} \right) = \sum_{k \geq 0} \beta_k z^{1-k} \, .
\end{align}
It follows that $\alpha_k = \beta_k = 0$ for $k > 1$, which leaves $\alpha_0 = \beta_1$ and $\alpha_1 = \beta_0.$ As such, every section $t$ is given by $t_0(z) = \alpha_0 z + \alpha_1$ and $t_1 = \alpha_0 + \alpha_1w.$ If $t$ also satisfies $t(0) = a_1$ and $t(1) = a_2$, then within the chart $U_0$ containing $0, 1 \in \mathbb{P}^1,$
\begin{align}
a_1 = t_0(0) = \alpha_1 \, , \qquad a_2 = t_0(1) = \alpha_0 + \alpha_1 = \alpha_0 + a_1 \, .
\end{align}
Thus, $t_0(z) = (a_2 - a_1)z + a_1$ and $t_1(w) = (a_2 - a_1) + a_1w$, which coincides with $s$.
\end{myproof}

By virtue of the above lemma, there is a unique way of gluing a local section over an exceptional component to local sections over the irreducible components at each end. This leads us to the following corollary.
\begin{corollary} \label{cor:CountingSections}
Let $C^\bullet$ be a connected nodal curve with irreducible components $C^\bullet_1, \dots, C^\bullet_k$. Let $L^\bullet$ be a line bundle on $C^\bullet$, and $n$ be an integer with $n \geq 2$ and $n | \deg(L^\bullet)$. For any limit $n$-th root $(C^\circ, P^\circ, \alpha)$ of $L^\bullet$,
\begin{align}
    h^0(C^\circ, P^\circ) = \sum^k_{i=1} h^0(C^\bullet_i, P^\circ|_{C^\bullet_i}).
\end{align}
\end{corollary}

\begin{myproof}
Let two irreducible components $C^\bullet_i$ and $C^\bullet_j$ intersect an exceptional component $\mathcal{E} \cong \mathbb{P}^1$ at $p_i \in C^\bullet_i$ and $p_j \in C^\bullet_j$ respectively. Set $Y = C^\bullet_i \cup \mathcal{E} \cup C^\bullet_j$. Then, we have
\begin{align} 
\begin{split}
h^0(Y, P^\circ) &\geq h^0(C^\bullet_i, P^\circ|_{C^\bullet_i}) + h^0(C^\bullet_j, P^\circ|_{C^\bullet_j}) + h^0(\mathcal{E}, P^\circ|_\mathcal{E}) \\
&\quad - h^0(C^\bullet_i \cap \mathcal{E}, P^\circ|_{C^\bullet_i \cap \mathcal{E}}) - h^0(C^\bullet_j \cap \mathcal{E}, P^\circ|_{C^\bullet_j \cap \mathcal{E}})\notag 
\end{split}
\\ 
&\geq  h^0(C^\bullet_i, P^\circ|_{C^\bullet_i}) + h^0(C_j, P^\circ|_{C^\bullet_j}) + 2 - 1-1  \notag \\ 
&\geq h^0(C^\bullet_i, P^\circ|_{C^\bullet_i}) + h^0(C^\bullet_j, P^\circ|_{C^\bullet_j}). 
\end{align}
It remains to prove equality. Recall that the number of independent conditions met at $p_i$ and $p_j$ is at most 2 -- the number of intersection points on $\mathcal{E}$. The previous lemma showed that there are exactly two independent conditions; one at each $p_i$ and $p_j$. Thus, equality holds. Since any two irreducible components of $C^\bullet$ either intersect a common exceptional component, or they do not in the full blow-up, the result follows.
\end{myproof}
Let us apply these results to the \emph{Holiday lights} $H^\bullet$, and count the global sections of the limit 3rd roots of $K_{H^\bullet}^2$. Recall that $H^\bullet$ is the union of a rational curve $\Gamma,$ and 10 elliptic curves $E_i$. Also, the limit 3rd root $P^\circ$ of $K_{H^\bullet}^2$ has multi-degree $(2, 0, \dots, 0, 1, \dots, 1)$. Since $\Gamma$ is rational, $h^0(\Gamma, P^\circ|_\Gamma) = h^0(\mathbb{P}^1, \mathcal{O}_{\mathbb{P}^1}(2)) = 3.$ By the above results, we have
\begin{align}
h^0 (H^\circ, P^\circ) = h^0( \Gamma, P^\circ|_{\Gamma}) + \sum_{i = 1}^{10}{h^0( E_i, P^\circ|_{E_i})} = 3 + \left\{ \begin{array}{c} 0 \\ 1 \end{array} \right\} + \dots + \left\{  \begin{array}{c} 0 \\ 1 \end{array} \right\} \,
\end{align}
The last term in the above expression means $0$ or $1$. This refers to the two cases described in \oref{prop:RootsElliptic} in which $P^\circ|_{E_i}$ is either non-trivial or trivial. 

This example highlights the general fact that a line bundle of degree $d$ over a smooth curve can have different $h^0$'s. Since counting the global sections of a limit root is \mbox{equivalent} to counting its local sections over the smooth irreducible components, we address the effect of this phenomenon on section-counting in the following corollary.

\begin{corollary}
Let $C^\bullet$ be a connected nodal curve with irreducible components $C^\bullet_1, \dots, C^\bullet_k$. Let $L^\bullet$ be a line bundle on $C^\bullet$, and $n$ be an integer with $n \geq 2$ and $n | \deg(L^\bullet)$. For any limit $n$-th root $(C^\circ, P^\circ, \alpha)$ of $L^\bullet$,
\begin{align}
    \sum^k_{i=1} \min  h^0(C^\bullet_i, P^\circ|_{C^\bullet_i}) \leq h^0(C^\circ, P^\circ) \leq \sum^k_{i=1} \max h^0(C^\bullet_i, P^\circ|_{C^\bullet_i}),
\end{align}
where for each $i$, the minimum and maximum are taken over all line bundles of degree $\deg(P^\circ|_{C_i^\bullet})$ over $C^\bullet_i$.
\end{corollary}
In the example of the \emph{Holiday lights} $H^\bullet$,
\begin{align}
   \min_{P^\circ|_\Gamma \in \mathrm{Pic}^2(\Gamma)} h^0(\Gamma, P^\circ|_\Gamma) &= 3, \quad \max_{P^\circ|_\Gamma \in \mathrm{Pic}^2(\Gamma)} h^0(\Gamma, P^\circ|_\Gamma) = 3, \\
   \min_{P^\circ|_{E_i} \in J(E_i)} h^0(E_i, P^\circ|_{E_i}) &= 0, \quad \max_{P^\circ|_{E_i} \in J(E_i)} h^0(E_i, P^\circ|_{E_i}) = 1,
\end{align}
for $i = 1, ..., 10$. Hence, $3 \leq h^0(H^\circ, P^\circ) \leq 13.$

Let $\mathrm{Roots}(n, L^\bullet)^\circ$ be the set of limit $n$-th roots of $L^\bullet$ on $C^\bullet.$ In a broader sense, we wish to understand the map,
\begin{align}
    h^0(C^\circ, \cdot): \mathrm{Roots}(n, L^\bullet)^\circ \rightarrow \mathbb{N} \cup \{0\}, \quad P^\circ \mapsto h^0(C^\circ, P^\circ),
\end{align}
For curves of compact type, every limit root comes from one weighted graph, and is constructed over the full blow-up. In this case, the global sections of the limit root are fully determined by the local sections over the irreducible components. Hence, we can compute $|h^0(C^\circ, \cdot)^{-1}(a)|$ for every $a \in \mathbb{N} \cup \{0\}$, i.e. the number of limit $n$-th roots with $h^0 = a$. We illustrate this with the \emph{Holiday lights}, which is a curve of compact type. Denote the number of elliptic curves on which $P^\circ|_{E_i}$ is non-trivial by $N_i$. Then, the number $N_{P^\circ}( h^0 )$ of limit 3rd roots with specific $h^0$ are as follows:

\begin{center}
\begin{adjustbox}{max width=\textwidth}
\begin{tabular}{c|ccccc|cccccc}
\toprule
$N_i$ & 10 & 9 & 8 & 7 & 6 & 5 & 4 & 3 & 2 & 1 & 0 \\
\midrule
$N_{P^\circ}( 3 )$ & $1$ & $\binom{1}{0} \cdot 8$ & $\binom{2}{0} \cdot 8^2$ & $\binom{3}{0} \cdot 8^3$ & $\binom{4}{0} \cdot 8^4$ & $\binom{5}{0} \cdot 8^5$ & $\binom{6}{0} \cdot 8^6$ & $\binom{7}{0} \cdot 8^7$ & $\binom{8}{0} \cdot 8^8$ & $\binom{9}{0} \cdot 8^9$ & $\binom{10}{0} \cdot 8^{10}$ \\[0.2em]
$N_{P^\circ}( 4 )$ &     & $1$ & $\binom{2}{1} \cdot 8^1$ & $\binom{3}{1} \cdot 8^2$ & $\binom{4}{1} \cdot 8^3$ & $\binom{5}{1} \cdot 8^4$ & $\binom{6}{1} \cdot 8^5$ & $\binom{7}{1} \cdot 8^6$ & $\binom{8}{1} \cdot 8^7$ & $\binom{9}{1} \cdot 8^8$ & $\binom{10}{1} \cdot 8^9$ \\[0.2em]
$N_{P^\circ}( 5 )$ &     &      & $1$ & $\binom{3}{2} \cdot 8^1$ & $\binom{4}{2} \cdot 8^2$ & $\binom{5}{2} \cdot 8^3$ & $\binom{6}{2} \cdot 8^4$ & $\binom{7}{2} \cdot 8^5$ & $\binom{8}{2} \cdot 8^6$ & $\binom{9}{2} \cdot 8^7$ & $\binom{10}{2} \cdot 8^8$ \\[0.2em]
$N_{P^\circ}( 6 )$ &     &      &     & $1$ & $\binom{4}{3} \cdot 8^1$ & $\binom{5}{3} \cdot 8^2$ & $\binom{6}{3} \cdot 8^3$ & $\binom{7}{3} \cdot 8^4$ & $\binom{8}{3} \cdot 8^5$ & $\binom{9}{3} \cdot 8^6$ & $\binom{10}{3} \cdot 8^7$ \\[0.2em]
$N_{P^\circ}( 7 )$ &     &      &     &    & $1$ & $\binom{5}{4} \cdot 8^1$ & $\binom{6}{4} \cdot 8^2$ & $\binom{7}{4} \cdot 8^3$ & $\binom{8}{4} \cdot 8^4$ & $\binom{9}{4} \cdot 8^5$ & $\binom{10}{4} \cdot 8^6$ \\[0.2em]
$N_{P^\circ}( 8 )$ &    &      &     &    &     & $1$ & $\binom{6}{5} \cdot 8^1$ & $\binom{7}{5} \cdot 8^2$ & $\binom{8}{5} \cdot 8^3$ & $\binom{9}{5} \cdot 8^4$ & $\binom{10}{5} \cdot 8^5$ \\[0.2em]
$N_{P^\circ}( 9 )$ &     &      &     &    &     &     & $1$ & $\binom{7}{6} \cdot 8^1$ & $\binom{8}{6} \cdot 8^2$ & $\binom{9}{6} \cdot 8^3$ & $\binom{10}{6} \cdot 8^4$ \\[0.2em]
$N_{P^\circ}( 10 )$ &    &      &     &    &     &     &    & $1$ & $\binom{8}{7} \cdot 8^1$ & $\binom{9}{7} \cdot 8^2$ & $\binom{10}{7} \cdot 8^3$ \\[0.2em]
$N_{P^\circ}( 11 )$ &     &      &     &    &     &     &    &   & $1$ & $\binom{9}{8} \cdot 8^1$ & $\binom{10}{8} \cdot 8^2$ \\[0.2em]
$N_{P^\circ}( 12 )$ &     &      &     &    &     &     &    &   &     & $1$ & $\binom{10}{9} \cdot 8^1$ \\[0.2em]
$N_{P^\circ}( 13 )$ &     &      &     &    &     &     &    &   &     &     & $1$ \\
\midrule
Factor & $3^{20}$ & $3^{18}$ & $3^{16}$ & $3^{14}$ & $3^{12}$ & $3^{10}$ & $3^{8}$ & $3^{6}$ & $3^{4}$ & $3^{3}$ & $3^{0}$ \\
\bottomrule
\end{tabular}
\end{adjustbox}
\end{center}
This table says that for $N_i = 10$, we find $N_{P^\circ}( 3 ) = 1 \cdot 3^{20}$ limit 3rd roots $P^\circ$ with $h^0 = 3$. Similarly, for $N_i = 4$, we find $N_{P^\circ}( 3 ) = \binom{6}{3} \cdot 8^3 \cdot 3^{8}$ limit 3rd roots $P^\circ$ with $h^0 = 6$. For ease of presentation, the overall factors are collected at the bottom of this table.

We would like to generalize this section-counting of limit roots for all curves, which may have multiple weighted subgraphs. Complications arise when counting the global sections of limits roots over partial blow-ups; namely, it is unclear what $h^0$ of a limit root is over a singular component of the curve, i.e., the component containing a singularity that has not been blown up. Although we will not discuss this direction in this paper, it presents an interesting problem which we hope to revisit in the future.

\section{Limit root applications in F-theory}\label{sec:ApplicationToMSSMs}

After the detailed exposition of root bundles and limit roots in the previous section, we now wish to apply these techniques to F-theory. We first outline how limit roots can be used to provide an explicit and oftentimes constructive argument for the absence of certain vector-like exotics. We demonstrate these ideas in one particular geometry among the largest class of \mbox{currently-known} globally consistent F-theory Standard Models without chiral exotics and gauge coupling unification \cite{Cvetic:2019gnh}. We will argue that there are solutions without vector-like exotics in the representations $C_{({\mathbf{3}},\mathbf{2})_{1/6}}$, $C_{(\overline{\mathbf{3}},\mathbf{1})_{-2/3}}$, $C_{(\overline{\mathbf{3}},\mathbf{1})_{1/3}}$ and $C_{({\mathbf{1}},\mathbf{1})_{1}}$.

\subsection{Absence of vector-like exotics} \label{subsec:AbsenceOfVectorlikes}

Let us look at an F-theory compactification to 4-dimensions on a space $Y_4$, which admits a smooth, flat, crepant resolution $\widehat{Y}_4$. As explained in \oref{subsec:RootBundleAppearance}, root bundles appear naturally in such settings when studying vector-like spectra. We found that the geometry determines a class $A^\prime = \widehat{\gamma}(\mathcal{A}^\prime) \in H^4_D( \widehat{Y}_4, \mathbb{Z}(2) )$ for some $\mathcal{A}^\prime \in \mathrm{CH}^2( \widehat{Y}_4, \mathbb{Z} )$, and an integer $\xi \in \mathbb{Z}_{> 0}$ such that $\mathcal{A}$ is subject to the two constraints:
\begin{align}
\gamma( \mathcal{A} ) = G_4 \, , \qquad \xi \cdot \widehat{\gamma}( \mathcal{A} ) \sim \widehat{\gamma} ( \mathcal{A}^\prime ) \, .
\end{align}
The condition $\gamma( \mathcal{A} ) = G_4$ immediately follows from \cref{equ:CommutativeDiagramLifts} and it means that $A = \widehat{\gamma}( \mathcal{A} )$ is an F-theory gauge potential for the given $G_4$-flux. The second condition ensures the absence of chiral exotics in the F-theory Standard Models \cite{Cvetic:2019gnh}. It follows that the line bundle on the matter curve $C_{\mathbf{R}}$ satisfies
\begin{align}
P_{\mathbf{R}} = \mathcal{O}_{C_{\mathbf{R}}} \left( D_{\mathbf{R}}( \mathcal{A} ) \right) \otimes_{\mathcal{O}_{C_{\mathbf{R}}}} \mathcal{O}_{C_{\mathbf{R}}} \left( D^{\text{spin}}_{C_{\mathbf{R}}} \right) \, ,
\end{align}
where $D_{\mathbf{R}}( \mathcal{A} )$ and $D^{\text{spin}}_{C_{\mathbf{R}}}$ are solutions to the root bundle constraints
\begin{align}
\xi \cdot D_{\mathbf{R}}( \mathcal{A} ) \sim D_{\mathbf{R}}( \mathcal{A}^\prime ) \, , \qquad 2 \cdot D^{\text{spin}}_{C_{\mathbf{R}}} \sim K_{{\mathbf{R}}} \, .
\end{align}
Recall from \oref{sec:LimitRoots} that these root bundle constraints have many solutions. In general, it cannot be expected that all solutions are realized from roots in $H^4_D( \widehat{Y}_4, \mathbb{Z}(2) )$ and spin$^\text{c}$-structures on the gauge surfaces. We reserve a detailed study of this interesting and challenging question for future works. In this article, we study all the $\xi$-th roots of $D_{\mathbf{R}} \left( \mathcal{A}^\prime \right)$ and all of the spin divisors $D^{\text{spin}}_{C_{\mathbf{R}}}$ systematically. Our goal is to identify roots $P_{\mathbf{R}}$ subject to the physical demand of absence/presence of vector-like pairs. In future works, we hope to identify which of these desired roots stem from F-theory gauge potentials in $H^4_D( \widehat{Y}_4, \mathbb{Z}(2) )$.

At special loci of the complex structure moduli space, massive vector-like pairs can be rendered massless. Mathematically, this is reflected in the fact that deformations of a line bundle can have higher cohomologies. For example, if we assume $\chi( P_{\mathbf{R}} ) \geq 0$, then we could have:
\begin{center}
\begin{tabular}{cc}
\toprule
Geometry of curves $C_{\mathbf{R}}$ & $\left( h^0( C_{\mathbf{R}}, P_{\mathbf{R}} ), h^1( C_{\mathbf{R}}, P_{\mathbf{R}} ) \right)$ \\
\midrule
Generic & $\left( \chi( P_{\mathbf{R}} ), 0 \right)$ \\
Less generic & $\left( \chi( P_{\mathbf{R}} ) + 1, 1 \right) \equiv \left( \chi( P_{\mathbf{R}} ), 0 \right) \oplus (1,1)$ \\
Even less generic & $\left( \chi( P_{\mathbf{R}} ) + 2, 2 \right) \equiv \left( \chi( P_{\mathbf{R}} ), 0 \right) \oplus (2,2)$ \\
\vdots & \vdots \\
\bottomrule
\end{tabular}
\end{center}
In \cite{Bies:2020gvf}, such cohomology jumps have been analyzed in large detail. In particular, it was explained that even on generic curves, line bundles with the same chiral index need not have the same cohomologies. This classic observation goes by the name of \emph{Brill-Noether theory} \cite{Brill1874} (see also \cite{Watari:2016lft} for another application of Brill-Noether theory to F-theory). This observation in particular applies to root bundles. In \oref{subsec:MthRootsOfDivisors}, we have explained that of the four spin structures on an elliptic curve, one has $h^0( E, \mathcal{O}^{\text{spin}}_{E} ) = 1$ and the other three have vanishing number of global sections. This is a special instance of the results in \cite{atiyah1971riemann,mumford1971theta}, which show that all odd spin structures have odd number of zero modes, while the remaining even spin structures have even number of zero modes. Generally speaking, different roots $P_{\mathbf{R}}$ will have different numbers of zero modes.

That said, our task is to construct root bundles $P_{\mathbf{R}}$ with the cohomologies that are physically desired. For simplicity, let us assume $\chi( P_{\mathbf{R}} ) \geq 0$. Inspired by physics, we should then distinguish the generic case $h^0( P_{\mathbf{R}} ) = \chi( P_{\mathbf{R}} )$ and the non-generic case $h^0( P_{\mathbf{R}} ) > \chi( P_{\mathbf{R}} )$. The former corresponds to the absence of exotic vector-like pairs, while the latter most prominently features on the Higgs curve in F-theory Standard Model constructions. In the latter case, for MSSM constructions, one wishes to achieve $h^0( P_{\mathbf{R}} ) = \chi( P_{\mathbf{R}} ) + 1$ so that the additional vector-like pair describes a Higgs field.

We approach the task of constructing such physically desired root bundles $P_{\mathbf{R}}$ by first considering a deformation $C_{\mathbf{R}} \to C^{\bullet}_{\mathbf{R}}$, where $C^{\bullet}_{\mathbf{R}}$ is a \emph{nodal} curve. Therefore, $P_{\mathbf{R}} \to P_{\mathbf{R}}^\bullet$ becomes a root bundle on the nodal curve $C^{\bullet}_{\mathbf{R}}$. We focus on roots $P_{\mathbf{R}}^\bullet$, which we can describe by limit roots $P_{\mathbf{R}}^\circ$ on the full blow-up $C^{\circ}_{\mathbf{R}}$ of $C^{\bullet}_{\mathbf{R}}$. For those limit roots, we can employ the technology described in \oref{sec:LimitRoots} in order to identify $h^0( C^{\circ}_{\mathbf{R}}, P_{\mathbf{R}}^\circ )$. This enables us to identify roots $P_{\mathbf{R}}^\circ$ with 
$h^0( C^\circ_{\mathbf{R}}, P^\circ_{\mathbf{R}} ) = \chi( P^\circ ) + \delta$ from simple combinatorics, where $\delta \in \mathbb{Z}_{\geq 0}$ is the physically desired offset.

\begin{figure}
\resizebox{\textwidth}{!}{%
\begin{tikzpicture}[scale=1]
    
    \def\s{1.2};
    \def\r{0.05};

    { 
    \draw (-2.5*\s,5*\s) -- (0.5*\s,5*\s);
    \draw (-2.5*\s,5*\s) -- (-2.5*\s,3*\s);
    \draw (0.5*\s,5*\s) -- (0.5*\s,3*\s);
    \draw (-2.5*\s,3*\s) -- (0.5*\s,3*\s);
}
   
    \draw[->] (0.8*\s,4.5*\s) -- (2.7*\s,4.5*\s);
    \node at (1.75*\s,4.5*\s) [above] {\small Deformation};
    \draw[<-] (0.8*\s,3.5*\s) -- (2.7*\s,3.5*\s);
    \node at (0.8*\s,3.8*\s) [right] {\small Upper SC};
    \node at (1.75*\s,3.5*\s) [below] {\small $h^0$ remains 3};
\node at (-1*\s,5.2*\s) [above] {\small Matter curve $C_{\mathbf{R}}$};
    \node at (4.5*\s,5.2*\s) [above] {\small Nodal curve $C^\bullet_{\mathbf{R}}$};
    
    \node at (-1*\s, 4.6*\s) {$P_{\mathbf{R}}$};
    \draw[->] (-1*\s, 4.3*\s) -- (-1*\s, 3.7*\s);
{
  \draw (3*\s,5*\s) -- (6*\s,5*\s);
  \draw (3*\s,3*\s) -- (6*\s,3*\s);
  \draw (3*\s,5*\s) -- (3*\s,3*\s);
  \draw (6*\s,5*\s) -- (6*\s,3*\s);

}
   
    \draw[->] (6.3*\s,4.5*\s) -- (8.2*\s,4.5*\s);
    \node at (7.25*\s,4.5*\s) [above] {\small Limit roots};
    \draw[<-] (6.3*\s,3.5*\s) -- (8.2*\s,3.5*\s);
    \node at (6.5*\s,3.8*\s) [right] {\small Pushforward};
    \node at (7.25*\s,3.5*\s) [below] {\small $h^0(P^\circ) = h^0(P^\bullet)$};
    \node at (10*\s,5.2*\s) [above] {\small Blow-up curve $C^\circ_{\mathbf{R}}$};
     \node at (4.5*\s, 4.6*\s) {$P^\bullet_{\mathbf{R}}$};
    \draw[->] (4.5*\s, 4.3*\s) -- (4.5*\s, 3.7*\s);
{
  \draw (8.5*\s,5*\s) -- (11.5*\s,5*\s);
  \draw (8.5*\s,3*\s) -- (11.5*\s,3*\s);
  \draw (8.5*\s,5*\s) -- (8.5*\s,3*\s);
  \draw (11.5*\s,5*\s) -- (11.5*\s,3*\s);

}

\draw [thick,red] (-2.2*\s,3.5*\s) to[out=30,in=150] (-1*\s,3.5*\s) to[out=-30,in=150] (0.2*\s,3.5*\s) ;

\draw [thick,red] (3.3*\s,3.3*\s) to[out=30,in=150] (5*\s,3.3*\s) ;
\draw [thick,red] (4*\s,3.3*\s) to[out=30,in=150] (5.7*\s,3.3*\s) ;
\draw[blue , fill] (4.5*\s,3.5*\s) circle (\r);

\draw [thick,red] (9*\s,3.3*\s) to[out=30,in=150] (10.8*\s,3.3*\s) ;
\draw[thick, blue] (10*\s,3.5*\s) to (10*\s, 3.9*\s);
\draw [thick,red] (9*\s,4.1*\s) to[out=30,in=150] (9.8*\s,3.9*\s) to[out=-30,in=150] (10.8*\s,4.1*\s) ;
\draw[blue , fill] (10*\s,3.55*\s) circle (\r);
\draw[blue , fill] (10*\s,3.85*\s) circle (\r);
 \node at (10.1*\s, 4.7*\s) {$P^\circ_{\mathbf{R}}$};
    \draw[->] (10*\s, 4.4*\s) -- (10*\s, 4.1*\s);
\end{tikzpicture}}
\caption{Roots $P_{\mathbf{R}}$ with $h^0( C_{\mathbf{R}}, P_{\mathbf{R}} ) = 3$ from roots $P^\bullet_{\mathbf{R}}$ on a nodal curve $C^\bullet_{\mathbf{R}}$ and limit roots $P^\circ_{\mathbf{R}}$ on its blow-up $C^\circ_{\mathbf{R}}$.}
\label{fig:ExistenceOfH03Roots}
\end{figure}
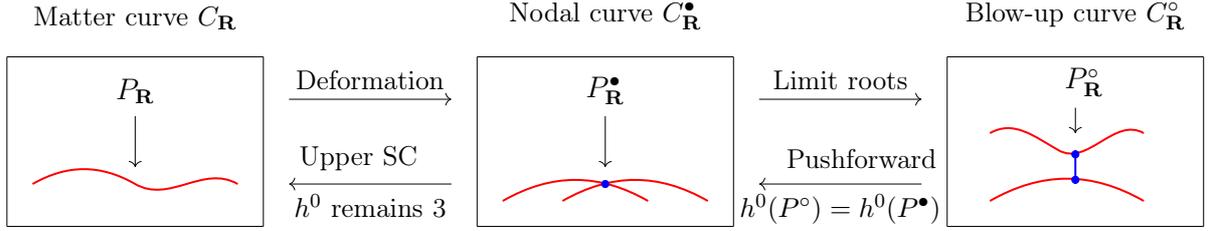

The pushforward of limit roots $P^\circ_{\mathbf{R}}$ along the blow-up map $\pi \colon C^\circ_{\mathbf{R}} \to C^\bullet_{\mathbf{R}}$ preserves the number of global sections, i.e. $h^0( C^\circ_{\mathbf{R}}, P^\circ_{\mathbf{R}} ) = h^0( C^\bullet_{\mathbf{R}}, P^\bullet_{\mathbf{R}} )$. We have thus identified the roots on $C^\bullet$ which have the physically desirable cohomologies. In theory, we can trace those roots $P^\bullet_{\mathbf{R}}$ along the deformation $C^\bullet_{\mathbf{R}} \to C_{\mathbf{R}}$ to find roots $P_{\mathbf{R}}$ on the original curve $C_{\mathbf{R}}$. Crucially though, such a deformation can change the number of sections (see e.g. \cite{Bies:2020gvf}). For the deformation $C^\bullet_{\mathbf{R}} \to C_{\mathbf{R}}$, which turns a nodal (i.e. singular and thus non-generic) curve into a smooth, irreducible curve, it is known that the number of sections is an upper semi-continuous function. This means that the number of sections either remains constant or decreases as we trace $P^\bullet_{\mathbf{R}}$ to $P_{\mathbf{R}}$ on $C_{\mathbf{R}}$:
\begin{align}
h^0( C_{\mathbf{R}}, P_{\mathbf{R}} ) \leq h^0( C^{\bullet}_{\mathbf{R}}, P^{\bullet}_{\mathbf{R}} ) = \chi( P_{\mathbf{R}} ) + \delta \, . \label{equ:Inequality}
\end{align}
The natural question is thus to look for roots $P^{\bullet}_{\mathbf{R}}$ for which equality holds. This happens in the generic case, i.e. the case $\delta = 0$. This is because the number of sections is then already minimal on $C^{\bullet}_{\mathbf{R}}$ and thus, it must remain constant along the deformation to $C^{\bullet}_{\mathbf{R}}$:
\begin{align}
h^0( C_{\mathbf{R}}, P_{\mathbf{R}} ) = h^0( C^\bullet_{\mathbf{R}}, P^{\bullet}_{\mathbf{R}} ) = \chi( P_{\mathbf{R}} ) \, .
\end{align}
The upshot of this strategy, which we summarize in \oref{fig:ExistenceOfH03Roots}, is that we can provide a lower bound to the number of roots $P_{\mathbf{R}}$ without vector-like exotics by studying the combinatorics of limit roots on the full blow-up $C^\circ_{\mathbf{R}}$ of the nodal curve $C^\bullet_{\mathbf{R}}$.\footnote{Recall that at least one of these roots stems from an F-theory gauge potential in $\mathrm{CH}^2( \widehat{Y}_4, \mathbb{Z} )$.}

In aiming for F-theory MSSMs, the non-generic case $\delta = 1$ is also fairly important for the Higgs curve. While it is not hard to construct limit roots on $C^{\circ}_{(\mathbf{1},\mathbf{2})_{-1/2}}$ with exactly 4 sections, the corresponding roots $P^\bullet_{(\mathbf{1},\mathbf{2})_{-1/2}}$ satisfy $h^0( C^{\bullet}_{(\mathbf{1},\mathbf{2})_{-1/2}}, P^{\bullet}_{(\mathbf{1},\mathbf{2})_{-1/2}} ) = 4$, which is larger than the minimal value $\chi( P_{(\mathbf{1},\mathbf{2})_{-1/2}} ) = 3$. Since the number of sections is non-minimal, we cannot conclude from upper semi-continuity that the number of sections remains constant. Rather, we expect some of those roots $P^\bullet_{(\mathbf{1},\mathbf{2})_{-1/2}}$ to lose a section when traced to $C_{(\mathbf{1},\mathbf{2})_{-1/2}}$. Currently, we do not know a sufficient discriminating property that allows us to identify the roots $P^\bullet_{(\mathbf{1},\mathbf{2})_{-1/2}}$ for which the number of sections remains constant. We reserve this interesting mathematical question for future work.

\subsection{Application to F-theory Standard Models} \label{subsec:RootBundlesF-theoryMSSMs}

We now continue the analysis initiated in \oref{subsec:RootBundlesInF-theoryMSSM}, where we summarized the geometry of the largest \mbox{currently-known} class of globally consistent F-theory Standard Models without chiral exotics and gauge coupling unification \cite{Cvetic:2019gnh}. The chiral index on all five matter curves
\begin{align}
C_{(\mathbf{3},\mathbf{2})_{1/6}} &= V( s_3, s_9 ) \, , & & C_{(\mathbf{1},\mathbf{2})_{-1/2}} = V \left( s_3, s_2 s_5^2 + s_1 ( s_1 s_9 - s_5 s_6 ) \right) \, , \\
C_{(\overline{\mathbf{3}},\mathbf{1})_{-2/3}} &= V( s_5, s_9 ) \, , & & C_{(\overline{\mathbf{3}},\mathbf{1})_{1/3}} = V \left( s_9, s_3 s_5^2 + s_6 ( s_1 s_6 - s_2 s_5 ) \right) \, , \\
C_{(\mathbf{1},\mathbf{1})_{1}} &= V( s_1, s_5 ) \, ,
\end{align}
is thus exactly three. We worked out the root bundle constraints (c.f. \oref{subsec:LineBundlesKbarCubeDifferent}). In aiming for an MSSM construction, which comes with exactly one Higgs pair, the vector-like spectrum is subject to the demand $4 = h^0( C_{(\mathbf{1},\mathbf{2})_{-1/2}}, P_{(\mathbf{1},\mathbf{2})_{-1/2}} )$. As explained above, since $4 = 1 + \chi( P_{(\mathbf{1},\mathbf{2})_{-1/2}} )$, our current technology does not allow us to tend to this case. However, we can address the absence of vector-like exotics on the remaining matter curves in a constructive way. That is, we can construct solutions to the constraint
\begin{align}
\begin{split}
3 &= h^0( C_{(\mathbf{3},\mathbf{2})_{1/6}}, P_{(\mathbf{3},\mathbf{2})_{1/6}} ) = h^0( C_{(\overline{\mathbf{3}},\mathbf{1})_{-2/3}}, P_{(\overline{\mathbf{3}},\mathbf{1})_{-2/3}} ) \\
  &= h^0( C_{(\overline{\mathbf{3}},\mathbf{1})_{1/3}}, P_{(\overline{\mathbf{3}},\mathbf{1})_{1/3}} ) = h^0( C_{(\mathbf{1},\mathbf{1})_{1}}, P_{(\mathbf{1},\mathbf{1})_{1}} ) \, . \label{equ:MinimalH0}
\end{split}
\end{align}
To outline these steps, let us first look at the quark-doublet curve $C_{(\mathbf{3},\mathbf{2})_{1/6}} = V( s_3, s_9 )$, where $s_3, s_9$ are generic sections of $\KB$. To make our construction explicit, let us focus on base spaces ${B}_3$ with $\KB^3 = 18$. It then follows from \oref{subsec:LineBundlesKbarCubeDifferent} that we are trying to argue for the existence of root bundles that satisfy
\begin{align}
P_{(\mathbf{3},\mathbf{2})_{1/6}}^{\otimes 24} \sim K_{(\mathbf{3},\mathbf{2})_{1/6}}^{\otimes 36} \, , \qquad h^0( C_{(\mathbf{3},\mathbf{2})_{1/6}} , P_{(\mathbf{3},\mathbf{2})_{1/6}} ) = 3 \, . \label{equ:OriginalTaskForConstruction}
\end{align}
For this, it suffices to argue that root bundles with the following properties exist
\begin{align}
P_{(\mathbf{3},\mathbf{2})_{1/6}}^{\otimes 3} \sim K_{(\mathbf{3},\mathbf{2})_{1/6}}^{\otimes 2} \, , \qquad h^0( C_{(\mathbf{3},\mathbf{2}_{1/6}} ), P_{(\mathbf{3},\mathbf{2})_{1/6}} ) = 3 \, . \label{equ:TaskForConstruction}
\end{align}
We achieve a proof of existence by studying a deformation $C_{(\mathbf{3},\mathbf{2})_{1/6}} \to C^\bullet_{(\mathbf{3},\mathbf{2})_{1/6}}$. Let us work with a concrete base geometry, we opt for the toric base space ${B}_3 = P_{39}$ with $\KB^3 = 18$, whose details are summarized in \oref{sec:ExampleGeometry}.

To describe the deformation $C_{(\mathbf{3},\mathbf{2})_{1/6}} \to C^\bullet_{(\mathbf{3},\mathbf{2})_{1/6}}$, we first notice that $s_3$ is a polynomial in the homogeneous coordinates $\left\{ x_i \right\}_{1 \leq i \leq 11}$ of $P_{39}$. Since $s_3$ is a section of $\overline{K}_{P_{39}}$, it contains the monomial $\prod_{i = 1}^{11}{x_i}$.\footnote{For any toric base space ${B}_3$ with homogeneous coordinates $x_i$, $\prod_{i}{x_i}$ is a section of $\KB \sim \sum_{i}{[x_i]}$.} This allows us to consider the deformation
\begin{align}
V( s_3, s_9 ) = C_{(\mathbf{3},\mathbf{2})_{1/6}} \to C_{(\mathbf{3},\mathbf{2})_{1/6}}^{\bullet} = V \left( \prod_{i = 1}^{11}{x_i}, s_9 \right) \, .
\end{align}
Since we assume \emph{generic} $s_9$, $C_{(\mathbf{3},\mathbf{2})_{1/6}}^{\bullet}$ is manifestly nodal in the K3-surface $V( s_9 )$ and the techniques of \oref{sec:LimitRoots} apply. To this end, we first identify the dual graph of $C_{(\mathbf{3},\mathbf{2})_{1/6}}^{\bullet}$, which has $17$ irreducible components:
\begin{align}
  \begin{array}{c|cccc}
      \toprule
      \text{curve} & \text{equation} & \text{genus} & \mathrm{deg} \left( 2 \cdot \left. K_{C^{\bullet}_{(\mathbf{3},\mathbf{2})_{1/6}}} \right|_{C_i} \right) \\
      \midrule
      C_1 & V( x_1, s_9 ) & 1 & 6 \\
      C_3 & V( x_3, s_9 ) & 1 & 6 \\
      C_6 & V( x_6, s_9 ) & 0 & 12 \\
      C_{11} & V( x_{11}, s_9 ) & 0 & 12 \\
      \midrule
      C_2 & V( x_2, s_9 ) & 0 & 0 \\
      \left\{ C_8^{(i)} \right\}_{1 \leq i \leq 6} & V( x_8, x_1 - \alpha_i x_3 ) & 0 & 0 \\
      \left\{ C_{10}^{(i)} \right\}_{1 \leq i \leq 6} & V( x_{10}, x_1 - \alpha_i x_3 ) & 0 & 0 \\
      \bottomrule
\end{array} \label{equ:CurveComponents}
\end{align}
For convenience, we list the degree of $2 \cdot K_{C^{\bullet}_{(\mathbf{3},\mathbf{2})_{1/6}}}$ on all irreducible components since \oref{equ:TaskForConstruction} instructs us to construct third roots of this bundle. By taking the Stanley-Reisner ideal of $P_{39}$ into account (see \oref{sec:ExampleGeometry}), one finds the dual graph of $C^{\bullet}_{(\mathbf{3},\mathbf{2})_{1/6}}$:
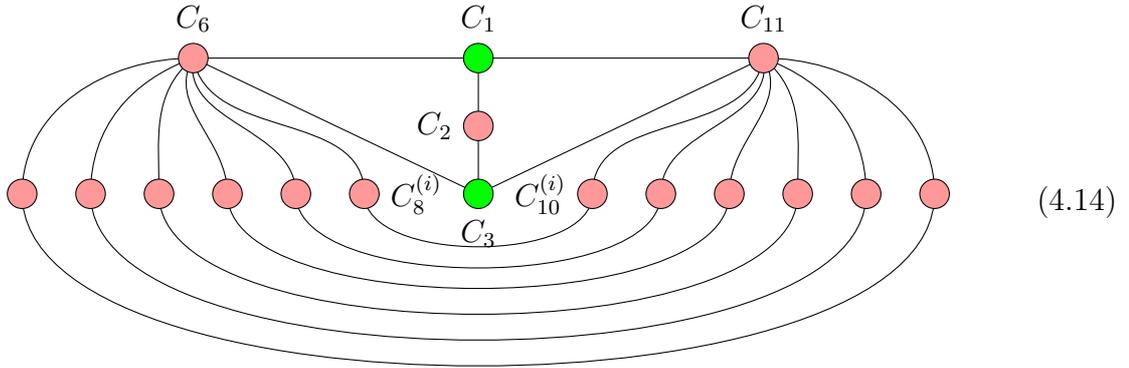
\begin{equation}
\begin{tikzpicture}[scale=0.6, baseline=(current  bounding  box.center)]
    
    \def\s{2.5};
    \def\h{1.0};
    
    \path[-, every node/.append style={fill=white}] (-2.5*\s,2.0*\h) edge (0,2*\h);
    \path[-, every node/.append style={fill=white}] (-2.5*\s,2.0*\h) edge (0,-1*\h);
    \path[-, every node/.append style={fill=white}] (0,2*\h) edge (0,0.5*\h);
    \path[-, every node/.append style={fill=white}] (0,0.5*\h) edge (0,-1*\h);
    \path[-, every node/.append style={fill=white}] (2.5*\s,2.0*\h) edge (0,-1*\h);
    \path[-, every node/.append style={fill=white}] (2.5*\s,2.0*\h) edge (0,2*\h);
    
    \path[-, every node/.append style={fill=white}, out = -90, in = -90, looseness = 0.8] (-1*\s,-1.0*\h) edge (1*\s,-1.0*\h);
    \path[-, every node/.append style={fill=white}, out = -90, in = -90, looseness = 0.7] (-1.6*\s,-1.0*\h) edge (1.6*\s,-1.0*\h);
    \path[-, every node/.append style={fill=white}, out = -90, in = -90, looseness = 0.65] (-2.2*\s,-1.0*\h) edge (2.2*\s,-1.0*\h);
    \path[-, every node/.append style={fill=white}, out = -90, in = -90, looseness = 0.65] (-2.8*\s,-1.0*\h) edge (2.8*\s,-1.0*\h);
    \path[-, every node/.append style={fill=white}, out = -90, in = -90, looseness = 0.65] (-3.4*\s,-1.0*\h) edge (3.4*\s,-1.0*\h);
    \path[-, every node/.append style={fill=white}, out = -90, in = -90, looseness = 0.65] (-4.0*\s,-1.0*\h) edge (4.0*\s,-1.*\h);
    
    \path[-, every node/.append style={fill=white},out=-80,in=90] (-2.5*\s,2.0*\h) edge (-1*\s,-1.0*\h);
    \path[-, every node/.append style={fill=white},out=-100,in=90] (-2.5*\s,2.0*\h) edge (-1.6*\s,-1.*\h);
    \path[-, every node/.append style={fill=white},out=-120,in=90] (-2.5*\s,2.0*\h) edge (-2.2*\s,-1.0*\h);
    \path[-, every node/.append style={fill=white},out=-140,in=90] (-2.5*\s,2.0*\h) edge (-2.8*\s,-1.0*\h);
    \path[-, every node/.append style={fill=white},out=-160,in=90] (-2.5*\s,2.0*\h) edge (-3.4*\s,-1.0*\h);
    \path[-, every node/.append style={fill=white},out=-180,in=90] (-2.5*\s,2.0*\h) edge (-4.0*\s,-1.0*\h);
    
    \path[-, every node/.append style={fill=white},out=-100,in=90] (2.5*\s,2.0*\h) edge (1*\s,-1.0*\h);
    \path[-, every node/.append style={fill=white},out=-80,in=90] (2.5*\s,2.0*\h) edge (1.6*\s,-1.0*\h);
    \path[-, every node/.append style={fill=white},out=-60,in=90] (2.5*\s,2.0*\h) edge (2.2*\s,-1.0*\h);
    \path[-, every node/.append style={fill=white},out=-40,in=90] (2.5*\s,2.0*\h) edge (2.8*\s,-1.0*\h);
    \path[-, every node/.append style={fill=white},out=-20,in=90] (2.5*\s,2.0*\h) edge (3.4*\s,-1.0*\h);
    \path[-, every node/.append style={fill=white},out=0,in=90] (2.5*\s,2.0*\h) edge (4.0*\s,-1.0*\h);
    
    \node at (-2.5*\s,2*\h) [stuff_fill_red, label=above:$C_6$]{};
    \node at (2.5*\s,2*\h)  [stuff_fill_red, label=above:$C_{11}$]{};
    \node at (0,2*\h) [stuff_fill_green, label=above:$C_1$] {};
    \node at (0,0.5*\h) [stuff_fill_red, label=left:$C_2$]{};
    \node at (0,-1*\h) [stuff_fill_green, label=below:$C_3$] {};
    
    \node at (-1*\s,-1.0*\h)  [stuff_fill_red,label=right:$C_{8}^{(i)}$]{};
    \node at (-1.6*\s,-1.0*\h)  [stuff_fill_red]{};
    \node at (-2.2*\s,-1.0*\h)  [stuff_fill_red]{};
    \node at (-2.8*\s,-1.0*\h)  [stuff_fill_red]{};
    \node at (-3.4*\s,-1.0*\h)  [stuff_fill_red]{};
    \node at (-4.0*\s,-1.0*\h)  [stuff_fill_red]{};
    
    \node at (1*\s,-1.0*\h)  [stuff_fill_red,label=left:$C_{10}^{(i)}$]{};
    \node at (1.6*\s,-1.0*\h)  [stuff_fill_red]{};
    \node at (2.2*\s,-1.0*\h)  [stuff_fill_red]{};
    \node at (2.8*\s,-1.0*\h)  [stuff_fill_red]{};
    \node at (3.4*\s,-1.0*\h)  [stuff_fill_red]{};
    \node at (4.0*\s,-1.0*\h) [stuff_fill_red]{};
    
\end{tikzpicture}
\end{equation}
We mark the $\mathbb{P}^1$s in pink and the elliptic curves in green. This diagram is easily extended to a weighted diagram, which encodes a 3rd root of $2 \cdot K_{C^{\bullet}_{(\mathbf{3},\mathbf{2})_{1/6}}}$. This involves placing weights $w_i \in \{ 1,2 \}$ subject to the following two rules (cf. \oref{subsec:LimitRoots}):
\begin{enumerate}
 \item Along each edge: The sum of weights is $3$.
 \item At each node: The sum of weights equals the degree in \oref{equ:CurveComponents} modulo $3$.
\end{enumerate}
It is readily verified that the following weighted diagram satisfies these rules: 
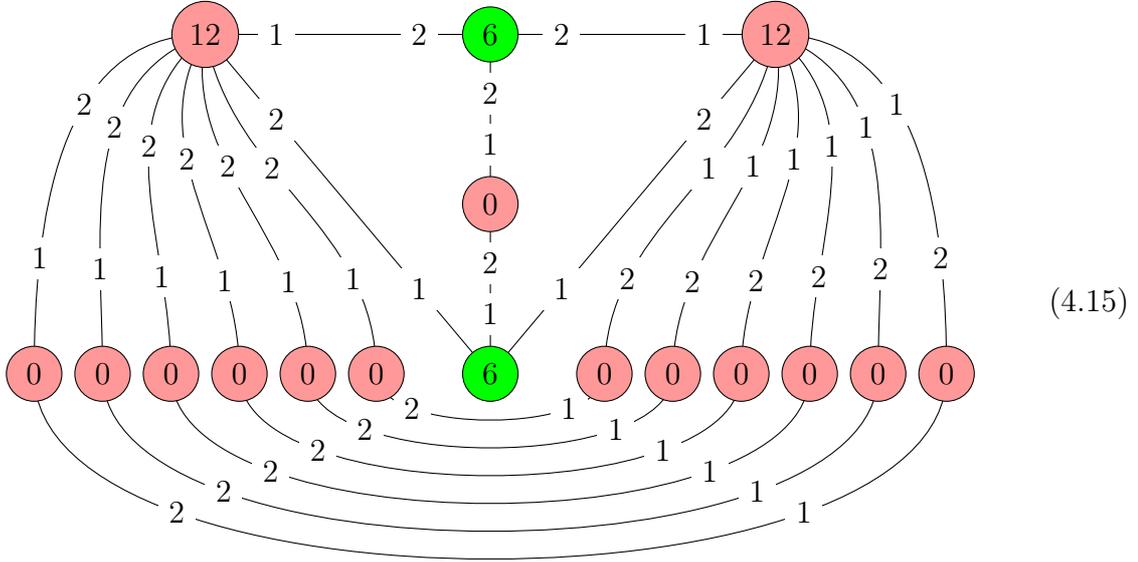
\begin{equation}
\begin{tikzpicture}[scale=0.6, baseline=(current  bounding  box.center)]
    
    \def\s{2.5};
    \def\h{2.5};
    
    \path[-, every node/.append style={fill=white}] (-2.5*\s,2.0*\h) edge node[pos=0.25] {1} node[pos=0.75] {2} (0,2*\h);
    \path[-, every node/.append style={fill=white}] (-2.5*\s,2.0*\h) edge node[pos=0.25] {2} node[pos=0.75] {1} (0,-1*\h);
    \path[-, every node/.append style={fill=white}] (0,2*\h) edge node[pos=0.35,] {2} node[pos=0.65] {1} (0,0.5*\h);
    \path[-, every node/.append style={fill=white}] (0,0.5*\h) edge node[pos=0.35,] {2} node[pos=0.65] {1} (0,-1*\h);
    \path[-, every node/.append style={fill=white}] (2.5*\s,2.0*\h) edge node[pos=0.25] {2} node[pos=0.75] {1} (0,-1*\h);
    \path[-, every node/.append style={fill=white}] (2.5*\s,2.0*\h) edge node[pos=0.25] {1} node[pos=0.75] {2} (0,2*\h);
    
    \path[-, every node/.append style={fill=white}, out = -90, in = -90, looseness = 0.7] (-1*\s,-1.0*\h) edge node[pos=0.25] {2} node[pos=0.75] {1} (1*\s,-1.0*\h);
    \path[-, every node/.append style={fill=white}, out = -90, in = -90, looseness = 0.7] (-1.6*\s,-1.0*\h) edge node[pos=0.25] {2} node[pos=0.75] {1} (1.6*\s,-1.0*\h);
    \path[-, every node/.append style={fill=white}, out = -90, in = -90, looseness = 0.7] (-2.2*\s,-1.0*\h) edge node[pos=0.25] {2} node[pos=0.75] {1} (2.2*\s,-1.0*\h);
    \path[-, every node/.append style={fill=white}, out = -90, in = -90, looseness = 0.7] (-2.8*\s,-1.0*\h) edge node[pos=0.25] {2} node[pos=0.75] {1} (2.8*\s,-1.0*\h);
    \path[-, every node/.append style={fill=white}, out = -90, in = -90, looseness = 0.7] (-3.4*\s,-1.0*\h) edge node[pos=0.25] {2} node[pos=0.75] {1} (3.4*\s,-1.0*\h);
    \path[-, every node/.append style={fill=white}, out = -90, in = -90, looseness = 0.7] (-4.0*\s,-1.0*\h) edge node[pos=0.25] {2} node[pos=0.75] {1} (4.0*\s,-1.*\h);
    
    \path[-, every node/.append style={fill=white},out=-80,in=90] (-2.5*\s,2.0*\h) edge node[pos=0.38] {2} node[pos=0.75] {1} (-1*\s,-1.0*\h);
    \path[-, every node/.append style={fill=white},out=-100,in=90] (-2.5*\s,2.0*\h) edge node[pos=0.38] {2} node[pos=0.75] {1} (-1.6*\s,-1.*\h);
    \path[-, every node/.append style={fill=white},out=-120,in=90] (-2.5*\s,2.0*\h) edge node[pos=0.38] {2} node[pos=0.75] {1} (-2.2*\s,-1.0*\h);
    \path[-, every node/.append style={fill=white},out=-140,in=90] (-2.5*\s,2.0*\h) edge node[pos=0.38] {2} node[pos=0.75] {1} (-2.8*\s,-1.0*\h);
    \path[-, every node/.append style={fill=white},out=-160,in=90] (-2.5*\s,2.0*\h) edge node[pos=0.38] {2} node[pos=0.75] {1} (-3.4*\s,-1.0*\h);
    \path[-, every node/.append style={fill=white},out=-180,in=90] (-2.5*\s,2.0*\h) edge node[pos=0.38] {2} node[pos=0.75] {1} (-4.0*\s,-1.0*\h);
    
    \path[-, every node/.append style={fill=white},out=-100,in=90] (2.5*\s,2.0*\h) edge node[pos=0.38] {1} node[pos=0.75] {2} (1*\s,-1.0*\h);
    \path[-, every node/.append style={fill=white},out=-80,in=90] (2.5*\s,2.0*\h) edge node[pos=0.38] {1} node[pos=0.75] {2} (1.6*\s,-1.0*\h);
    \path[-, every node/.append style={fill=white},out=-60,in=90] (2.5*\s,2.0*\h) edge node[pos=0.38] {1} node[pos=0.75] {2} (2.2*\s,-1.0*\h);
    \path[-, every node/.append style={fill=white},out=-40,in=90] (2.5*\s,2.0*\h) edge node[pos=0.38] {1} node[pos=0.75] {2} (2.8*\s,-1.0*\h);
    \path[-, every node/.append style={fill=white},out=-20,in=90] (2.5*\s,2.0*\h) edge node[pos=0.38] {1} node[pos=0.75] {2} (3.4*\s,-1.0*\h);
    \path[-, every node/.append style={fill=white},out=0,in=90] (2.5*\s,2.0*\h) edge node[pos=0.38] {1} node[pos=0.75] {2} (4.0*\s,-1.0*\h);
    
    \node at (-2.5*\s,2*\h) [stuff_fill_red]{$12$};
    \node at (2.5*\s,2*\h)  [stuff_fill_red]{$12$};
    \node at (0,2*\h) [stuff_fill_green] {$6$};
    \node at (0,0.5*\h) [stuff_fill_red]{$0$};
    \node at (0,-1*\h) [stuff_fill_green] {$6$};
    
    \node at (-1*\s,-1.0*\h)  [stuff_fill_red]{$0$};
    \node at (-1.6*\s,-1.0*\h)  [stuff_fill_red]{$0$};
    \node at (-2.2*\s,-1.0*\h)  [stuff_fill_red]{$0$};
    \node at (-2.8*\s,-1.0*\h)  [stuff_fill_red]{$0$};
    \node at (-3.4*\s,-1.0*\h)  [stuff_fill_red]{$0$};
    \node at (-4.0*\s,-1.0*\h)  [stuff_fill_red]{$0$};
    
    \node at (1*\s,-1.0*\h)  [stuff_fill_red]{$0$};
    \node at (1.6*\s,-1.0*\h)  [stuff_fill_red]{$0$};
    \node at (2.2*\s,-1.0*\h)  [stuff_fill_red]{$0$};
    \node at (2.8*\s,-1.0*\h)  [stuff_fill_red]{$0$};
    \node at (3.4*\s,-1.0*\h)  [stuff_fill_red]{$0$};
    \node at (4.0*\s,-1.0*\h) [stuff_fill_red]{$0$};
    
\end{tikzpicture}
\end{equation}
We then study the limit roots $P^{\circ}_{(\mathbf{3},\mathbf{2})_{1/6}}$ on the full blow-up $C^{\circ}_{(\mathbf{3},\mathbf{2})_{1/6}}$ of $C^{\bullet}_{(\mathbf{3},\mathbf{2})_{1/6}}$, which are encoded by this diagram. The degree of each such limit root $P^{\circ}_{(\mathbf{3},\mathbf{2})_{1/6}}$ is as follows:
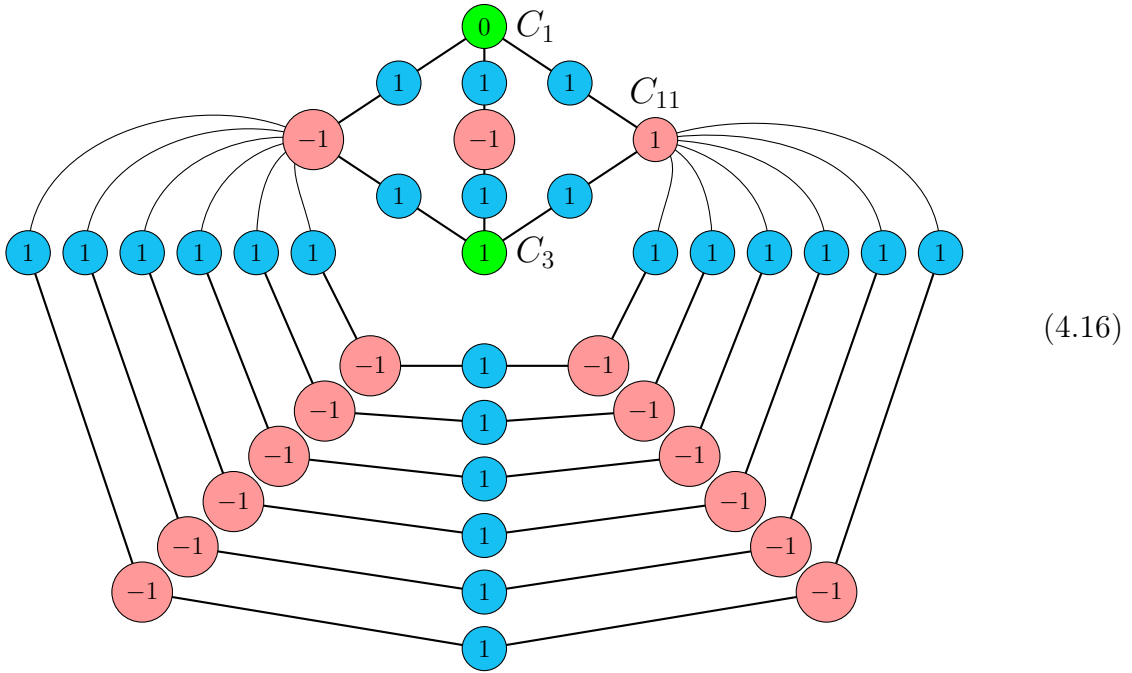
\begin{equation}
\begin{tikzpicture}[scale=0.6, every node/.style={scale=0.8}, baseline=(current  bounding  box.center)]
    
    \def\s{2.5};
    \def\h{2.5};
    
    \draw[thick] (0,1.5*\h)--(0,-0.5*\h);
    \draw[thick] (-1.5*\s,0.5*\h)--(0,-0.5*\h);
    \draw[thick] (1.5*\s,0.5*\h)--(0,-0.5*\h);
    \draw[thick] (-1.5*\s,0.5*\h)--(0,1.5*\h);
    \draw[thick] (1.5*\s,0.5*\h)--(0,1.5*\h);
    
    \draw[thick] (-1.5*\s,-0.5*\h) -- (-1*\s,-1.5*\h) -- (0,-1.5*\h) -- (1*\s,-1.5*\h) -- (1.5*\s,-0.5*\h);
    \draw[thick] (-2.0*\s,-0.5*\h) -- (-1.4*\s,-1.9*\h) -- (0,-2.0*\h) -- (1.4*\s,-1.9*\h) -- (2.0*\s,-0.5*\h);
    \draw[thick] (-2.5*\s,-0.5*\h) -- (-1.8*\s,-2.3*\h) -- (0,-2.5*\h) -- (1.8*\s,-2.3*\h) -- (2.5*\s,-0.5*\h);
    \draw[thick] (-3.0*\s,-0.5*\h) -- (-2.2*\s,-2.7*\h) -- (0,-3.0*\h) -- (2.2*\s,-2.7*\h) -- (3.0*\s,-0.5*\h);
    \draw[thick] (-3.5*\s,-0.5*\h) -- (-2.6*\s,-3.1*\h) -- (0,-3.5*\h) -- (2.6*\s,-3.1*\h) -- (3.5*\s,-0.5*\h);
    \draw[thick] (-4.0*\s,-0.5*\h) -- (-3.0*\s,-3.5*\h) -- (0,-4.0*\h) -- (3.0*\s,-3.5*\h) -- (4.0*\s,-0.5*\h);
    
    \path[-, every node/.append style={fill=white},out= 150,in=90] (-1.5*\s,0.5*\h) edge (-4.0*\s,-0.5*\h);
    \path[-, every node/.append style={fill=white},out= 160,in=90] (-1.5*\s,0.5*\h) edge (-3.5*\s,-0.5*\h);
    \path[-, every node/.append style={fill=white},out= 170,in=90] (-1.5*\s,0.5*\h) edge (-3.0*\s,-0.5*\h);
    \path[-, every node/.append style={fill=white},out= -180,in=90] (-1.5*\s,0.5*\h) edge (-2.5*\s,-0.5*\h);
    \path[-, every node/.append style={fill=white},out= -170,in=90] (-1.5*\s,0.5*\h) edge (-2.0*\s,-0.5*\h);
    \path[-, every node/.append style={fill=white},out= -160,in=90] (-1.5*\s,0.5*\h) edge (-1.5*\s,-0.5*\h);
    
    \path[-, every node/.append style={fill=white},out=20,in=90] (1.5*\s,0.5*\h) edge (4.0*\s,-0.5*\h);
    \path[-, every node/.append style={fill=white},out=10,in=90] (1.5*\s,0.5*\h) edge (3.5*\s,-0.5*\h);
    \path[-, every node/.append style={fill=white},out=0,in=90] (1.5*\s,0.5*\h) edge (3.0*\s,-0.5*\h);
    \path[-, every node/.append style={fill=white},out=-10,in=90] (1.5*\s,0.5*\h) edge (2.5*\s,-0.5*\h);
    \path[-, every node/.append style={fill=white},out=-20,in=90] (1.5*\s,0.5*\h) edge (2.0*\s,-0.5*\h);
    \path[-, every node/.append style={fill=white},out=-30,in=90] (1.5*\s,0.5*\h) edge (1.5*\s,-0.5*\h);
    
    \node at (-1.5*\s,0.5*\h) [stuff_fill_red]{$-1$};
    \node at (1.5*\s,0.5*\h)  [stuff_fill_red,label=above:\Large{$C_{11}$}]{$1$};
    \node at (0,1.5*\h) [stuff_fill_green,label=right:\Large{$C_1$}] {$0$};
    \node at (0,0.5*\h) [stuff_fill_red]{$-1$};
    \node at (0,-0.5*\h) [stuff_fill_green,label=right:\Large{$C_3$}] {$1$};
    
    \node at (-1*\s,-1.5*\h)  [stuff_fill_red]{$-1$};
    \node at (-1.4*\s,-1.9*\h)  [stuff_fill_red]{$-1$};
    \node at (-1.8*\s,-2.3*\h)  [stuff_fill_red]{$-1$};
    \node at (-2.2*\s,-2.7*\h)  [stuff_fill_red]{$-1$};
    \node at (-2.6*\s,-3.1*\h)  [stuff_fill_red]{$-1$};
    \node at (-3.0*\s,-3.5*\h)  [stuff_fill_red]{$-1$};
    
    \node at (1*\s,-1.5*\h)  [stuff_fill_red]{$-1$};
    \node at (1.4*\s,-1.9*\h)  [stuff_fill_red]{$-1$};
    \node at (1.8*\s,-2.3*\h)  [stuff_fill_red]{$-1$};
    \node at (2.2*\s,-2.7*\h)  [stuff_fill_red]{$-1$};
    \node at (2.6*\s,-3.1*\h)  [stuff_fill_red]{$-1$};
    \node at (3.0*\s,-3.5*\h) [stuff_fill_red]{$-1$};
    
    \node at (0,-1.5*\h)  [stuff_fill_blue]{$1$};
    \node at (0,-2.0*\h)  [stuff_fill_blue]{$1$};
    \node at (0,-2.5*\h)  [stuff_fill_blue]{$1$};
    \node at (0,-3.0*\h)  [stuff_fill_blue]{$1$};
    \node at (0,-3.5*\h)  [stuff_fill_blue]{$1$};
    \node at (0,-4.0*\h)  [stuff_fill_blue]{$1$};
    
    \node at (1.5*\s,-0.5*\h)  [stuff_fill_blue]{$1$};
    \node at (2.0*\s,-0.5*\h)  [stuff_fill_blue]{$1$};
    \node at (2.5*\s,-0.5*\h)  [stuff_fill_blue]{$1$};
    \node at (3.0*\s,-0.5*\h)  [stuff_fill_blue]{$1$};
    \node at (3.5*\s,-0.5*\h)  [stuff_fill_blue]{$1$};
    \node at (4.0*\s,-0.5*\h)  [stuff_fill_blue]{$1$};
    
    \node at (-1.5*\s,-0.5*\h)  [stuff_fill_blue]{$1$};
    \node at (-2.0*\s,-0.5*\h)  [stuff_fill_blue]{$1$};
    \node at (-2.5*\s,-0.5*\h)  [stuff_fill_blue]{$1$};
    \node at (-3.0*\s,-0.5*\h)  [stuff_fill_blue]{$1$};
    \node at (-3.5*\s,-0.5*\h)  [stuff_fill_blue]{$1$};
    \node at (-4.0*\s,-0.5*\h)  [stuff_fill_blue]{$1$};
    
    \node at (0,1.0*\h) [stuff_fill_blue] {$1$};
    \node at (0,0*\h) [stuff_fill_blue] {$1$};
    \node at (-0.75*\s,0*\h) [stuff_fill_blue] {$1$};
    \node at (0.75*\s,0*\h) [stuff_fill_blue] {$1$};
    \node at (-0.75*\s,1*\h) [stuff_fill_blue] {$1$};
    \node at (0.75*\s,1*\h) [stuff_fill_blue] {$1$};
    
\end{tikzpicture} \label{fig:LimitRootsQuarkDoublet}
\end{equation}
Note that we denote the blow-up $\mathbb{P}^1$s in blue and that, by construction of the limit roots, we consider a degree $1$ line bundle on each of these. It follows that the (total) degree of each such limit root $P^{\circ}_{({\mathbf{3}},\mathbf{2})_{1/6}}$ is $12$. This is expected from \oref{equ:TaskForConstruction} since it is equivalent to $\chi( P^{\circ}_{({\mathbf{3}},\mathbf{2})_{1/6}} ) = 3$. Here, we claim even more, namely that some of these limit roots have exactly three global sections.

To see this, recall from \oref{subsec:SectionTechnology} that the number of global sections of a limit root $P^{\circ}_{({\mathbf{3}},\mathbf{2})_{1/6}}$ is simply given by the sum of the sections on each irreducible component of $C^{\bullet}_{({\mathbf{3}},\mathbf{2})_{1/6}}$. Hence, we have to add the number of sections on the green and pink components in \oref{fig:LimitRootsQuarkDoublet}. From the degrees, it follows that only $C_1$, $C_3$ and $C_{11}$ support a non-zero number of sections, namely
\begin{align}
h^0 \left( C_1, P^{\circ}_{({\mathbf{3}},\mathbf{2})_{1/6}} \right) =  \left\{  \begin{array}{c} 0 \\ 1 \end{array} \right\}, \quad h^0 \left( C_3, P^{\circ}_{({\mathbf{3}},\mathbf{2})_{1/6}} \right) = 1 \, , \quad h^0 \left( C_{11}, P^{\circ}_{({\mathbf{3}},\mathbf{2})_{1/6}} \right) = 2 \, .
\end{align}
The notation for $C_1$ reminds us of the fact that on an elliptic curve, a line bundle with vanishing degree can either have $0$ or $1$ global section. Moreover, recall from \oref{subsec:LimitRoots} that the limit roots on $C_1$ are actually the 3rd roots of a line bundle of vanishing degree. In anticipation of this situation, we have already given a detailed exposition of exactly those root bundles on elliptic curves in \oref{subsec:MthRootsOfDivisors}. In particular, it follows from \oref{prop:RootsElliptic} that at least $8$ of the $9$ 3rd roots on $C_1$ satisfy $h^0 ( C_1, P^{\circ}_{({\mathbf{3}},\mathbf{2})_{1/6}} ) = 0$.

Note that also $C_3$ admits $9 = 3^{2 \cdot 1}$ different roots. However, in contrast to $C_1$, all of these roots are in the Kodaira stable regime and have exactly one section. Therefore, we conclude that we found at least $8 \cdot 9 = 72$ limit roots $P^{\circ}_{({\mathbf{3}},\mathbf{2})_{1/6}}$ with
\begin{align}
3 = h^0 \left( C^{\circ}_{({\mathbf{3}},\mathbf{2})_{1/6}}, P^{\circ}_{({\mathbf{3}},\mathbf{2})_{1/6}} \right) \, .
\end{align}
It therefore follows from our discussion in \oref{subsec:AbsenceOfVectorlikes} that there are at least 72 solutions to \oref{equ:TaskForConstruction} and consequently, also to \oref{equ:OriginalTaskForConstruction}. Let us emphasize that this analysis does not guarantee that one of these 72 solutions stems from an F-theory gauge potential in $H^4_D( \widehat{Y}_4, \mathbb{Z}(2) )$. This top-down study is reserved for future work.

Along exactly the same lines, we can argue that also $C_{(\overline{\mathbf{3}},\mathbf{1})_{-2/3}} = V( s_5, s_9 )$ and the singlet curve $C_{(\mathbf{1},\mathbf{1})_{1}} = V( s_1, s_5 )$ admit at least 72 solutions to the root bundle constraints with exactly three global sections. This leaves us to discuss the vector-like spectrum on
\begin{align}
C_{(\overline{\mathbf{3}},\mathbf{1})_{1/3}} = V \left( s_9, s_3 s_5^2 + s_6 ( s_1 s_6 - s_2 s_5 ) \right) \, .
\end{align}
On this curve we look for root bundles with (c.f. \oref{subsec:LineBundlesKbarCubeDifferent})
\begin{align}
P_{(\overline{\mathbf{3}},\mathbf{1})_{1/3}}^{\otimes 36} \sim \KB_{(\overline{\mathbf{3}},\mathbf{1})_{1/3}}^{\otimes 66} \otimes \mathcal{O}_{C_{(\overline{\mathbf{3}},\mathbf{1})_{1/3}}} \left( -30 \cdot Y_3 \right) \, , \qquad h^0( C_{(\overline{\mathbf{3}},\mathbf{1})_{1/3}}, P_{(\overline{\mathbf{3}},\mathbf{1})_{1/3}} ) = 3 \, , \label{equ:OriginalTaskForConstructionII}
\end{align}
where $Y_3 = V( s_3, s_6, s_9 )$. For this, it suffices to find solutions to
\begin{align}
P_{(\overline{\mathbf{3}},\mathbf{1})_{1/3}}^{\otimes 6} \sim \KB_{(\overline{\mathbf{3}},\mathbf{1})_{1/3}}^{\otimes 11} \otimes \mathcal{O}_{C_{(\overline{\mathbf{3}},\mathbf{1})_{1/3}}} \left( -5 \cdot Y_3 \right) \, , \qquad h^0( C_{(\overline{\mathbf{3}},\mathbf{1})_{1/3}}, P_{(\overline{\mathbf{3}},\mathbf{1})_{1/3}} ) = 3 \, . \label{equ:TaskForConstructionII}
\end{align}
To this end we consider the deformation $C_{(\overline{\mathbf{3}},\mathbf{1})_{1/3}} \to C^{\bullet}_{(\overline{\mathbf{3}},\mathbf{1})_{1/3}}$ with
\begin{align}
C^\bullet_{(\overline{\mathbf{3}},\mathbf{1})_{1/3}} &= V ( s_9, s_5 - s_6 ) \cup V( s_9, s_3 - s_6 ) \cup V( s_9, s_5 + s_6 ) \equiv Q_1 \cup Q_2 \cup Q_3 \, ,
\end{align}
which is obtained from
\begin{align}
s_1 \to s_6 - s_3 \, , \qquad s_2 \to s_5 - \prod_{i = 1}^{11}{x_i} \, , \qquad s_9 \to \prod_{i = 1}^{11}{x_i} \, ,
\end{align}
and \emph{generic} $s_3$, $s_5$, $s_6$. Therefore, $C_{(\overline{\mathbf{3}},\mathbf{1})_{1/3}} \to C_{(\overline{\mathbf{3}},\mathbf{1})_{1/3}}^\bullet$ turns this matter curve into three nodal curves, each of which looks like the curve $C^{\bullet}_{(\mathbf{3},\mathbf{2})_{1/6}}$ that we discussed above. From this point on, we can again employ the limit root techniques. On a technical level, the only distinction to the constructions presented for $C^{\bullet}_{(\mathbf{3},\mathbf{2})_{1/6}}$ is that we have to carefully take into account the line bundle contributions from the Yukawa point $Y_3$. Also, the resulting weighted diagrams become very large since the nodal curve $C^\bullet_{(\overline{\mathbf{3}},\mathbf{1})_{1/3}}$ has 51 irreducible components. For these reasons, it suffices to state that we can argue for at least $36^2\cdot 35^4$ solutions to \oref{equ:TaskForConstructionII}. Details are provided in \oref{sec:ExampleGeometry}. We reserve a detailed top-down study of which root bundles arise from an F-theory gauge potential in $H^4_D( \widehat{Y}_4, \mathbb{Z}(2) )$ for the future.

\section{Conclusion and Outlook} \label{sec:SummaryAndOutlook}

This work is motivated by the frequent appearance of fractional powers of line bundles when studying vector-like spectra of globally consistent 4d F-theory Standard Models with three chiral families and gauge coupling unification \cite{Cvetic:2019gnh}. In these models, the vector-like spectra on the low-genus matter curves are naively encoded in cohomologies of a line bundle that is identified with a fractional power of the canonical bundle. On high-genus curves, these fractional powers of the canonical bundle are further modified by contributions from Yukawa points. In order to understand these fractional bundles, we have analyzed their origin and nature. 

First, in \cref{subsec:RootBundleAppearance}, we analyzed the origin of such fractional powers of line bundles. We recalled that the vector-like spectra are not specified by a $G_4$-flux, but rather by its associated gauge potential in the Deligne cohomology $H^4_D( \widehat{Y}_4, \mathbb{Z}(2) )$ \cite{Bies:2014sra,Bies:2017fam,Bies:2018uzw}. In fact, a given $G_4$-flux has many such gauge potentials. To see this, recall that in the dual M-theory picture, \mbox{$G_4 = d C_3$}, where $C_3$ is the internal M-theory 3-form potential. Any other 3-form potential $C_3^\prime$ with closed $C_3^\prime - C_3$ still has $G_4$ as its field strength. Such closed 3-form potentials are encoded by the 
intermediate Jacobian $J^2( \widehat{Y}_4 )$ in the F-theory geometry. While it is well-defined in theory, it can be very challenging to associate even a single gauge potential in $H^4_D( \widehat{Y}_4, \mathbb{Z}(2) )$ to a given $G_4$-flux in practice. We were able to tie the appearance of fractional powers of line bundles to exactly this challenge.

For an F-theory model, we need an F-theory gauge potential, i.e., a class in the Deligne cohomology group $A \in H^4_D ( \widehat{Y}_4, \mathbb{Z} )$. This will be specified as $ \widehat{\gamma} (\mathcal{A})$ for some \linebreak ``potential" $\mathcal{A} \in \mathrm{CH}^2( \widehat{Y}_4, \mathbb{Z} )$. We found that the geometry determines a class \linebreak $A^\prime = \widehat{\gamma}(\mathcal{A}^\prime) \in H^4_D( \widehat{Y}_4, \mathbb{Z}(2) )$ and an integer $\xi \in \mathbb{Z}_{> 0}$ such that $\mathcal{A}$ is subject to the two constraints:
\begin{align}
\gamma( \mathcal{A} ) = G_4 \, , \qquad \xi \cdot \widehat{\gamma}( \mathcal{A} ) \sim \widehat{\gamma}( \mathcal{A}^\prime ) \, . \label{equ:ConditionsII}
\end{align}
The condition $\gamma( \mathcal{A} ) = G_4$ immediately follows from \cref{equ:CommutativeDiagramLifts} and it means that $A = \widehat{\gamma}( \mathcal{A} )$ is an F-theory gauge potential for the given $G_4$-flux. In the dual M-theory picture, it states that the 3-form potential $C_3$ satisfies $d C_3 = G_4$.  We illustrated with several examples that the absence of chiral exotics in the F-theory Standard Models \cite{Cvetic:2019gnh} boils down to the second constraint.

It is important to notice that the gauge potential $A = \widehat{\gamma}( \mathcal{A} )$ specified by the two conditions in \cref{equ:ConditionsII} is in general not unique. The collection of all $\xi$-th roots of $\widehat{\gamma}( \mathcal{A}^\prime )$ (if non-empty) is a coset of the group of all $\xi$-th roots of $0$. In particular, the number of solutions is $\xi^{2 \cdot \mathrm{dim}_{\mathbb{C}} \left( J^2( \widehat{Y}_4 ) \right)}$. All these solutions lead to the same chiral spectrum  \eqref{equ:ChiralIndex} since they all have the same degree when restricted to the curves $C_{\mathbf{R}}$, and hence, the same index. However, they could differ in their actual spectrum \eqref{actual}. This extra flexibility is the key tool that we intend to use to produce a desirable spectrum such as the MSSM.

In theory, we could proceed by studying gauge potentials $A = \widehat{\gamma}( \mathcal{A} ) \in H^4_D( \widehat{Y}_4, \mathbb{Z}(2) )$ subject to \cref{equ:ConditionsII}. However, in practice it seems more efficient to proceed with the algebraic cycle $\mathcal{A}^\prime$, which we could construct explicitly in the largest currently-known class of globally consistent F-theory Standard Models without chiral exotics and gauge coupling unification \cite{Cvetic:2019gnh}. Hence, we have a sufficient level of arithmetic control over $A^\prime = \widehat{\gamma}( \mathcal{A}^\prime )$. In particular, we can identify the $\mathbb{Z}$-Cartier divisor $D_{\mathbf{R}}( \mathcal{A}^\prime )$ induced from $\mathcal{A}^\prime$ on the matter curve $C_{\mathbf{R}}$. It follows that
\begin{align}
\xi \cdot D_{\mathbf{R}}( \mathcal{A} ) \sim D_{\mathbf{R}}( \mathcal{A}^\prime ) \, . \label{equ:ConclusionRoots}
\end{align}
Divisors $D_{\mathbf{R}}( \mathcal{A} )$, which solve this equation for given $D_{\mathbf{R}}( \mathcal{A}^\prime )$ and $\xi$, are called \emph{root divisors} and their associated line bundles are \emph{root bundles}. They exist if and only if $\xi$ divides the degree of $D_{\mathbf{R}}( \mathcal{A}^\prime )$. Such root bundles are by no means unique. For example, spin bundles on a genus $g$ matter curve $C_{\mathbf{R}}$ are 2nd roots of the canonical bundle $K_{{\mathbf{R}}}$ and there are $2^{2g}$ such roots. Similarly, on a genus $g$-curve, \oref{equ:ConclusionRoots} admits $\xi^{2g}$ solutions (if they exist).


It is well-known that not all spin bundles have the same number of global sections. Rather, roughly half of the spin bundles on a curve $C_{\mathbf{R}}$ have an odd number of global sections and the remaining ones have an even number \cite{atiyah1971riemann,mumford1971theta}. More generally, we can therefore expect that the gauge potentials $A = \widehat{\gamma}( \mathcal{A} )$ subject to \cref{equ:ConditionsII} lead to different vector-like spectra. This mirrors the physical expectation that inequivalent F-theory gauge potentials --- equivalently, in the dual M-theory picture, two 3-form potentials $C_3$ and $C_3^\prime$ that differ by a closed 3-form --- will in general lead to different vector-like spectra. This was anticipated e.g. in \cite{Bies:2014sra,Bies:2017fam,Bies:2018uzw}.

In general, only a subset of the root divisors in \cref{equ:ConclusionRoots} are induced from F-theory gauge potentials in $H^4_D( \widehat{Y}_4, \mathbb{Z}(2) )$. While this work does not answer the important \mbox{question} of which root divisors are induced from F-theory potentials, we hope that this work initiates and facilitates this study by providing a systematic analysis of all root bundles and spin bundles on the matter curves. Our goal in this work was to identify combinations of root bundles and spin bundles on the matter curves, such that their global sections satisfy the physical demand of the presence/absence of vector-like pairs.

While we expect that our techniques apply more generally, we have focused on the largest currently-known class of globally consistent F-theory Standard Models with \mbox{realistic} chiral spectra \cite{Cvetic:2019gnh}, which emphasizes the genuine appearance of root \mbox{bundles} in vector-like spectra of F-theory compactifications. It should be mentioned that the \mbox{background} $G_4$-flux in these F-theory Standard Models models does not only lead to realistic chiral spectra, but also allows cancelation of the D3-tadpole and ensures \mbox{masslessness} of the $U(1)$-gauge boson. We summarize the involved technical steps in the derivation of these root bundle constraints in \cref{sec:LineBundleComputations}. This derivation heavily relies on a detailed understanding of the elliptically fibered 4-fold F-theory geometry $\widehat{Y}_4$, including intersection numbers in the fiber over the Yukawa points. We supplement the earlier works \cite{Klevers:2014bqa,Cvetic:2015txa,Cvetic:2019gnh} by providing a complete list of all fiber intersection numbers in \cref{sec:FibreStructure}.

Our approach to identifying root and spin bundles on the matter curves, whose \mbox{cohomologies} are physically desired for the presence/absence of vector-like pairs, is \mbox{inspired} by the work in \cite{2004math4078C}, which gives a diagrammatic description of root bundles on nodal curves $C_{\mathbf{R}}^\bullet$. More explicitly, it relates these roots with so-called \emph{limit roots} on (partial) blow-ups $C_{\mathbf{R}}^\circ$ of $C_{\mathbf{R}}^\bullet$. We summarized these ideas in \cref{sec:LimitRoots}, and then introduced counting procedures for the global sections. In order to fully appreciate this finding, recall from \cite{Bies:2020gvf} that in general one will merely find a lower bound. The \mbox{argument} that we provide in this work is stronger -- it provides an exact count of the global sections of limit roots on \emph{full} blow-ups of $C_{\mathbf{R}}^\bullet$. This observation may be interesting in its own right since it provides a combinatoric access to Brill-Noether theory of limit roots. We demonstrated this for a nodal curve -- the \emph{Holiday lights} $H^\bullet$. This curve is of compact type and its only blow-up that is to be considered for the limit root is its full blow-up. Our approach then allowed us to identify exactly how many limit roots possess a certain number of global sections. It will be an interesting mathematical question to extend these ideas to partial blow-ups. We reserve this analysis for future work.

Given these insights on root bundles on nodal curves $C_{\mathbf{R}}^\bullet$, it remained to extract information on root bundles on actual matter curves $C_{\mathbf{R}}$ in F-theory compactifications. As the latter are typically smooth, it is natural to wonder what we can say about (limit) roots when traced along a deformation $C_{\mathbf{R}}^\bullet \to C_{\mathbf{R}}$. In particular, we can wonder if there are deformations of $C_\mathbf{R}$ that are conducive for a more fruitful analysis. As we have already mentioned, curves of compact type, such as the \emph{Holiday lights}, are prime candidates. The lack of cycles in their dual graph limits the number of possible weighted graphs, so much so that we have a complete understanding of the limit roots and their global sections. In contrast, the dual graphs of the deformed matter curves $C^\bullet_\mathbf{R}$ in explicit geometries are more complex in which there are multiple weighted subgraphs, and limit roots over partial blow-ups. In particular, some singularities on the curve still remain in its partial blow-up, and it is therefore far more challenging to count the sections. It would be useful to compare these two examples in more depth and to determine exactly what features of the dual graph allow for better section-counting. One obvious feature is the cyclomatic number, which happens to be the first Betti number of a graph when viewed as a 1-dimensional simplicial complex. Curves of compact type have zero cyclomatic number, and thus, are topologically simple. Subsequently, we can explore possible ways of deforming $C_\mathbf{R}$ to a nodal curve whose dual graph has these desirable features.

In this work, we have focused on deformations $C_{\mathbf{R}} \to C_{\mathbf{R}}^\bullet$ which arise naturally by modifying the defining polynomials in a concrete base geometry $B_3$. Most curves that we encountered in this way had planar dual graphs. Still, for the most involved matter curve discussed in this article, the dual graph is non-planar. The subject of planarity raises many interesting questions and applications in graph theory \cite{Wagner1937, MacLane1937, 2009JAMS...22..309G, CHMEISS199761, felsner2012geometric}. However, the geometric significance for a nodal curve to have a non-planar dual graph is not mentioned in the literature to our knowledge \cite{busonero2006combinatorial, arbarello2011algebraic}. It is possible that planarity does not play a role in the geometry of nodal curves. Indeed, the curve associated to the well-known non-planar graph $K_{3, 3}$ is quite ordinary. Nevertheless, it would be useful to explore this feature as it raises the question of whether there are better ways to represent a given dual graph.

For a physical application, we have studied vector-like spectra of F-theory Standard Models without chiral exotics in \cref{sec:ApplicationToMSSMs}. In aiming for MSSM constructions, we should wonder what we can say about the global sections of a root $P_{\mathbf{R}}^\bullet$ as we trace it to a root $P_{\mathbf{R}}$ along a deformation \mbox{$C_{\mathbf{R}}^\bullet \to C_{\mathbf{R}}$}. In this work, we did not attempt to provide a complete answer to this question. Rather, we recalled that a certain behavior of the cohomologies along such a deformation is known. This is called \emph{upper semi-continuity} and it means that the number of global sections cannot increase when tracing a root $P_{\mathbf{R}}^\bullet$ on $C_{\mathbf{R}}^\bullet$ to a root $P_{\mathbf{R}}$ on $C_{\mathbf{R}}$. Put differently,
\begin{align}
h^0 \left( C_{\mathbf{R}}, P_{\mathbf{R}} \right) \leq h^0 \left( C_{\mathbf{R}}^\bullet, P_{\mathbf{R}}^\bullet \right) \, .
\end{align}
For F-theory MSSM constructions, it is important to understand (limit) roots on the Higgs curve with $h^0 ( C_{(\mathbf{1},\mathbf{2})_{-1/2}}, P_{(\mathbf{1},\mathbf{2})_{-1/2}} ) = 4 = 1 + \chi( P_{(\mathbf{1},\mathbf{2})_{-1/2}} )$. While we can construct roots $P_{(\mathbf{1},\mathbf{2})_{-1/2}}^\bullet$ with $h^0 ( C_{(\mathbf{1},\mathbf{2})_{-1/2}}^\bullet, P_{(\mathbf{1},\mathbf{2})_{-1/2}}^\bullet ) = 4$, upper semi-continuity does then not guarantee that $P_{(\mathbf{1},\mathbf{2})_{-1/2}}^\bullet \to P_{(\mathbf{1},\mathbf{2})_{-1/2}}$ along $C_{(\mathbf{1},\mathbf{2})_{-1/2}}^\bullet \to C_{(\mathbf{1},\mathbf{2})_{-1/2}}$ yields roots with 4 global sections. Rather, the roots could lose sections along this transition (cf. \cite{Bies:2020gvf}). To our knowledge, a sufficient criterion that identifies the Higgs roots $P^\bullet_{(\mathbf{1},\mathbf{2})_{-1/2}}$ that do not lose sections is currently unknown. However, given the physical significance of such a condition, we hope to return to this interesting question in the future.

Even a subset of (limit) roots that do not lose sections along $C^\bullet_{\mathbf{R}} \to C_{\mathbf{R}}$ is valuable. We identified a family of such roots $P^\bullet_{\mathbf{R}}$. Namely, for a root with \mbox{$h^0 \left( C_{\mathbf{R}}^\bullet, P_{\mathbf{R}}^\bullet \right) = \chi( P_{\mathbf{R}} ) \geq 0$}, it follows from upper semi-continuity that $h^0 \left( C_{\mathbf{R}}, P_{\mathbf{R}} \right) = h^0 \left( C_{\mathbf{R}}^\bullet, P^\bullet_{\mathbf{R}} \right)$. Any such root thus satisfies $h^i \left( C_{\mathbf{R}}, P_{\mathbf{R}} \right) = \left( \chi( P_{\mathbf{R}} ), 0 \right)$, which means it describes a zero mode spectrum on $C_{\mathbf{R}}$ without vector-like pairs. For example, in the F-theory MSSM constructions, this is a desired feature for the representations $C_{({\mathbf{3}},\mathbf{2})_{1/6}}$, $C_{(\overline{\mathbf{3}},\mathbf{1})_{-2/3}}$, $C_{(\overline{\mathbf{3}},\mathbf{1})_{1/3}}$, $C_{({\mathbf{1}},\mathbf{1})_{1}}$ for which vector-like pairs are exotic, i.e. have thus far not been observed in particle accelerators.

We have applied these techniques to a particular F-theory geometry among the largest \mbox{currently-known} class of globally consistent F-theory Standard Model constructions without chiral exotics and gauge coupling unification \cite{Cvetic:2019gnh}. To this end, we worked with the base space ${{B}_3 = P_{39}}$. This 3-fold is one of the triangulations of the 39-th \mbox{polyhedron} of the Kreuzer-Skarke list\cite{Kreuzer:1998vb}, hence the name. In this space, we have \mbox{explicitly} \mbox{deformed} the matter curves $C_{\mathbf{R}}$ to nodal curves $C_{\mathbf{R}}^\bullet$. On those nodal curves, we could then easily construct limit roots on the full blow-up $C_{\mathbf{R}}^\circ$ of $C_{\mathbf{R}}^\bullet$ which have exactly 3 sections. We collect details on the base space ${{B}_3 = P_{39}}$ and limit roots on the blow-up of a genus \mbox{$g = 82$} matter curve in \cref{sec:ExampleGeometry}. In future works, we hope to investigate which of these desired root bundles are realized from F-theory gauge potentials in $H^4_D( \widehat{Y}_4, \mathbb{Z}(2) )$.

To fully appreciate these findings, let us point out that this task cannot be \mbox{performed} with state-of-the-art algorithms such as \cite{ToricVarietiesProject} unless one explicitly specifies the line \mbox{bundle} divisor in question. In past works \cite{Bies:2017fam, Bies:2018uzw}, such constructions were described. A \mbox{computer} model of such line bundles (by dualizing the corresponding ideal sheaf) \mbox{requires} Gröbner basis computations. Even by the use of state-of-the-art algorithms such as \cite{DGPS}, the \mbox{involved} geometries resulted in excessively long runtimes and heavy memory \mbox{consumption}. By approaching root bundles from limit roots on full blow-ups, these complications are circumvented at the cost of studying deformation theory.

This work provides a constructive approach to identifying limit root bundles on full blow-ups of a nodal curve with specific number of global sections. Since our approach is completely constructive, we anticipate a computer implementation which can find all such limit roots. For this, one would work out all of the weighted diagrams associated to the dual graph of a nodal curve $C_{\mathbf{R}}^\bullet$, and then identify the limit roots with the desired number of global sections. In generalizing this approach even further, we anticipate a scan over many of the F-theory Standard model geometries in \cite{Cvetic:2019gnh}. By employing \mbox{state-of-the-art} data-science and machine learning techniques, it can be expected that such a scan will lead to a more refined understanding of F-theory Standard Model constructions. We hope to return to this fascinating question in the near future.

\paragraph{Acknowledgements}
We thank Gavril Farkas, Iñaki García-Etxebarria, Ling Lin and Claire Voisin for valuable discussions. M.B., R.D. and M.O.~are partially supported by NSF grant DMS 2001673 and by the Simons Foundation Collaboration grant \#390287 on ``Homological Mirror Symmetry''. The work of M.C.~and M.L.~ is supported by DOE Award DE-SC0013528Y. M.B.~and M.C.~further acknowledges support by the Simons Foundation Collaboration grant \#724069 on ``Special Holonomy in Geometry, Analysis and Physics''. M.C.~thanks the Slovenian Research Agency No.~P1-0306 and the Fay R.~and Eugene L.~Langberg Chair for their support.

\appendix

\section{Fiber structure of F-theory Standard Models} \label{sec:FibreStructure}

In this section, we investigate the fiber structure of the resolved 4-fold with \linebreak${SU ( 3 ) \times SU ( 2 ) \times U(1)_Y}$ gauge symmetry as employed in the largest currently-known class of globally consistent F-theory Standard Models without chiral exotics and gauge coupling unification \cite{Cvetic:2019gnh}. We work out the intersection numbers in the fibers over generic points of the gauge divisors, matter curves and Yukawa loci. The knowledge of the fiber structure determines the vector-like spectrum in this F-theory vacuum.
\subsection{Away from Matter Curves}

\subsubsection{\texorpdfstring{$SU(2)$}{SU(2)} Gauge Divisor}

This gauge divisor is $V( s_3 )$. Here, the defining equation of $P_{F_{11}}$ factors as 
\begin{equation}\label{eq:F11Factors}
  \resizebox{0.85\textwidth}{!}{%
$\begin{aligned}
  p_{F_{11}} = e_1 \left( e_1 e_2^2 e_3 e_4^4 s_1 u^3 + e_2^2 e_3^2 e_4^2 s_2 u^2 v + e_1 e_2 e_4^3 s_5 u^2 w + e_2 e_3 e_4 s_6 u v w + s_9 v w^2 \right) \, .
\end{aligned}$%
}
\end{equation}
The Cartan divisors are therefore as follows
\begin{equation}
  \resizebox{0.89\textwidth}{!}{%
$\begin{aligned}
  &D_0^{SU(2)} = V \left( e_1 e_2^2 e_3 e_4^4 s_1 u^3 + e_2^2 e_3^2 e_4^2 s_2 u^2 v + e_1 e_2 e_4^3 s_5 u^2 w + e_2 e_3 e_4 s_6 u v w + s_9 v w^2, s_3 \right)\, ,\\
  &D_1^{SU(2)} = V \left(e_1,s_3\right)\, .
\end{aligned}$%
}
\end{equation}
The intersection numbers in the fiber over a generic base point $p \in V( s_3 )$ are:
\begin{align}
  \adjustbox{max width=0.72\textwidth}{
\begin{tabular}{c|cc|c}
\toprule
$D_i^{SU(2)} \cdot D_j^{SU(2)} \cdot \hat{\pi}^{-1} \left( p \right)$ & $D_0^{SU(2)}$ & $D_1^{SU(2)}$ & $U(1)_Y$ \\
\midrule
$D_0^{SU(2)}$ & -2 & 2 & 0 \\
$D_1^{SU(2)}$ & 2 & -2 & 0 \\
\bottomrule
\end{tabular}}
\end{align}

\subsubsection{\texorpdfstring{$SU(3)$}{SU(3)} Gauge Divisor}

This $SU(3)$ gauge divisor $V( s_9 )$ relates to the Cartan divisors as follows:
\begin{equation}
  \resizebox{0.89\textwidth}{!}{%
$\begin{aligned}
  &D_0^{SU(3)} = V \left( e_1^2 e_2 e_3 e_4^4 s_1 u^2 + e_1 e_2 e_3^2 e_4^2 s_2 u v + e_2 e_3^3 s_3 v^2 + e_1^2 e_4^3 s_5 u w + e_1 e_3 e_4 s_6 v w, s_9 \right) \, , &&\\
&D_1^{SU(3)} = V \left( e_2, s_9 \right) \, , \qquad D_2^{SU(3)} = V \left( u, s_9 \right) \, .&&
\end{aligned}$%
}
\end{equation}
The intersection numbers in the fiber over a generic base point $p \in V( s_9 )$ are:
\begin{align}
  \adjustbox{max width=0.83\textwidth}{
\begin{tabular}{c|ccc|c}
\toprule
$D_i^{SU(3)} \cdot D_j^{SU(3)} \cdot \hat{\pi}^{-1} \left( p \right)$ & $D_0^{SU(3)}$ & $D_1^{SU(3)}$ & $D_3^{SU(3)}$ & $U(1)_Y$ \\
\midrule
$D_0^{SU(3)}$ & -2 & 1 & 1 & 0 \\
$D_1^{SU(3)}$ & 1 & -2 & 1 & 0 \\
$D_2^{SU(3)}$ & 1 &  1 & -2 & 0 \\
\bottomrule
\end{tabular}}
\end{align}

\subsection{Over Matter Curves}\label{app_Mattersurfaces}

\subsubsection*{Intersection Structure over \texorpdfstring{$\mathbf{C_{(\mathbf{3},\mathbf{2})_{1/6}}}$}{C3216} away from Yukawa Loci}

Over the matter curves, singularity enhancements occur. They are geometrically related to the presence of new $\mathbb{P}^1$-fibrations, of which linear combinations eventually serve as matter surfaces. Over $C_{(\mathbf{3},\mathbf{2})_{1/6}} = V( s_3, s_9 )$ the following $\mathbb{P}^1$-fibrations are present:
\begin{align}
\begin{split}
  &\mathbb{P}_{0}^1 \left( (\mathbf{3},\mathbf{2})_{1/6} \right) = V( s_3, s_9, e_1 e_2 e_3 e_4^3 s_1 u^2 + e_2 e_3^2 e_4 s_2 u v + e_1 e_4^2 s_5 u w + e_3 s_6 v w ) \, , \\
  &\mathbb{P}_{1}^1 \left( (\mathbf{3},\mathbf{2})_{1/6} \right) = V( s_3, s_9, e_1 ) \, , \qquad \mathbb{P}_{2}^1 \left( (\mathbf{3},\mathbf{2})_{1/6} \right) = V( s_3, s_9, e_4 ) \, , \\
  &\mathbb{P}_{3}^1 \left( (\mathbf{3},\mathbf{2})_{1/6} \right) = V( s_3, s_9, u ) \, , 
\qquad \hspace{0.3em} \mathbb{P}_{4}^1 \left( (\mathbf{3},\mathbf{2})_{1/6} \right) = V( s_3, s_9, e_2 ) \, .
\end{split}
\end{align}
These $\mathbb{P}^1$-fibrations relate to restrictions of the $SU(3)$ and $SU(2)$ Cartan divisors:
\begin{align}\label{table:BifundCartanSplit}
  \adjustbox{max width=0.8\textwidth}{
\begin{tabular}{c|c}
\toprule
Original & Split components over $C_{\mathbf{R}}$ \\
\midrule
$D_0^{SU(2)}$ & $\mathbb{P}_{0}^1 \left( (\mathbf{3},\mathbf{2})_{1/6} \right) + \mathbb{P}_{2}^1 \left( (\mathbf{3},\mathbf{2})_{1/6} \right) + \mathbb{P}_{3}^1 \left( (\mathbf{3},\mathbf{2})_{1/6} \right) + \mathbb{P}_{4}^1 \left( (\mathbf{3},\mathbf{2})_{1/6} \right)$ \\
$D_1^{SU(2)}$ & $\mathbb{P}_{1}^1 \left( (\mathbf{3},\mathbf{2})_{1/6} \right)$ \\
\midrule
$D_0^{SU(3)}$ & $\mathbb{P}_{0}^1 \left( (\mathbf{3},\mathbf{2})_{1/6} \right) + \mathbb{P}_{1}^1 \left( (\mathbf{3},\mathbf{2})_{1/6} \right) + \mathbb{P}_{2}^1 \left( (\mathbf{3},\mathbf{2})_{1/6} \right)$ \\
$D_1^{SU(3)}$ & $\mathbb{P}_{4}^1 \left( (\mathbf{3},\mathbf{2})_{1/6} \right)$ \\
$D_2^{SU(3)}$ & $\mathbb{P}_{3}^1 \left( (\mathbf{3},\mathbf{2})_{1/6} \right)$ \\
\bottomrule
\end{tabular}}
\end{align}
Over $p \in C_{(\mathbf{3},\mathbf{2})_{1/6}}$ which is not a Yukawa point, these $\mathbb{P}^1$-fibrations intersect as follows:
\begin{align}\label{table:BifundP1Int}
\adjustbox{max width=0.9\textwidth}{
\begin{tabular}{c|ccccc}
\toprule
 & $\mathbb{P}_{0}^1 \left( (\mathbf{3},\mathbf{2})_{1/6} \right)$ & $\mathbb{P}_{1}^1 \left( (\mathbf{3},\mathbf{2})_{1/6} \right)$ & $\mathbb{P}_{2}^1 \left( (\mathbf{3},\mathbf{2})_{1/6} \right)$ & $\mathbb{P}_{3}^1 \left( (\mathbf{3},\mathbf{2})_{1/6} \right)$ & $\mathbb{P}_{4}^1 \left( (\mathbf{3},\mathbf{2})_{1/6} \right)$ \\
\midrule
$\mathbb{P}_{0}^1 \left( (\mathbf{3},\mathbf{2})_{1/6} \right)$ & -2 & 1 & 0 & 0 & 1 \\
$\mathbb{P}_{1}^1 \left( (\mathbf{3},\mathbf{2})_{1/6} \right)$ & 1 & -2 & 1 & 0 & 0 \\
$\mathbb{P}_{2}^1 \left( (\mathbf{3},\mathbf{2})_{1/6} \right)$ & 0 & 1 & -2 & 1 & 0 \\
$\mathbb{P}_{3}^1 \left( (\mathbf{3},\mathbf{2})_{1/6} \right)$ & 0 & 0 & 1 & -2 & 1 \\
$\mathbb{P}_{4}^1 \left( (\mathbf{3},\mathbf{2})_{1/6} \right)$ & 1 & 0 & 0 & 1 & -2 \\
\bottomrule
\end{tabular}}
\end{align}
The intersection numbers between the $\mathbb{P}^1$-fibrations over $C_{(\mathbf{3},\mathbf{2})_{1/6}}$ and the pullbacks of the Cartan divisors are readily computed as follows:
\begin{align}
\adjustbox{max width=0.9\textwidth}{
\begin{tabular}{c|cc|ccc|c}
\toprule
 & $D_0^{SU(2)}$ & $D_1^{SU(2)}$ & $D_0^{SU(3)}$ & $D_1^{SU(3)}$ & $D_2^{SU(3)}$ & $U(1)_Y$ \\
\midrule
$\mathbb{P}_{0}^1 \left( (\mathbf{3},\mathbf{2})_{1/6} \right)$ & -1 & 1 & -1 & 1 & 0 & -1/6 \\
$\mathbb{P}_{1}^1 \left( (\mathbf{3},\mathbf{2})_{1/6} \right)$ & 2 & -2 & 0 & 0 & 0 & 0 \\
$\mathbb{P}_{2}^1 \left( (\mathbf{3},\mathbf{2})_{1/6} \right)$ & 0 & 1 & 0 & 0 & 1 & 1/6 \\
$\mathbb{P}_{3}^1 \left( (\mathbf{3},\mathbf{2})_{1/6} \right)$ & 0 & 0 & 1 & 1 & -2 & 0 \\
$\mathbb{P}_{4}^1 \left( (\mathbf{3},\mathbf{2})_{1/6} \right)$ & 0 & 0 & 1 & -2 & 1 & 0 \\
\bottomrule
\end{tabular}}
\end{align}
The matter surfaces $S_{(\mathbf{3},\mathbf{2})_{1/6}}^{(a)}$ over $C_{(\mathbf{3},\mathbf{2})_{1/6}}$ are linear combinations of the above \mbox{$\mathbb{P}^1$-fibrations}. We use $\mathbf{P}$ to denote such a linear combination compactly. Explicitly, $\mathbf{P}$ is a list of the multiplicities with which these $\mathbb{P}^1$-fibrations appear in the above order:
\begin{equation}
  \resizebox{0.65\textwidth}{!}{%
$\begin{aligned}
  \mathbf{P} = \left( 0, 1, 0, 4, 0 \right) \quad \leftrightarrow \quad 1 \cdot \mathbb{P}^1_{1} \left( (\mathbf{3},\mathbf{2})_{1/6} \right) + 4 \cdot \mathbb{P}^1_{3} \left( (\mathbf{3},\mathbf{2})_{1/6} \right) \, . \nonumber
\end{aligned}$%
}
\end{equation}
We apply $\mathbf{\beta}$ to indicate the Cartan charges of such a linear combination, these notations will also be adopted for the other matter curves. All that said, the matter surfaces over $C_{(\mathbf{3},\mathbf{2})_{1/6}}$ take the following form:
\begin{equation}
  \resizebox{0.8\textwidth}{!}{%
  \begin{tabular}{ccc||ccc}
\toprule
Label & $\vec{P}$ & $\mathbf{\beta}$ & Label & $\vec{P}$ & $\mathbf{\beta}$ \\
\midrule
$S_{(\mathbf{3},\mathbf{2})_{1/6}}^{(1)}$ & $\left( 0,0,1,0,0 \right)$ & $\left( 1 \right) \otimes \left( 0, 1 \right)$ 
& $S_{(\mathbf{3},\mathbf{2})_{1/6}}^{(4)}$ & $\left( 0,1,1,1,0 \right)$ & $\left( -1 \right) \otimes \left( 1, -1 \right)$ \\
$S_{(\mathbf{3},\mathbf{2})_{1/6}}^{(2)}$ & $\left( 0,1,1,0,0 \right)$ & $\left( -1 \right) \otimes \left( 0, 1 \right)$ 
& $S_{(\mathbf{3},\mathbf{2})_{1/6}}^{(5)}$ & $\left( 0,0,1,1,1 \right)$ & $\left( 1 \right) \otimes \left( -1, 0 \right)$\\
$S_{(\mathbf{3},\mathbf{2})_{1/6}}^{(3)}$ & $\left( 0,0,1,1,0 \right)$ & $\left( 1 \right) \otimes \left( 1, -1 \right)$ 
& $S_{(\mathbf{3},\mathbf{2})_{1/6}}^{(6)}$ & $\left( 0,1,1,1,1 \right)$ & $\left( -1 \right) \otimes \left( -1, 0 \right)$ \\
\bottomrule
\end{tabular}%
}
\end{equation}

\subsubsection*{Intersection Structure over \texorpdfstring{$\mathbf{C_{(\mathbf{1},\mathbf{2})_{-1/2}}}$}{C12M12} away from Yukawa Loci}
For convenience, we employ $p_i^H$ to denote the following polynomials:
\begin{equation}
  \resizebox{0.89\textwidth}{!}{%
$\begin{aligned}
  & p_1^H=s_1 e_2 e_3 e_4 u + s_5 w \, , \qquad p_2^H=s_1^2 e_1 e_2 e_4^3 u^2 + s_1 s_2 e_2 e_3 e_4 u v - s_2 s_5 v w + s_1 s_6 v w \, ,\\
  &p_3^H=s_2 e_2^2 e_3^2 e_4^2 u^2 + s_6 e_2 e_3 e_4 u w + s_9 w^2 \, ,  \qquad  \hspace{0.4em} p_4^H=s_2 s_5 e_2 e_3 e_4 u + s_5 s_6 w - s_1 s_9 w\, ,\\
  & p_5^H=s_2 s_5^2 + s_1^2 s_9 - s_1 s_5 s_6 \, ,\qquad \hspace{1.7em} p_6^H=s_1 s_5 e_1 e_2 e_4^3 u^2 + s_2 s_5 e_2 e_3 e_4 u v + s_1 s_9 v w \, ,\\
  & p_7^H=s_5^2 e_1 e_2 e_4^3 u^2 + s_5 s_6 e_2 e_3 e_4 u v - s_1 s_9 e_2 e_3 e_4 u v + s_5 s_9 v w\, .
\end{aligned}$%
}
\end{equation}
Over $C_{(\mathbf{1},\mathbf{2})_{-1/2}}$ which is not a Yukawa point, the following $\mathbb{P}^1$-fibrations are present:
\begin{equation}
  \resizebox{0.67\textwidth}{!}{%
$\begin{aligned}
  \mathbb{P}_{0}^1 \left( (\mathbf{1},\mathbf{2})_{-1/2} \right) &= V( s_3, p_1^H, p_3^H, p_4^H, p_5^H) \, , \\
  \mathbb{P}_{1}^1 \left( (\mathbf{1},\mathbf{2})_{-1/2} \right) &= V( s_3, p_2^H, p_5^H,  p_6^H, p_7^H, p_1^H\cdot e_1e_2e_4^3u^2 + p_3^H\cdot v ) \, , \\
  \mathbb{P}_{2}^1 \left( (\mathbf{1},\mathbf{2})_{-1/2} \right) &= V( s_3, p_5^H, e_1 ) \, .
\end{aligned}$%
}
\end{equation}
Equivalently, $\mathbb{P}_{0}^1=V(s_3,p_1^H,p_3^H)$ arises from the analysis of a primary decomposition. The above $\mathbb{P}^1$-fibrations relate to restrictions of the $SU(2)$ Cartan divisors as follows:
\begin{align}
  \adjustbox{max width=0.89\textwidth}{
\begin{tabular}{cc||cc}
\toprule
Original & Split components over $C_{\mathbf{R}}$ & Original & Split components over $C_{\mathbf{R}}$ \\
\midrule
$D_0^{SU(2)}$ & $\mathbb{P}_{0}^1 \left( (\mathbf{1},\mathbf{2})_{-1/2} \right) + \mathbb{P}_{1}^1 \left( (\mathbf{1},\mathbf{2})_{-1/2} \right)$ 
& $D_1^{SU(2)}$ & $\mathbb{P}_{2}^1 \left( (\mathbf{1},\mathbf{2})_{-1/2} \right)$ \\
\bottomrule
\end{tabular}}
\end{align}
Over $p \in C_{(\mathbf{1},\mathbf{2})_{-1/2}}$ which is not a Yukawa point, these $\mathbb{P}^1$-fibrations correspond to the representation state at the right column and intersect each other as follows:
\begin{align}\adjustbox{max width=0.89\textwidth}{
\begin{tabular}{c|ccc}
\toprule
 & $\mathbb{P}_{0}^1 \left( (\mathbf{1},\mathbf{2})_{-1/2} \right)$ & $\mathbb{P}_{1}^1 \left( (\mathbf{1},\mathbf{2})_{-1/2} \right)$ & $\mathbb{P}_{2}^1 \left( (\mathbf{1},\mathbf{2})_{-1/2} \right)$ \\
\midrule
$\mathbb{P}_{0}^1 \left( (\mathbf{1},\mathbf{2})_{-1/2} \right)$ & -2 & 1 & 1 \\
$\mathbb{P}_{1}^1 \left( (\mathbf{1},\mathbf{2})_{-1/2} \right)$ & 1 & -2 & 1 \\
$\mathbb{P}_{2}^1 \left( (\mathbf{1},\mathbf{2})_{-1/2} \right)$ & 1 & 1 & -2 \\
\bottomrule
\end{tabular}}
\end{align}
The matter surfaces are
\begin{align}
\begin{tabular}{ccc||ccc}
\toprule
Label & $\vec{P}$ & $\mathbf{\beta}$ & Label & $\vec{P}$ & $\mathbf{\beta}$ \\
\midrule
$S_{(\mathbf{1},\mathbf{2})_{-1/2}}^{(1)}$ & $\left( 1,0,0 \right)$ & $\left( 1 \right)$ &
$S_{(\mathbf{1},\mathbf{2})_{-1/2}}^{(2)}$ & $\left( 1,0,1 \right)$ & $\left( -1 \right)$ \\
\bottomrule
\end{tabular}
\end{align}

\subsubsection*{Intersection Structure over \texorpdfstring{$\mathbf{C_{(\mathbf{\overline{3}},\mathbf{1})_{-2/3}}}$}{C3113} away from Yukawa Loci}

Over $C_{(\mathbf{\overline{3}},\mathbf{1})_{-2/3}} = V( s_5, s_9 )$ the following $\mathbb{P}^1$-fibrations are present:
\begin{equation}
  \adjustbox{max width=0.85\textwidth}{%
$\begin{aligned}
  \mathbb{P}_{0}^1 \left( (\mathbf{\overline{3}},\mathbf{1})_{-2/3} \right) &= V( s_5, s_9, e_1^2 e_2 e_4^4 s_1 u^2 + e_1 e_2 e_3 e_4^2 s_2 u v + e_2 e_3^2 s_3 v^2 + e_1 e_4 s_6 v w ) \, , \\
  \mathbb{P}_{1}^1 \left( (\mathbf{\overline{3}},\mathbf{1})_{-2/3} \right) &= V( s_5, s_9, u ) \, , \qquad 
  \mathbb{P}_{2}^1 \left( (\mathbf{\overline{3}},\mathbf{1})_{-2/3} \right) = V( s_5, s_9, e_2 ) \, , \\
  \mathbb{P}_{3}^1 \left( (\mathbf{\overline{3}},\mathbf{1})_{-2/3} \right) &= V( s_5, s_9, e_3 ) \, .
\end{aligned}$%
}
\end{equation}
These $\mathbb{P}^1$-fibrations relate to restrictions of the $SU(3)$ Cartan divisors as follows:
\begin{align}
  \adjustbox{max width=0.87\textwidth}{
\begin{tabular}{cc||cc}
\toprule
Original & Split components over $C_{\mathbf{R}}$ & Original & Split components over $C_{\mathbf{R}}$ \\
\midrule
$D_0^{SU(3)}$ & $\mathbb{P}_{0}^1 \left( (\mathbf{\overline{3}},\mathbf{1})_{-2/3} \right) + \mathbb{P}_{3}^1 \left( (\mathbf{\overline{3}},\mathbf{1})_{-2/3} \right)$ 
& $D_1^{SU(3)}$ & $\mathbb{P}_{2}^1 \left( (\mathbf{\overline{3}},\mathbf{1})_{-2/3} \right)$ \\
$D_2^{SU(3)}$ & $\mathbb{P}_{1}^1 \left( (\mathbf{\overline{3}},\mathbf{1})_{-2/3} \right)$ \\
\bottomrule
\end{tabular}}
\end{align}
Over $p \in C_{(\mathbf{\overline{3}},\mathbf{1})_{-2/3}}$ which is not a Yukawa point, these $\mathbb{P}^1$-fibrations intersect as follows:
\begin{align}
  \adjustbox{max width=0.89\textwidth}{
\begin{tabular}{c|cccc}
\toprule
 & $\mathbb{P}_{0}^1 \left( (\mathbf{\overline{3}},\mathbf{1})_{-2/3} \right)$ & $\mathbb{P}_{1}^1 \left( (\mathbf{\overline{3}},\mathbf{1})_{-2/3} \right)$ & $\mathbb{P}_{2}^1 \left( (\mathbf{\overline{3}},\mathbf{1})_{-2/3} \right)$ & $\mathbb{P}_{3}^1 \left( (\mathbf{\overline{3}},\mathbf{1})_{-2/3} \right)$ \\
\midrule
$\mathbb{P}_{0}^1 \left( (\mathbf{\overline{3}},\mathbf{1})_{-2/3} \right)$ & -2 & 1 & 0 & 1 \\
$\mathbb{P}_{1}^1 \left( (\mathbf{\overline{3}},\mathbf{1})_{-2/3} \right)$ & 1 & -2 & 1 & 0 \\
$\mathbb{P}_{2}^1 \left( (\mathbf{\overline{3}},\mathbf{1})_{-2/3} \right)$ & 0 & 1 & -2 & 1 \\
$\mathbb{P}_{3}^1 \left( (\mathbf{\overline{3}},\mathbf{1})_{-2/3} \right)$ & 1 & 0 & 1 & -2 \\
\bottomrule
\end{tabular}}
\end{align}

The matter surfaces $S_{(\mathbf{\overline{3}},\mathbf{1})_{-2/3}}^{(a)}$ take the following form:
\begin{equation}
  \adjustbox{max width=0.89\textwidth}{
\begin{tabular}{ccc||ccc||ccc}
\toprule
Label & $\vec{P}$ & $\mathbf{\beta}$ & Label & $\vec{P}$ & $\mathbf{\beta}$   & Label & $\vec{P}$ & $\mathbf{\beta}$ \\
\midrule
$S_{(\mathbf{\overline{3}},\mathbf{1})_{-2/3}}^{(1)}$ & $\left( 0,0,0,1 \right)$ & $\left( 1, 0 \right)$ 
&$S_{(\mathbf{\overline{3}},\mathbf{1})_{-2/3}}^{(2)}$ & $\left( 0,0,1,1 \right)$ & $\left( -1, 1 \right)$ 
&
$S_{(\mathbf{\overline{3}},\mathbf{1})_{-2/3}}^{(3)}$ & $\left( 0,1,1,1 \right)$ & $\left( 0, -1 \right)$ \\
\bottomrule
\end{tabular}}
\end{equation}

\subsubsection*{Intersection Structure over \texorpdfstring{$\mathbf{C_{(\mathbf{\overline{3}},\mathbf{1})_{1/3}}}$}{C31M23} away from Yukawa Loci}
Over $C_{(\mathbf{\overline{3}},\mathbf{1})_{1/3}}$ which is not a Yukawa point, the following $\mathbb{P}^1$-fibrations are present:
\begin{equation}
  \resizebox{0.89\textwidth}{!}{%
$\begin{aligned}
  \mathbb{P}_{0}^1 \left( (\mathbf{\overline{3}},\mathbf{1})_{1/3} \right) &= V( s_9, s_3 s_5^2 - s_2 s_5 s_6 + s_1 s_6^2, s_5 e_1 e_4^2 u + s_6 e_3 v, \\
  & \qquad s_1 s_6 e_1 e_4^2 u - s_3 s_5 e_3 v + s_2 s_6 e_3 v, s_1 e_1^2 e_4^4 u^2 + s_2 e_1 e_3 e_4^2 u v + s_3 e_3^2 v^2 ) \, , \\
  \mathbb{P}_{1}^1 \left( (\mathbf{\overline{3}},\mathbf{1})_{1/3} \right) &= V( s_9, s_3 s_5^2 - s_2 s_5 s_6 + s_1 s_6^2, s_1 s_6 e_1 e_2 e_3 e_4^2 u + s_3 s_5 e_2 e_3^2 v + s_5 s_6 e_1 e_4 w, \\
  & \qquad s_1 e_1^2 e_2 e_3 e_4^4 u^2 + s_2 e_1 e_2 e_3^2 e_4^2 u v + s_5 e_1^2 e_4^3 u w + s_3 e_2 e_3^3 v^2 + s_6 e_1 e_3 e_4 v w  ,  \\ 
  & \qquad s_3 s_5 e_1 e_2 e_3 e_4^2 u - s_2 s_6 e_1 e_2 e_3 e_4^2 u - s_3 s_6 e_2 e_3^2 v - s_6^2 e_1 e_4 w, \\
  & \qquad s_1 s_5 e_1 e_2 e_3 e_4^2 u + s_2 s_5 e_2 e_3^2 v - s_1 s_6 e_2 e_3^2 v + s_5^2 e_1 e_4 w )\,  ,\\
  \mathbb{P}_{2}^1 \left( (\mathbf{\overline{3}},\mathbf{1})_{1/3} \right) &= V( s_9, s_3 s_5^2 + s_1 s_6^2 - s_2 s_5 s_6, u ) \, , \\
  \mathbb{P}_{3}^1 \left( (\mathbf{\overline{3}},\mathbf{1})_{1/3} \right) &= V( s_9, s_3 s_5^2 + s_1 s_6^2 - s_2 s_5 s_6, e_2 ) \, .
\end{aligned}$%
}
\end{equation}
Due to primary decomposition analysis, $\mathbb{P}_{0}^1 \left( (\mathbf{\overline{3}},\mathbf{1})_{1/3} \right)$ can be rewritten as
\begin{equation}
  \resizebox{0.89\textwidth}{!}{%
$\begin{aligned}
  \mathbb{P}_{0}^1 \left( (\mathbf{\overline{3}},\mathbf{1})_{1/3} \right) &= V(s_9,u s_5e_1e_4^2+vs_6e_3, s_1 s_6 e_1 e_4^2 u - s_3 s_5 e_3 v + s_2 s_6 e_3 v)- V (s_9,s_5,s_6) \, .
\end{aligned}$%
}
\end{equation}
The above $\mathbb{P}^1$-fibrations relate to restrictions of the $SU(3)$ Cartan divisors as follows:
\begin{align}
  \adjustbox{max width=0.89\textwidth}{
\begin{tabular}{cc||cc}
\toprule
Original & Split components over $C_{(\mathbf{\overline{3}},\mathbf{1})_{1/3}}$ & Original & Split components over $C_{(\mathbf{\overline{3}},\mathbf{1})_{1/3}}$ \\
\midrule
$D_0^{SU(3)}$ & $\mathbb{P}_{0}^1 \left( (\mathbf{\overline{3}},\mathbf{1})_{1/3} \right) + \mathbb{P}_{1}^1 \left( (\mathbf{\overline{3}},\mathbf{1})_{1/3} \right)$ 
& $D_1^{SU(3)}$ & $\mathbb{P}_{3}^1 \left( (\mathbf{\overline{3}},\mathbf{1})_{1/3} \right)$ \\
$D_2^{SU(3)}$ & $\mathbb{P}_{2}^1 \left( (\mathbf{\overline{3}},\mathbf{1})_{1/3} \right)$ \\
\bottomrule
\end{tabular}}
\end{align}
Over $p \in C_{(\mathbf{\overline{3}},\mathbf{1})_{1/3}}$ which is not a Yukawa point, these $\mathbb{P}^1$-fibrations intersect as follows:
\begin{align}
  \adjustbox{max width=0.85\textwidth}{
\begin{tabular}{c|cccc}
\toprule
 & $\mathbb{P}_{0}^1 \left( (\mathbf{\overline{3}},\mathbf{1})_{1/3} \right)$ & $\mathbb{P}_{1}^1 \left( (\mathbf{\overline{3}},\mathbf{1})_{1/3} \right)$ & $\mathbb{P}_{2}^1 \left( (\mathbf{\overline{3}},\mathbf{1})_{1/3} \right)$ & $\mathbb{P}_{3}^1 \left( (\mathbf{\overline{3}},\mathbf{1})_{1/3} \right)$ \\
\midrule
$\mathbb{P}_{0}^1 \left( (\mathbf{\overline{3}},\mathbf{1})_{1/3} \right)$ & -2 & 1 & 0 & 1 \\
$\mathbb{P}_{1}^1 \left( (\mathbf{\overline{3}},\mathbf{1})_{1/3} \right)$ & 1 & -2 & 1 & 0 \\
$\mathbb{P}_{2}^1 \left( (\mathbf{\overline{3}},\mathbf{1})_{1/3} \right)$ & 0 & 1 & -2 & 1 \\
$\mathbb{P}_{3}^1 \left( (\mathbf{\overline{3}},\mathbf{1})_{1/3} \right)$ & 1 & 0 & 1 & -2 \\
\bottomrule
\end{tabular}}
\end{align}
The matter surfaces $S_{(\mathbf{\overline{3}},\mathbf{1})_{1/3}}^{(a)}$ take the following form:
\begin{equation}
  \adjustbox{max width=0.89\textwidth}{
\begin{tabular}{ccc||ccc||ccc}
\toprule
Label & $\vec{P}$ & $\mathbf{\beta}$ & Label & $\vec{P}$ & $\mathbf{\beta}$ & Label & $\vec{P}$ & $\mathbf{\beta}$ \\
\midrule
$S_{(\mathbf{\overline{3}},\mathbf{1})_{1/3}}^{(1)}$ & $\left( 1,0,0,0 \right)$ & $\left( 1, 0 \right)$
& $S_{(\mathbf{\overline{3}},\mathbf{1})_{1/3}}^{(2)}$ & $\left( 1,0,0,1 \right)$ & $\left( -1, 1 \right)$
& $S_{(\mathbf{\overline{3}},\mathbf{1})_{1/3}}^{(3)}$ & $\left( 1,0,1,1 \right)$ & $\left( 0, -1 \right)$ \\
\bottomrule
\end{tabular}}
\end{equation}

\subsubsection*{Intersection Structure over \texorpdfstring{$\mathbf{C_{(\mathbf{\overline{1}},\mathbf{1})_{1}}}$}{C31M23} away from Yukawa Loci}

Over the singlet curve $C_{(\mathbf{1}, \mathbf{1})_{1}} = V \left( s_1, s_5 \right)$ the following two $\mathbb{P}^1$-fibrations are present:
\begin{equation}
  \resizebox{0.82\textwidth}{!}{%
$\begin{aligned}
  \mathbb{P}_{0}^1 \left( (\mathbf{1},\mathbf{1})_{1} \right) &= V \left( s_1, s_5, e_1 e_2^2 e_3^2 e_4^2 s_2 u^2 + e_2^2 e_3^3 s_3 u v + e_1 e_2 e_3 e_4 s_6 u w + e_1 s_9 w^2 \right) \, , \\
  \mathbb{P}_{1}^1 \left( (\mathbf{1},\mathbf{1})_{1} \right) &= V \left( s_1, s_5, v \right) \, .
\end{aligned}$%
}
\end{equation}
These fibrations intersect as follows:
\begin{align}
  \adjustbox{max width=0.7\textwidth}{
\begin{tabular}{c|cc|c}
\toprule
 & $\mathbb{P}_{0}^1 \left( (\mathbf{1},\mathbf{1})_{1} \right)$ & $\mathbb{P}_{1}^1 \left( (\mathbf{1},\mathbf{1})_{1} \right)$ & $U(1)_Y$ \\
\midrule
$\mathbb{P}_{0}^1 \left( (\mathbf{1},\mathbf{1})_{1} \right)$ & {-2} & 2 & -1 \\
$\mathbb{P}_{1}^1 \left( (\mathbf{1},\mathbf{1})_{1} \right)$ & 2 & -2 & 1 \\
\bottomrule
\end{tabular}}
\end{align}
We use $\mathbb{P}_{1}^1 \left( (\mathbf{1},\mathbf{1})_{1} \right)$ as matter surface for the singlet state with $q_{U(1)_Y} = 1$.

\subsection{Over Yukawa Loci} \label{app_Yukawas}

\subsubsection*{Intersection Structure over Yukawa Locus \texorpdfstring{$\mathbf{Y_1}$}{Y1}}

Over the Yukawa point $Y_1 = V ( s_3, s_5, s_9 )$ the following $\mathbb{P}^1$-fibrations are present:
\begin{equation}
  \resizebox{0.89\textwidth}{!}{%
$\begin{aligned}
  & \mathbb{P}_{0}^1 \left( Y_1 \right) = V \left( s_3, s_5, s_9, e_1 \right) \, , 
  \quad \mathbb{P}_{1}^1 \left( Y_1 \right) = V \left( s_3, s_5, s_9, e_2 \right) \, , 
  \quad \mathbb{P}_{2}^1 \left( Y_1 \right) = V \left( s_3, s_5, s_9, e_3 \right)\, , 
   \\
 &\mathbb{P}_{3}^1 \left( Y_1 \right) = V \left( s_3, s_5, s_9, e_4 \right) \, , 
  \quad \mathbb{P}_{4}^1 \left( Y_1 \right) = V \left( s_3, s_5, s_9, u \right) \, , \\
  &\mathbb{P}_{5}^1 \left( Y_1 \right) = V \left( s_3, s_5, s_9, e_1 e_2 e_4^3 s_1 u^2 + e_2 e_3 e_4 s_2 u v + s_6 v w \right) \, .
\end{aligned}$%
}
\end{equation}
The intersection numbers in the fiber over $Y_1$ are as follows:
\begin{align}
  \adjustbox{max width=0.7\textwidth}{
\begin{tabular}{c|cccccc}
\toprule
 & $\mathbb{P}_{0}^1 \left( Y_1 \right)$ & $\mathbb{P}_{1}^1 \left( Y_1 \right)$ & $\mathbb{P}_{2}^1 \left( Y_1 \right)$ & $\mathbb{P}_{3}^1 \left( Y_1 \right)$ & $\mathbb{P}_{4}^1 \left( Y_1 \right)$ & $\mathbb{P}_{5}^1 \left( Y_1 \right)$\\
\midrule
$\mathbb{P}_{0}^1 \left( Y_1 \right)$ & -2  &  0 &  0 & 1  & 0  & 1 \\
$\mathbb{P}_{1}^1 \left( Y_1 \right)$ &  0  & -2 &  1 & 0  & 1  & 0 \\
$\mathbb{P}_{2}^1 \left( Y_1 \right)$ &  0  &  1 & -2 & 0  & 0  & 1 \\
$\mathbb{P}_{3}^1 \left( Y_1 \right)$ &  1  &  0 &  0 & -2 & 1  & 0 \\
$\mathbb{P}_{4}^1 \left( Y_1 \right)$ &  0  &  1 &  0 & 1  & -2 & 0 \\
$\mathbb{P}_{5}^1 \left( Y_1 \right)$ &  1  &  0 &  1 & 0  & 0  & -2 \\
\bottomrule
\end{tabular}}
\end{align}
Restrictions of the fibrations over the matter curves relate to the $\mathbb{P}_{i}^1 \left( Y_1 \right)$ as follows:
\begin{align}
  \adjustbox{max width=0.89\textwidth}{
\begin{tabular}{cc||cc}
\toprule
Split $\mathbb{P}^1$ over $C_{\mathbf{R}}$ & Split $\mathbb{P}^1$ over $Y_1$ & Split $\mathbb{P}^1$ over $C_{\mathbf{R}}$ & Split $\mathbb{P}^1$ over $Y_1$ \\
\midrule
$\mathbb{P}_{0}^1 \left( (\mathbf{3},\mathbf{2})_{1/6} \right)$ & $\mathbb{P}_{2}^1 \left( Y_1 \right)+\mathbb{P}_{5}^1 \left( Y_1 \right)$ & $\mathbb{P}_{1}^1 \left( (\mathbf{1},\mathbf{2})_{-1/2} \right)$ & $\mathbb{P}_{5}^1 \left( Y_1 \right)$ \\
$\mathbb{P}_{1}^1 \left( (\mathbf{3},\mathbf{2})_{1/6} \right)$ & $\mathbb{P}_{0}^1 \left( Y_1 \right)$ & $\mathbb{P}_{2}^1 \left( (\mathbf{1},\mathbf{2})_{-1/2} \right)$ & $\mathbb{P}_{0}^1 \left( Y_1 \right)$ \\
$\mathbb{P}_{2}^1 \left( (\mathbf{3},\mathbf{2})_{1/6} \right)$ & $\mathbb{P}_{3}^1 \left( Y_1 \right)$ & $\mathbb{P}_{0}^1 \left( (\mathbf{\overline{3}},\mathbf{1})_{-2/3} \right)$ & $\mathbb{P}_{0}^1 \left( Y_1 \right)+\mathbb{P}_{3}^1 \left( Y_1 \right)+\mathbb{P}_{5}^1 \left( Y_1 \right)$ \\
$\mathbb{P}_{3}^1 \left( (\mathbf{3},\mathbf{2})_{1/6} \right)$ & $\mathbb{P}_{4}^1 \left( Y_1 \right)$ & $\mathbb{P}_{1}^1 \left( (\mathbf{\overline{3}},\mathbf{1})_{-2/3} \right)$ & $\mathbb{P}_{4}^1 \left( Y_1 \right)$ \\
$\mathbb{P}_{4}^1 \left( (\mathbf{3},\mathbf{2})_{1/6} \right)$ & $\mathbb{P}_{1}^1 \left( Y_1 \right)$ & $\mathbb{P}_{2}^1 \left( (\mathbf{\overline{3}},\mathbf{1})_{-2/3} \right)$ & $\mathbb{P}_{1}^1 \left( Y_1 \right)$ \\
\midrule
{$\mathbb{P}_{0}^1 \left( (\mathbf{1},\mathbf{2})_{-1/2} \right)$ } & $\sum_{i=1}^4\mathbb{P}_{i}^1 \left( Y_1 \right)$ 
& {$\mathbb{P}_{3}^1 \left( (\mathbf{\overline{3}},\mathbf{1})_{-2/3} \right)$} & {$\mathbb{P}_{2}^1 \left( Y_1 \right)$}\\
\bottomrule
\end{tabular}}
\end{align}

\subsubsection*{Intersection Structure over Yukawa Locus \texorpdfstring{$\mathbf{Y_2}$}{Y2}}

Over the Yukawa point $Y_2 = V ( s_3, s_9, s_2 s_5 - s_1 s_6 )$ the following $\mathbb{P}^1$-fibrations are present:
\begin{equation}
  \resizebox{0.8\textwidth}{!}{%
$\begin{aligned}
  \mathbb{P}_{0}^1 \left( Y_2 \right) &= V \left( s_9, s_3, s_2 s_5 - s_1 s_6, s_5 e_1 e_4^2 u + s_6 e_3 v, s_1 e_1 e_4^2 u + s_2 e_3 v \right) \, , \\
  \mathbb{P}_{1}^1 \left( Y_2 \right) &= V \left( s_9, s_3, s_2 s_5 - s_1 s_6, e_1 \right) \, , 
  \qquad \quad \mathbb{P}_{2}^1 \left( Y_2 \right) = V \left( s_9, s_3, s_2 s_5 - s_1 s_6, e_2 \right) \, , \\
  \mathbb{P}_{3}^1 \left( Y_2 \right) &= V \left( s_9, s_3, s_2 s_5 - s_1 s_6, e_4 \right) \, , 
  \qquad \quad \mathbb{P}_{4}^1 \left( Y_2 \right) = V \left( s_9, s_3, s_2 s_5 - s_1 s_6, u \right) \, , \\
  \mathbb{P}_{5}^1 \left( Y_2 \right) &= V \left( s_9, s_3, s_2 s_5 - s_1 s_6, s_2 e_2 e_3 e_4 u + s_6 w, s_1 e_2 e_3 e_4 u + s_5 w \right) \, .
\end{aligned}$%
}
\end{equation}
The intersection numbers in the fiber over $Y_2$ are as follows:
\begin{align}
  \adjustbox{max width=0.7\textwidth}{
\begin{tabular}{c|cccccc}
\toprule
 & $\mathbb{P}_{0}^1 \left( Y_2 \right)$ & $\mathbb{P}_{1}^1 \left( Y_2 \right)$ & $\mathbb{P}_{2}^1 \left( Y_2 \right)$ & $\mathbb{P}_{3}^1 \left( Y_2 \right)$ & $\mathbb{P}_{4}^1 \left( Y_2 \right)$ & $\mathbb{P}_{5}^1 \left( Y_2 \right)$\\
\midrule
$\mathbb{P}_{0}^1 \left( Y_2 \right)$ & -2 & 0  & 1  & 0  & 0  & 1 \\
$\mathbb{P}_{1}^1 \left( Y_2 \right)$ & 0  & -2 & 0  & 1  & 0  & 1 \\
$\mathbb{P}_{2}^1 \left( Y_2 \right)$ & 1  & 0  & -2 & 0  & 1  & 0 \\
$\mathbb{P}_{3}^1 \left( Y_2 \right)$ & 0  & 1  & 0  & -2 & 1  & 0 \\
$\mathbb{P}_{4}^1 \left( Y_2 \right)$ & 0  & 0  & 1  & 1  & -2 & 0 \\
$\mathbb{P}_{5}^1 \left( Y_2 \right)$ & 1  & 1  & 0  & 0  & 0  & -2 \\
\bottomrule
\end{tabular}}
\end{align}
Restrictions of the fibrations over the matter curves relate to the $\mathbb{P}_{i}^1 \left( Y_2 \right)$ as follows:
\begin{align}
  \adjustbox{max width=0.89\textwidth}{
\begin{tabular}{cc||cc}
\toprule
Split $\mathbb{P}^1$ over $C_{\mathbf{R}}$ & Split $\mathbb{P}^1$ over $Y_2$ & Split $\mathbb{P}^1$ over $C_{\mathbf{R}}$ & Split $\mathbb{P}^1$ over $Y_2$ \\
\midrule
$\mathbb{P}_{0}^1 \left( (\mathbf{3},\mathbf{2})_{1/6} \right)$ & $\mathbb{P}_{0}^1 \left( Y_2 \right)+\mathbb{P}_{5}^1 \left( Y_2 \right)$ &
$\mathbb{P}_{1}^1 \left( (\mathbf{1},\mathbf{2})_{-1/2} \right)$ & $\mathbb{P}_{0}^1 \left( Y_2 \right) + \sum_{i=2}^{4}\mathbb{P}_{i}^1 \left( Y_2 \right)$ \\
$\mathbb{P}_{1}^1 \left( (\mathbf{3},\mathbf{2})_{1/6} \right)$ & $\mathbb{P}_{1}^1 \left( Y_2 \right)$ &
$\mathbb{P}_{2}^1 \left( (\mathbf{1},\mathbf{2})_{-1/2} \right)$ & $\mathbb{P}_{1}^1 \left( Y_2 \right)$ \\
$\mathbb{P}_{2}^1 \left( (\mathbf{3},\mathbf{2})_{1/6} \right)$ & $\mathbb{P}_{3}^1 \left( Y_2 \right)$ &
$\mathbb{P}_{0}^1 \left( (\mathbf{\overline{3}},\mathbf{1})_{1/3} \right)$ & $\mathbb{P}_{0}^1 \left( Y_2 \right)$ \\
$\mathbb{P}_{3}^1 \left( (\mathbf{3},\mathbf{2})_{1/6} \right)$ & $\mathbb{P}_{4}^1 \left( Y_2 \right)$ &
$\mathbb{P}_{1}^1 \left( (\mathbf{\overline{3}},\mathbf{1})_{1/3} \right)$ & $\mathbb{P}_{1}^1 \left( Y_2 \right) + \mathbb{P}_{3}^1 \left( Y_2 \right) + \mathbb{P}_{5}^1 \left( Y_2 \right)$ \\
$\mathbb{P}_{4}^1 \left( (\mathbf{3},\mathbf{2})_{1/6} \right)$ & $\mathbb{P}_{2}^1 \left( Y_2 \right)$ &
$\mathbb{P}_{2}^1 \left( (\mathbf{\overline{3}},\mathbf{1})_{1/3} \right)$ & $\mathbb{P}_{4}^1 \left( Y_2 \right)$ \\
\midrule
$\mathbb{P}_{0}^1 \left( (\mathbf{1},\mathbf{2})_{-1/2} \right)$ & $\mathbb{P}_{5}^1( Y_2 )$ &
$\mathbb{P}_{3}^1 \left( (\mathbf{\overline{3}},\mathbf{1})_{1/3} \right)$ & $\mathbb{P}_{2}^1 \left( Y_2 \right)$ \\
\bottomrule
\end{tabular}}
\end{align}

\subsubsection*{Intersection Structure over Yukawa Locus \texorpdfstring{$\mathbf{Y_3}$}{Y3}}

Over the Yukawa point $Y_3 = V ( s_3, s_6, s_9 )$, we use $A_i^1$ to denote the reduced $\mathbb{P}^1$-fibrations such that the following structure is presented:
\begin{equation}
  \resizebox{0.85\textwidth}{!}{%
$\begin{aligned}
  \mathbb{P}_{0}^1 \left( Y_3 \right) &=A_{0}^1 \left( Y_3 \right) = V \left( s_3, s_6, s_9, e_1 \right) \, , 
  \quad  \hspace{0.65em} \mathbb{P}_{1}^1 \left( Y_3 \right) =A_{1}^1 \left( Y_3 \right) = V \left( s_3, s_6, s_9, e_2 \right) \, , \\
  \mathbb{P}_{2}^1 \left( Y_3 \right) &=2A_{2}^1 \left( Y_3 \right) = V \left( s_3, s_6, s_9, e_4^2 \right) \, , 
   \quad  \mathbb{P}_{3}^1 \left( Y_3 \right) =2A_{3}^1 \left( Y_3 \right) = V \left( s_3, s_6, s_9, u^2 \right) \, , \\
  \mathbb{P}_{4}^1 \left( Y_3 \right) &=A_{4}^1 \left( Y_3 \right) = V \left( s_3, s_6, s_9, u s_1e_1e_2e_3e_4^2+vs_2e_2e_3^2+ws_5e_1e_4 \right) \, .
\end{aligned}$%
}
\end{equation}
Restrictions of the fibrations over the matter curves relate to the $\mathbb{P}_{i}^1 \left( Y_3 \right)$ as follows:
\begin{align}
  \adjustbox{max width=0.89\textwidth}{
  \begin{tabular}{cc||cc}
  \toprule
  Split $\mathbb{P}^1$ over $C_{\mathbf{R}}$ & Split $\mathbb{P}^1$ over $Y_3$ & Split $\mathbb{P}^1$ over $C_{\mathbf{R}}$ & Split $\mathbb{P}^1$ over $Y_3$ \\
  \midrule
  $\mathbb{P}_{0}^1 \left( (\mathbf{3},\mathbf{2})_{1/6} \right)$ & $\sum_{i=2}^{4}A_{i}^1 \left( Y_3 \right)$ &
  $\mathbb{P}_{0}^1 \left( (\mathbf{\overline{3}},\mathbf{1})_{1/3} \right)$ & $A_{0}^1 \left( Y_3 \right) + A_{3}^1 \left( Y_3 \right) + A_{2}^1 \left( Y_3 \right)$ \\
  $\mathbb{P}_{1}^1 \left( (\mathbf{3},\mathbf{2})_{1/6} \right)$ & $A_{0}^1 \left( Y_3 \right)$ &
  $\mathbb{P}_{1}^1 \left( (\mathbf{\overline{3}},\mathbf{1})_{1/3} \right)$ & $A_{4}^1 \left( Y_3 \right)$ \\
  $\mathbb{P}_{2}^1 \left( (\mathbf{3},\mathbf{2})_{1/6} \right)$ & $A_{2}^1 \left( Y_3 \right)$ &
  $\mathbb{P}_{2}^1 \left( (\mathbf{\overline{3}},\mathbf{1})_{1/3} \right)$ & $A_{3}^1 \left( Y_3 \right)$ \\
  $\mathbb{P}_{3}^1 \left( (\mathbf{3},\mathbf{2})_{1/6} \right)$ & $A_{3}^1 \left( Y_3 \right)$ &
  $\mathbb{P}_{3}^1 \left( (\mathbf{\overline{3}},\mathbf{1})_{1/3} \right)$ & $A_{1}^1 \left( Y_3 \right)$ \\
  \midrule
  $\mathbb{P}_{4}^1 \left( (\mathbf{3},\mathbf{2})_{1/6} \right)$ & $A_{1}^1 \left( Y_3 \right)$ & & \\
  \bottomrule
  \end{tabular}}
  \end{align}
Their intersection numbers are slightly away from standard, namely
\begin{align}
  \adjustbox{max width=0.62\textwidth}{
\begin{tabular}{c|ccccc}
\toprule
 & $A_{0}^1 \left( Y_3 \right)$ & $A_{1}^1 \left( Y_3 \right)$ & $A_{2}^1 \left( Y_3 \right)$ & $A_{3}^1 \left( Y_3 \right)$ & $A_{4}^1 \left( Y_3 \right)$ \\
\midrule
$A_{0}^1 \left( Y_3 \right)$ & -2 &  0  & 1  & 0  & 0 \\
$A_{1}^1 \left( Y_3 \right)$ & 0  & -2  & 0  & 1  & 0 \\
$A_{2}^1 \left( Y_3 \right)$ & 1  &  0  & -$\frac{3}{2}$ & 1  & 0 \\
$A_{3}^1 \left( Y_3 \right)$ & 0  &  1  & 1  & -2 & 1 \\
$A_{4}^1 \left( Y_3 \right)$ & 0  &  0  & 0  & 1  & -2 \\
\bottomrule
\end{tabular}} \label{equ:HalfIntegerIntersectionOverY3}
\end{align}
The meaning of $\left( A_{2}^1 \left( Y_3 \right) \right)^2 = - \frac{3}{2}$ becomes clear once we draw the associated diagram:
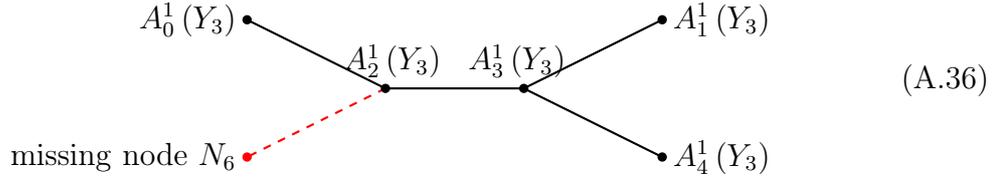
\begin{equation}
\begin{tikzpicture}[scale=0.7, baseline=(current  bounding  box.center)]

\def\w{1.3};

\draw[thick] (0,0) -- (-2*\w,\w)  ;
\draw[thick,red,dashed] (0,0) -- (-2*\w,-\w)  ;
\draw[thick] (0,0) -- (2*\w,0) -- (4*\w,\w) ;
\draw[thick] (2*\w,0) -- (4*\w,-\w) ;

\fill (0,0) circle[radius=2.5pt];
\fill (-2*\w,\w) circle[radius=2.5pt];
\fill (2*\w,0) circle[radius=2.5pt];
\fill (4*\w,\w) circle[radius=2.5pt];
\fill (4*\w,-\w) circle[radius=2.5pt];
\fill[red] (-2*\w,-\w) circle[radius=2.5pt];

\node at (0.1*\w,0) [above] {$A_2^1 \left( Y_3 \right)$};
\node at (-2*\w,\w) [left] {$A_0^1 \left( Y_3 \right)$};
\node at (1.9*\w,0) [above] {$A_3^1 \left( Y_3 \right)$};
\node at (4*\w,\w) [right] {$A_1^1 \left( Y_3 \right)$};
\node at (4*\w,-\w) [right] {$A_4^1 \left( Y_3 \right)$};
\node at (-2*\w,-\w) [left] {missing node $N_6$};

\end{tikzpicture}
\end{equation}
Consequently, we see that the node $N_6$ is missing and it holds $\mathbb{P}_2^1( Y_3 ) = 2 \cdot N_2 + N_6 $, where $N_2$ is the standard node that ordinarily appear instead of $\mathbb{P}_2^1( Y_3 )$. It follows
\begin{equation}
  \resizebox{0.72\textwidth}{!}{%
$\begin{aligned}
  \left( \mathbb{P}_2^1( Y_3 ) \right)^2 = 4 \cdot N_2^2 + 4 N_2 N_6 + N_6^2 = 4 \cdot (-2) + 4 \cdot 1 + (-2) = -6 \, .
\end{aligned}$%
}
\end{equation}
Likewise, $A_2^1( Y_3 ) = N_2 + \frac{1}{2} \cdot N_6$ leads to the half-integer intersection in \oref{equ:HalfIntegerIntersectionOverY3}.

\subsubsection*{Intersection Structure over Yukawa Locus \texorpdfstring{$\mathbf{Y_4}$}{Y4}}

Over the Yukawa point $Y_4 = V ( s_1, s_3, s_5 )$ the following $\mathbb{P}^1$-fibrations are present:
\begin{equation}
  \resizebox{0.67\textwidth}{!}{%
$\begin{aligned}
  \mathbb{P}_{0}^1 \left( Y_4 \right) &= V \left( s_1, s_3, s_5, e_1 \right) \, , 
  \qquad \quad \mathbb{P}_{1}^1 \left( Y_4 \right) = V \left( s_1, s_3, s_5, v \right) \, , \\
  \mathbb{P}_{2}^1 \left( Y_4 \right) &= V \left( s_1, s_3, s_5, s_2 e_2^2 e_3^2 e_4^2 u^2 + s_6 e_2 e_3 e_4 u w + s_9 w^2 \right) \, .
\end{aligned}$%
}
\end{equation}
Restrictions of the fibrations over the matter curves relate to the $\mathbb{P}_{i}^1 \left( Y_4 \right)$ as follows:
\begin{align}
  \adjustbox{max width=0.82\textwidth}{
  \begin{tabular}{cc||cc}
  \toprule
  Split $\mathbb{P}^1$ over $C_{\mathbf{R}}$ & Split $\mathbb{P}^1$ over $Y_4$ & Split $\mathbb{P}^1$ over $C_{\mathbf{R}}$ & Split $\mathbb{P}^1$ over $Y_4$ \\
  \midrule
  $\mathbb{P}_{0}^1 \left( (\mathbf{1},\mathbf{2})_{1/2} \right)$ & $\mathbb{P}_{2}^1 \left( Y_4 \right)$ &
  $\mathbb{P}_{1}^1 \left( (\mathbf{1},\mathbf{1})_{0} \right)$ & $\mathbb{P}_{1}^1 \left( Y_4 \right) + \mathbb{P}_{2}^1 \left( Y_4 \right)$ \\
  $\mathbb{P}_{1}^1 \left( (\mathbf{1},\mathbf{2})_{1/2} \right)$ & $\mathbb{P}_{1}^1 \left( Y_4 \right)$ &
  $\mathbb{P}_{2}^1 \left( (\mathbf{1},\mathbf{1})_{0} \right)$ & $\mathbb{P}_{0}^1 \left( Y_4 \right)$ \\
  \midrule
  $\mathbb{P}_{2}^1 \left( (\mathbf{1},\mathbf{2})_{1/2} \right)$ & $\mathbb{P}_{0}^1 \left( Y_4 \right)$ && \\
  \bottomrule
  \end{tabular}}
  \end{align}
The intersection numbers in the fiber over $Y_4$ are as follows:
\begin{align}
\begin{tabular}{c|ccc}
\toprule
 & $\mathbb{P}_{0}^1 \left( Y_4 \right)$ & $\mathbb{P}_{1}^1 \left( Y_4 \right)$ & $\mathbb{P}_{2}^1 \left( Y_4 \right)$ \\
\midrule
$\mathbb{P}_{0}^1 \left( Y_4 \right)$ & -2 &  1  & 1  \\
$\mathbb{P}_{1}^1 \left( Y_4 \right)$ & 1  & -2  & 1  \\
$\mathbb{P}_{2}^1 \left( Y_4 \right)$ & 1  &  1  & -2 \\
\bottomrule
\end{tabular}
\end{align}

\subsubsection*{Intersection Structure over Yukawa Locus \texorpdfstring{$\mathbf{Y_5}$}{Y5}}

Over the Yukawa point $Y_5 = V ( s_5, s_6^2, s_9 )$ the following $\mathbb{P}^1$-fibrations are present:
\begin{equation}
  \resizebox{0.89\textwidth}{!}{%
$\begin{aligned}
  \mathbb{P}_{0}^1 \left( Y_5 \right) &= V \left( s_5, s_6^2, s_9, e_2^3, s_6e_2^2, u^2s_1e_1^2e_2^2e_4^4+uvs_2e_1e_2^2e_3e_4^2+v^2s_3e_2^2e_3^2+vws_6e_1e_2e_4 \right) \\
  &= n_0 \cdot V \left( s_5, s_6, s_9, e_2 \right) = n_0 \cdot A_0 ( Y_5 ) \, , \\
  \mathbb{P}_{1}^1 \left( Y_5 \right) &= V \left( s_5, s_6^2, s_9, e_3 \right)=n_1 \cdot V \left( s_5, s_6, s_9, e_3 \right)  = n_1 \cdot A_1 ( Y_5 ) \, , \\
  \mathbb{P}_{2}^1 \left( Y_5 \right) &= V \left( s_5, s_6^2, s_9, u \right) =n_2 \cdot V \left( s_5, s_6, s_9, u \right) =n_2\cdot A_2 ( Y_5 ) \, , \\
  \mathbb{P}_{3}^1 \left( Y_5 \right) &= V ( s_5, s_6^2, s_9, u^2s_1e_1^2e_2e_4^4+uvs_2e_1e_2e_3e_4^2+v^2s_3e_2e_3^2+vws_6e_1e_4,\\
  &\qquad u^4s_1^2e_1^4e_4^8+2u^3vs_1s_2e_1^3e_3e_4^6+u^2v^2s_2^2e_1^2e_3^2e_4^4+2u^2v^2s_1s_3e_1^2e_3^2e_4^4\\
  &\qquad +2uv^3s_2s_3e_1e_3^3e_4^2+v^4s_3^2e_3^4, u^2s_1s_6e_1^2e_4^4+uvs_2s_6e_1e_3e_4^2+v^2s_3s_6e_3^2 ) \\
  & =n_3\cdot V \left( s_5, s_6, s_9, u^2s_1e_1^2e_4^4+uvs_2e_1e_3e_4^2+v^2s_3e_3^2  \right) =n_3\cdot A_3 ( Y_5 ) \, .
\end{aligned}$%
}
\end{equation}
The total elliptic fiber over $Y_5$ is given 
\begin{equation}\label{eq:T2onY5}
  \resizebox{0.8\textwidth}{!}{%
$\begin{aligned}
  T^2 \left( Y_5 \right) = \sum_{i=0}^{3}{\mathbb{P}_{i}^1 \left( Y_5 \right)} = n_0 \cdot A_0 ( Y_5 ) + n_1 \cdot A_1 ( Y_5 ) + n_2 \cdot A_2 ( Y_5 ) + n_3 \cdot A_3 ( Y_5 ) \, .
\end{aligned}$%
}
\end{equation}
Restrictions of the fibrations over the gauge divisor $ D_i^{SU(3)}$ are:
\begin{equation}
  \resizebox{0.52\textwidth}{!}{%
$\begin{aligned}
  &D_0^{SU(3)}|_{Y_5} = n_3 \cdot A_3 ( Y_5 ) + n_1 \cdot A_1 ( Y_5 ) + n_4 \cdot A_0 ( Y_5 ) \, ,\\
  &D_1^{SU(3)}|_{Y_5} = n_5 \cdot A_0 ( Y_5 ) \, ,\\
  &D_2^{SU(3)}|_{Y_5} = n_2 \cdot A_2 ( Y_5 ) \, .
\end{aligned}$%
}
\end{equation}
The total elliptic fiber can be obtained from the total torus over the gauge divisor $ D_i^{SU(3)}$:
\begin{equation}\label{eq:T2onSU3Y5}
  \resizebox{0.8\textwidth}{!}{%
$\begin{aligned}
  T^2 \left( D_i^{SU(3)}|_{Y_5} \right)  = (n_4+n_5)\cdot A_0 ( Y_5 ) + n_1\cdot A_1 ( Y_5 )+ n_2\cdot A_2 ( Y_5 )+ n_3\cdot A_3 ( Y_5 ) \, .
\end{aligned}$%
}
\end{equation}
Since this must recover (\oref{eq:T2onY5}), we conclude that
\begin{align}
n_0=n_4+n_5
\end{align}
Restrictions of the fibrations over the matter curves $ (\mathbf{\overline{3}},\mathbf{1})_{1/3})$ are as follows:
\begin{equation}
  \resizebox{0.89\textwidth}{!}{%
$\begin{aligned}
  \mathbb{P}_{0}^1 \left( (\mathbf{\overline{3}},\mathbf{1})_{1/3} \right)|_{Y_5} &= V \left( s_5, s_6, s_9, u^2s_1e_1^2e_4^4+uvs_2e_1e_3e_4^2+v^2s_3e_3^2  \right) =A_3( Y_5 ) \, ,\\
  \mathbb{P}_{1}^1 \left( (\mathbf{\overline{3}},\mathbf{1})_{1/3} \right)|_{Y_5} &= V \left( s_5, s_6, s_9, u^2s_1e_1^2e_4^4+uvs_2e_1e_3e_4^2+v^2s_3e_3^2  \right) + \mathbb{P}_{1}^1 \left( Y_5 \right)+  \\
  & V ( s_5, s_6^2, s_9, e_2^2, s_6e_2, u^2s_1e_1^2e_2^2e_4^4+uvs_2e_1e_2^2e_3e_4^2+v^2s_3e_2^2e_3^2+vws_6e_1e_2e_4 )\\
  &=A_3( Y_5 )+n_1\cdot A_1( Y_5 ) + n_4\cdot A_0( Y_5 )  \, ,\\
  \mathbb{P}_{2}^1 \left( (\mathbf{\overline{3}},\mathbf{1})_{1/3} \right)|_{Y_5} &= V \left( s_5, s_6^2, s_9, e_2  \right) =n_5\cdot A_0( Y_5 ) \, ,\\
  \mathbb{P}_{3}^1 \left( (\mathbf{\overline{3}},\mathbf{1})_{1/3} \right)|_{Y_5} &= V \left( s_5, s_6^2, s_9, u  \right) =n_2\cdot A_2( Y_5 ) \, .
\end{aligned}$%
}
\end{equation}
The fibrations over the matter curves $ (\mathbf{\overline{3}},\mathbf{1})_{1/3}$ give the total elliptic fiber over $Y_5$ as:
\begin{equation}
  \resizebox{0.8\textwidth}{!}{%
$\begin{aligned}
  T^2 \left(  (\mathbf{\overline{3}},\mathbf{1})_{1/3}|_{Y_5} \right)  =(n_4+n_5) \cdot A_0( Y_5 ) + n_1\cdot A_1( Y_5 )+ n_2\cdot A_2( Y_5 )+ 2\cdot A_3( Y_5 ) \, .
\end{aligned}$%
}
\end{equation}
Restrictions of the fibrations over the matter curves $ (\mathbf{\overline{3}},\mathbf{1})_{-2/3}$ are:
\begin{equation}
  \resizebox{0.89\textwidth}{!}{%
$\begin{aligned}
  \mathbb{P}_{0}^1 \left( (\mathbf{\overline{3}},\mathbf{1})_{-2/3} \right)|_{Y_5} &= V ( s_5, s_6^2, s_9, e_2^2, s_6e_2, u^2s_1e_1^2e_2^2e_4^4+uvs_2e_1e_2^2e_3e_4^2+v^2s_3e_2^2e_3^2+vws_6e_1e_2e_4 ) \\
& + \mathbb{P}_{3}^1 \left( Y_5 \right) =n_3\cdot A_3( Y_5 ) + n_4\cdot A_0( Y_5 )\, ,\\
\mathbb{P}_{1}^1 \left( (\mathbf{\overline{3}},\mathbf{1})_{-2/3} \right)|_{Y_5} &=V \left( s_5, s_6^2, s_9, u  \right)=n_2\cdot  A_2( Y_5 )  \, ,\\
\mathbb{P}_{2}^1 \left( (\mathbf{\overline{3}},\mathbf{1})_{-2/3} \right)|_{Y_5} &= V \left( s_5, s_6^2, s_9, e_2  \right) =n_5\cdot A_0( Y_5 ) \, ,\\
\mathbb{P}_{3}^1 \left( (\mathbf{\overline{3}},\mathbf{1})_{-2/3} \right)|_{Y_5} &= V \left( s_5, s_6^2, s_9, e_3  \right) =n_1\cdot A_1( Y_5 ) \, .
\end{aligned}$%
}
\end{equation}
The fibrations over the matter curves $ (\mathbf{\overline{3}},\mathbf{1})_{-2/3}$ give the total elliptic fiber over $Y_5$ as:
\begin{equation}
  \resizebox{0.85\textwidth}{!}{%
$\begin{aligned}
  T^2 \left(  (\mathbf{\overline{3}},\mathbf{1})_{-2/3}|_{Y_5} \right)  =(n_4+n_5) \cdot A_0( Y_5 ) + n_1\cdot A_1( Y_5 )+ n_2\cdot A_2( Y_5 )+ n_3\cdot A_3( Y_5 ) \, .
\end{aligned}$%
}
\end{equation}
We conclude that the restriction from the two triplet matter curves to the Yukawa point $Y_5 $ preserve the elliptic fiber structure as presented in (\oref{eq:T2onY5}) iff $n_3 = 2$.

Before we continue our discussion of the factors $n_i$, let us look at the intersection numbers among the $A_i( Y_5 )$ as follows:
\begin{align}
  \adjustbox{max width=0.6\textwidth}{
\begin{tabular}{c|cccc}
\toprule
 & $A_{0}^1 \left( Y_5 \right)$ & $A_{1}^1 \left( Y_5 \right)$ & $A_{2}^1 \left( Y_5 \right)$ & $A_{3}^1 \left( Y_5 \right)$ \\
\midrule
$A_{0}^1 \left( Y_5 \right)$ & -2 &  1 &  1 & 2 \\
$A_{1}^1 \left( Y_5 \right)$ &  1 & -2 &  0 & 0 \\
$A_{2}^1 \left( Y_5 \right)$ &  1 &  0 & -2 & 0 \\
$A_{3}^1 \left( Y_5 \right)$ &  2 &  0 &  0 & -2 \\
\bottomrule
\end{tabular}}
\end{align}
Let us return to the factors $n_i$ we can fix $n_1 = n_2 = 2$ intuitively. Then, by infering that $-2 = \left( \left. P_0( (\overline{\mathbf{3}}, \mathbf{1} )_{-2/3} ) \right|_{Y_5} \right)^2 $, we find $n_4 \in \{ 1, 3 \}$. Intuitively, we discard $n_4 = 1$ and pick $n_4 = 3$ instead. By accepting all these steps above, we are then left to conclude
\begin{equation}
  \resizebox{0.6\textwidth}{!}{%
$\begin{aligned}
  n_0 = 5, \qquad n_1 = n_2 = n_3 = 2 \, , \qquad n_4 = 3 \, , \qquad n_5 = 2 \, .
\end{aligned}$%
}
\end{equation}
This finally, completes our understanding of the fiber structure over $Y_5$.

\subsubsection*{Intersection Structure over Yukawa Locus \texorpdfstring{$\mathbf{Y_6}$}{Y6}}
Over the Yukawa point $Y_6 = V ( s_1, s_5, s_9 )$ the following $\mathbb{P}^1$-fibrations are present:
\begin{equation}
  \resizebox{0.89\textwidth}{!}{%
$\begin{aligned}
  & \mathbb{P}_{0}^1 \left( Y_6 \right) = V \left( s_1, s_5, s_9, e_2 \right) \, , 
  \quad \mathbb{P}_{1}^1 \left( Y_6 \right) = V \left( s_1, s_5, s_9, e_3 \right) \, , \quad \mathbb{P}_{2}^1 \left( Y_6 \right) = V \left( s_1, s_5, s_9, u \right) \, ,\\
  & \mathbb{P}_{3}^1 \left( Y_6 \right) = V \left( s_1, s_5, s_9, v \right) \, ,\quad \hspace{0.3em}  \mathbb{P}_{4}^1 \left( Y_6 \right) = V \left( s_1, s_5, s_9, s_2 e_1 e_2 e_3 e_4^2 u + s_3 e_2 e_3^2 v + s_6 e_1 e_4 w \right) \, .
\end{aligned}$%
}
\end{equation}
Restrictions of the fibrations over the matter curves relate to the $\mathbb{P}_{i}^1 \left( Y_6 \right)$ as follows:
\begin{align}
  \adjustbox{max width=0.89\textwidth}{
  \begin{tabular}{cc||cc}
  \toprule
  Split $\mathbb{P}^1$ over $C_{\mathbf{R}}$ &   Split $\mathbb{P}^1$ over $Y_6$ & Split $\mathbb{P}^1$ over $C_{\mathbf{R}}$ &   Split $\mathbb{P}^1$ over $Y_6$  \\
  \midrule
  $\mathbb{P}_{0}^1 \left( (\mathbf{\overline{3}},\mathbf{1})_{1/3} \right)$ & $\mathbb{P}_{1}^1 \left( Y_6 \right) + \mathbb{P}_{3}^1 \left( Y_6 \right)$ &
  $\mathbb{P}_{1}^1 \left( (\mathbf{\overline{3}},\mathbf{1})_{-2/3} \right)$ & $\mathbb{P}_{2}^1 \left( Y_6 \right)$ \\
  $\mathbb{P}_{1}^1 \left( (\mathbf{\overline{3}},\mathbf{1})_{1/3} \right)$ & $\mathbb{P}_{4}^1 \left( Y_6 \right)$ &
  $\mathbb{P}_{2}^1 \left( (\mathbf{\overline{3}},\mathbf{1})_{-2/3} \right)$ & $\mathbb{P}_{0}^1 \left( Y_6 \right)$ \\
  $\mathbb{P}_{2}^1 \left( (\mathbf{\overline{3}},\mathbf{1})_{1/3} \right)$ & $\mathbb{P}_{2}^1 \left( Y_6 \right)$ &
  $\mathbb{P}_{3}^1 \left( (\mathbf{\overline{3}},\mathbf{1})_{-2/3} \right)$ & $\mathbb{P}_{1}^1 \left( Y_6 \right)$ \\
  $\mathbb{P}_{3}^1 \left( (\mathbf{\overline{3}},\mathbf{1})_{1/3} \right)$ & $\mathbb{P}_{0}^1 \left( Y_6 \right)$ &
  $\mathbb{P}_{0}^1 \left( (\mathbf{\overline{3}},\mathbf{1})_{-2/3} \right)$ & $\mathbb{P}_{3}^1 \left( Y_6 \right) + \mathbb{P}_{4}^1 \left( Y_6 \right)$\\
  \midrule
  $\mathbb{P}_{1}^1 \left( (\mathbf{1},\mathbf{1})_{1} \right)$ & $\sum_{i=0}^2\mathbb{P}_{i}^1 \left( Y_6 \right) + \mathbb{P}_{4}^1 \left( Y_6 \right)$ &
  $\mathbb{P}_{2}^1 \left( (\mathbf{1},\mathbf{1})_{1} \right)$ & $\mathbb{P}_{3}^1 \left( Y_6 \right)$ \\
  \bottomrule
  \end{tabular}}
  \end{align}
The intersection numbers in the fiber over $Y_6$ are as follows:
\begin{align}\adjustbox{max width=0.6\textwidth}{
\begin{tabular}{c|ccccc}
\toprule
 & $\mathbb{P}_{0}^1 \left( Y_6 \right)$ & $\mathbb{P}_{1}^1 \left( Y_6 \right)$ & $\mathbb{P}_{2}^1 \left( Y_6 \right)$ & $\mathbb{P}_{3}^1 \left( Y_6 \right)$ & $\mathbb{P}_{4}^1 \left( Y_6 \right)$ \\
\midrule
$\mathbb{P}_{0}^1 \left( Y_6 \right)$ & -2 &  1 &  1 &  0 & 0 \\
$\mathbb{P}_{1}^1 \left( Y_6 \right)$ &  1 & -2 &  0 &  1 & 0 \\
$\mathbb{P}_{2}^1 \left( Y_6 \right)$ &  1 &  0 & -2 &  0 & 1 \\
$\mathbb{P}_{3}^1 \left( Y_6 \right)$ &  0 &  1 &  0 & -2 & 1 \\
$\mathbb{P}_{4}^1 \left( Y_6 \right)$ &  0 &  0 &  1 &  1 & -2 \\
\bottomrule
\end{tabular}}
\end{align}

\section{Induced line bundles in F-theory Standard Models} \label{sec:LineBundleComputations}

In this section, we give details on how we identify the root bundles in the largest \mbox{currently-known} class of globally consistent F-theory Standard Model constructions without chiral exotics and gauge coupling unification \cite{Cvetic:2019gnh}. More details can be found in the earlier works \cite{Klevers:2014bqa,Cvetic:2015txa}. We provide details on the employed $G_4$-flux in \oref{subsec:FluxDetails}. Subsequently, we outline our computational techniques in \oref{subsec:ComputationalStrategies} and summarize the resulting root bundle constraints in \oref{subsec:LineBundlesKbarCubeDifferent}. Finally, we \mbox{construct} root bundle solutions in compactifications over a particular 3-fold base space $B_3$ in \mbox{\oref{sec:ExampleGeometry}}.

\subsection{\texorpdfstring{$G_4$}{G4}-flux and matter surfaces} \label{subsec:FluxDetails}

\paragraph{$\mathbf{U(1)}$-flux}
We associate to the section $s_1 = V( e_4 )$ a $U(1)$-flux. To this end, we employ the Shioda map to turn $s_1$ into $\sigma \in H^{(1,1)} ( \widehat{Y}_4 )$:
\begin{align}
\sigma = \left. \left( \left[ e_4 \right] - \left[ v \right] - \left[ \widehat{\pi}^\ast \left( \KB \right) \right] + \frac{1}{2} \left[ e_1 \right] + \frac{1}{3} \left[ e_2 \right] + \frac{2}{3} \left[ u \right] \right) \right|_{\hat{Y}_4} \, .
\end{align}
In this expression, $[ e_4 ] = \gamma(V(e_4)) \in H^{(1,1)} ( \widehat{Y}_4 )$ denotes the image of the divisor \linebreak\mbox{$V( e_4 ) \subseteq X_5$} under the cycle map $\gamma$. Furthermore, recall that $\widehat{\pi} \colon \widehat{Y}_4 \twoheadrightarrow {B}_3$. The $U(1)$-flux is then given by
\begin{align}
G_4^{U(1)} \equiv \omega \wedge \sigma \in H^{(2,2)}( \widehat{Y}_4 ) \, , \qquad \omega \in \pi^\ast( H^{(1,1)}( {B}_3 )) \, .
\end{align}

\paragraph{Matter surface flux}

To the matter surface $S^{(1)}_{(\mathbf{3},\mathbf{2})_{1/6}}$ over the quark-doublet curve $C_{(\mathbf{3},\mathbf{2})_{1/6}}$ (cf. \oref{sec:FibreStructure}) one can associate a \emph{gauge invariant} flux
\begin{equation}
  \resizebox{0.9\textwidth}{!}{%
$\begin{aligned}
  G_4^{(\three, \two)_{1/6}} \equiv \left[ S_{(\three, \two)_{1/6}}^{(1)} \right] + \frac{1}{2} \cdot \left[ \mathbb{P}^1_1( (\three, \two)_{1/6}) \right]  + \frac{1}{3} \cdot \left[ \mathbb{P}^1_3( (\three, \two)_{1/6}) \right] + \frac{2}{3} \cdot \left[ \mathbb{P}^1_4( (\three, \two)_{1/6}) \right]  \, .
\end{aligned}$%
}
\end{equation}

\paragraph{Total flux expression}

One can now consider a linear combination of these fluxes
\begin{align}
G_4( a, \omega ) = a \cdot G_4^{(\three, \two)_{1/6}} + \omega \wedge \sigma \in H^{(2,2)}( \widehat{Y}_4 ) \, .
\end{align}
The parameters $a \in \mathbb{Q}$ and $\omega \in \pi^\ast \left( H^{(1,1)}( B_3 ) \right)$ are subject to flux quantization, \mbox{$D_3$-tadpole} cancelation, masslessness of the $U(1)$-gauge boson and exactly three \mbox{chiral} families on all matter curves. These conditions are solved base-independently by
\begin{align}
\omega = \frac{3}{\overline{K}_{B_3}^3} \cdot \overline{K}_{B_3} \, , \qquad a = \frac{15}{\overline{K}_{B_3}^3} \, .
\end{align}
This leads to
\begin{equation}
  G_4 = \frac{-3}{\KB^3} \cdot \big( 5 [ e_1 ] \wedge [ e_4 ]- 
  \KB \wedge ( 3 [ e_1 ]  - 2 [ e_2 ] - 6 [ e_4 ]  + \KB  - 4 u + v) \big) \, .
\end{equation}
For this $G_4$, it was verified in \cite{Cvetic:2019gnh}, that the integral over all matter surfaces $S_{\mathbf{R}}$ and complete intersections of toric divisors is integral. This is a necessary condition, for this algebraic cycle to be integral. A sufficient check is computationally very demanding and currently beyond our arithmetic abilities. Therefore, we proceed under the assumption that this $G_4$-flux candidate \cref{equ:G4-FieldStrength}, is indeed integral and thus a proper $G_4$-flux.

We next look at
\begin{align}
\begin{split}
  \mathcal{A}^\prime = -3 \cdot & \left( 5 V( e_1, e_4 ) - 3 V( e_1, t_1 ) - 2 V( e_2, t_2 ) - 6 V( e_4, t_3 ) \right. \\
                            & \qquad \qquad \left. \left. + V( t_4, t_5 ) - 4 V( t_6, u ) + V( t_6, v ) \right) \right|_{\widehat{Y}_4} \in \mathrm{CH}^2( \widehat{Y}_4, \mathbb{Z} ) \, ,
                            \label{equ:IntegralFlux}
\end{split}
\end{align}
where $t_i \in H^0( X_5, \alpha^\ast( \KB ) )$ and $\alpha \colon X_5 = {B}_3 \times \mathbb{P}_{F_{11}} \twoheadrightarrow {B}_3$. Note that $\gamma( \mathcal{A}^\prime ) = \KB^3 \cdot G_4$. Therefore, this gauge potential would induce chiral exotics, unless we ''devide`` it by $\xi = \KB^3$. Hence, we are led to consider gauge potentials $A = \widehat{\gamma}( \mathcal{A} ) \in H^4_D( \widehat{Y}_4, \mathbb{Z}(2) )$ with
\begin{align}
\gamma( \mathcal{A} ) = G_4 \, , \qquad \xi \cdot \widehat{\gamma} ( \mathcal{A} ) \sim \widehat{\gamma}( \mathcal{A}^\prime ) \, .
\end{align}
Hence, we can infer that the line bundles induced from $A = \widehat{\gamma} ( \mathcal{A} )$ are $\overline{K}_{B_3}^3$-th roots of the ones induced from $A^\prime = \widehat{\gamma}( \mathcal{A}^\prime )$. The divisors $D_{\mathbf{R}}( \mathcal{A}^\prime )$ are then $\KB^3$-th roots of the $D_{\mathbf{R}}( \mathcal{A}^\prime )$. In the following, we outline the arithmetic identification of the divisors $D_{\mathbf{R}}( \mathcal{A}^\prime )$.

\paragraph{Matter surfaces}
As \oref{equ:IntegralFlux} is gauge invariant, it suffices to focus on the following matter surfaces (cf. \oref{sec:FibreStructure})
\begin{align}\label{eq:MatterSurfaces}
\begin{split}
&S_{(\three, \two)_{1/6}}^{(1)}=V(s_3,s_9,e_4) \, , \quad S_{(\overline{\three}, \one)_{-2/3}}^{(1)}=V(s_5,s_9,e_3) \, , \quad S_{(\one, \one)_{1}}^{(1)}=V(s_1,s_5,v) \, ,\\
&S_{(\one, \two)_{-1/2}}^{(1)}= V( s_3, s_2 s_5^2 + s_1^2 s_9 - s_1 s_5 s_6, s_1 e_2 e_3 e_4 u + s_5 w, \\
& \qquad s_2 s_5 e_2 e_3 e_4 u + s_5 s_6 w - s_1 s_9 w, s_2 e_2^2 e_3^2 e_4^2 u^2 + s_6 e_2 e_3 e_4 u w + s_9 w^2 ) \, , \\
&S_{(\overline{\three}, \one)_{1/3}}^{(1)}= V( s_9, s_3 s_5^2 - s_2 s_5 s_6 + s_1 s_6^2, s_5 e_1 e_4^2 u + s_6 e_3 v, \\
& \qquad s_1 s_6 e_1 e_4^2 u - s_3 s_5 e_3 v + s_2 s_6 e_3 v, s_1 e_1^2 e_4^4 u^2 + s_2 e_1 e_3 e_4^2 u v + s_3 e_3^2 v^2 ) \, .
\end{split}
\end{align}
Note that $S_{(\one, \two)_{-1/2}}^{(1)}$ and $S_{(\overline{\three}, \one)_{1/3}}^{(1)}$ are not complete intersections. In \oref{subsec:TopIntersectionNonCompleteIntersection}, we explain how one can compute topological intersection numbers of cycles with them. Moreover, we can simplify the expressions for those two matter surfaces.
\begin{equation}
  \resizebox{0.9\textwidth}{!}{%
$\begin{aligned}
  &S_{(\one, \two)_{-1/2}}^{(1)} =V(s_3,us_1e_2e_3e_4+ws_5, u^2s_2e_2^2e_3^2e_4^2+uws_6e_2e_3e_4+w^2s_9) \, ,\\
  &S_{(\overline{\three}, \one)_{1/3}}^{(1)} =V(s_9,u s_5e_1e_4^2+vs_6e_3,u^2s_1e_1^2e_4^4+uvs_2e_1e_3e_4^2+v^2s_3e_3^2) - V(s_9,s_5,s_6)\, .
\end{aligned}$%
}
\end{equation}
Therefore, we can express all matter surfaces as pullbacks from elements in $\mathrm{CH}^2( X_5 )$:
\begin{equation}
  \resizebox{0.75\textwidth}{!}{%
$\begin{aligned}
  &S_{(\three, \two)_{1/6}}^{(1)} = \left. V(s_3, e_4) - V(e_1, e_4) \right|_{\hat{Y}_4} \, ,\\
  &S_{(\one, \two)_{-1/2}}^{(1)} = \left. V(s_3, p_1) - V(e_1, p_1) - V(s_3, v) \right|_{\hat{Y}_4} \, ,\\
&S_{(\overline{\three}, \one)_{-2/3}}^{(1)}= \left. V(s_5, e_3) - V(v, e_3) \right|_{\hat{Y}_4} \, ,\\
&S_{(\overline{\three}, \one)_{1/3}}^{(1)}= \left. V(s_9, q_1) + V(e_2, e_3) - V(e_2^2, q_1)- V(e_3, s_9) - V(u, q_1) \right|_{\hat{Y}_4} \, ,\\
&S_{(\one, \one)_{1}}^{(1)}= \left. V(s_1, v) - V(v,w) \right|_{\hat{Y}_4} \, ,
\end{aligned}$%
}
\end{equation}
where $p_1=us_1e_2e_3e_4+ws_5$ and $q_1=u s_5e_1e_4^2+vs_6e_3$. We exploit this in \oref{subsec:LineBundleFromChowX5} to compute the actual intersection loci.

Finally in \oref{subsec:LineBundleFromFibre}, we make use of the fact that we know that the matter surfaces are particular $\mathbb{P}^1$-fibrations over the matter curves. The matter surface flux originates from the matter surface $S_{(\three, \two)_{1/6}}$. This allows us to derive the divisors $D_{\mathbf{R}}( \mathcal{A}^\prime )$ intuitively from intersections in the fiber and intersections in the base. The former is facilitated by knowledge of the intersection numbers listed in \oref{sec:FibreStructure}.

\subsection{Computational strategies} \label{subsec:ComputationalStrategies}

\subsubsection{Euler characteristic of structure sheaf of intersection variety} \label{subsec:TopIntersectionNonCompleteIntersection}

\paragraph{The twisted cubic -- a non-complete intersection}

Let us start with a simple and instructive example that involves a non-complete intersection. We consider $\mathbb{P}^3$ with homogeneous coordinates $[ x: y: z: w]$ and focus on the hypersurface $Y = V( xw - yz )$. In this hypersurface $Y$, we consider the twisted cubic
\begin{align}
S = V( xz - y^2, yw - z^2 ) \cap Y = V( xz - y^2, yw -z^2, xw - yz ) \cong \mathbb{P}^1 \, ,
\end{align}
and a union of two lines
\begin{align}
\mathcal{A} = V( x ) \cap Y = V( x, xw -yz ) = V( x,y ) \cup V( x, z ) \, .
\end{align}
Crucially, note that $S$ is not a complete intersection and cannot be expressed as any sort of pullback from $\mathbb{P}^3$. In order to compute the topological intersection number $S \cdot \mathcal{A}$, we notice that this intersection number coincides with the Euler characteristic of the structure sheaf of the intersection variety $V( x, xz - y^2, yw -z^2, xw - yz )$.

Let us denote the coordinate ring of $\mathbb{P}^3$ by $R$. Then an f.p. graded (left) $R$-module, which sheafifies to the structure sheaf in question, is given by
\begin{align}
R(-2)^{\oplus 3} \oplus R(-1) \xrightarrow{\left( 
  \adjustbox{max width=0.32\textwidth}{$\begin{array}{cccc} xw - yz  & xz - y^2 & yw -z^2 & x \end{array}$} 
\right)^T} R \twoheadrightarrow O_{S \cdot \mathcal{A}} \to 0 \, .
\end{align}
Denote this sequence by $F_1 \xrightarrow{M_1} R \twoheadrightarrow O_{S \cdot \mathcal{A}} \to 0$. A minimal free resolution is given by
\begin{align}
0 \to F_3 \xrightarrow{M_3} F_2 \xrightarrow{M_2} F_1 \xrightarrow{M_1} R \twoheadrightarrow O_{S \cdot \mathcal{A}} \to 0 \, ,
\end{align}
where $F_2 = R( -3 )^{ \oplus 5}$, $F_3 = R( -4 )^{\oplus 2}$ and
\begin{align}
M_2 &= \left(  \adjustbox{max width=0.25\textwidth}{$\begin{array}{cccc}
-w  z & y & 0 \\
-y & w & x & 0 \\
-y & w & 0 &   y z-w^2 \\
-x & 0 &   0 &   x z-y w \\
0 &    x & 0 &   y^2-x w
\end{array}$} \right) \, , \qquad
M_3 = \left(  
  \adjustbox{max width=0.25\textwidth}{$\begin{array}{ccccc}
0 &   0 &    x & -y & -w \\
x & -y & y & -w & -z 
\end{array}$} 
\right) \, .
\end{align}
The vector bundles $\widetilde{F}_i$ has the following sheaf cohomologies:
\begin{align} \adjustbox{max width=0.9\textwidth}{
\begin{tabular}{ccccc}
\toprule
& $\mathcal{O}_{\mathbb{P}^3}( -4 )^{\oplus 2}$ & $\mathcal{O}_{\mathbb{P}^3}( -3 )^{ \oplus 5}$ & $\mathcal{O}_{\mathbb{P}^3}(-2)^{\oplus 3} \oplus \mathcal{O}_{\mathbb{P}^3}(-1)$ & $\widetilde{F}_0 \equiv \widetilde{R} \cong \mathcal{O}_{\mathbb{P}^3}$ \\
\midrule
$h^0$ & 0 & 0 & 0 & 1 \\
$h^1$ & 0 & 0 & 0 & 0 \\
$h^2$ & 0 & 0 & 0 & 0 \\
$h^3$ & 2 & 0 & 0 & 0 \\
\bottomrule
\end{tabular}}
\end{align}
It follows that $h^i( \mathbb{P}^3, \mathcal{O}_{S \cdot \mathcal{A}} ) = (3,0,0,0)$, i.e. $S \cdot \mathcal{A} = \chi( \mathcal{O}_{S \cdot \mathcal{A}} ) = 3$. Equivalently, we find
\begin{align}
V( xw - yz, xz - y^2, yw -z^2, x ) = V( x, y, z ) \, ,
\end{align}
which allows us to conclude $S \cdot \mathcal{A} = 3 \cdot V( x,y,z )$.

\paragraph{Application to Higgs matter surface}

We employ this technique as a consistency check on intersections with the non-complete matter surfaces. For instance, let us work out the topological intersection number of $B = V( e_1, e_4, p_{F_{11}} )$ with (cf. \oref{eq:MatterSurfaces} for $p_i$)
\begin{align}
\begin{split}
S \equiv S_{(\one, \two)_{-1/2}}^{(1)} = V( p_1, p_2, p_3, p_4, p_5 ) \, .
\end{split}
\end{align}
in the elliptic fibration $\widehat{Y}_4$ over the base space ${B}_3 = P_{39}$ (cf. \oref{sec:ExampleGeometry}). To construct the structure sheaf of the variety $V( p_1, p_2, p_3, p_4, p_5, e_1, e_4 )$, we model the coordinate ring of $X_5$ as $R = \mathbb{Q} \left[ s_1, s_2, s_3, s_5, s_6, s_9, u, v, w, e_1, e_2, e_3, e_4 \right]$ with $\mathbb{Z}^6$-grading\footnote{We could use the actual coordinate ring for the fibration over $P_{39}$. This ring has $18$ indeterminates and is $\mathbb{Z}_{14}$-graded. As a consequence, the resulting computations take longer than the ones performed with the simpler ring. Both lead to the same result.}
\begin{align}
\begin{tabular}{c|cccccc|ccc|cccc}
\toprule
& $s_1$ & $s_2$ & $s_3$ & $s_5$ & $s_6$ & $s_9$ & u & v & w & $e_1$ & $e_2$ & $e_3$ & $e_4$ \\
\midrule
$\overline{K}_{{B}_3}$ & 1 & 1 & 1 & 1 & 1 & 1 & & & \\
\midrule
H & & & & & & & 1 & 1 & 1 \\
\midrule
$E_1$ & & & & & & & -1 &    & -1 & 1 \\
$E_2$ & & & & & & & -1 & -1 &    &    & 1 \\
$E_3$ & & & & & & &    & -1 &    &    & -1 & 1 \\
$E_4$ & & & & & & & -1 &    &    & -1 &    &   & 1 \\
\bottomrule
\end{tabular}
\end{align}
Then, an f.p. graded (left) $R$-module $O_{S \cdot B}$ which sheafifies to $\mathcal{O}_{S \cdot B}$ is given by
\begin{align}
  \left(
    \adjustbox{max height=0.1\textwidth}{$
\begin{array}{c}
R( - \overline{K}_{{B}_3} ) \\ 
R( - 3 \overline{K}_{{B}_3} ) \\ 
R( - \overline{K}_{{B}_3} - H + E_1 ) \\
R( - 2 \overline{K}_{{B}_3} - H + E_1 ) \\
R( - \overline{K}_{{B}_3} - 2 H + 2 E_1 ) \\
R( -E_1 + E_4 ) \\
R( - E_4 ) 
\end{array}$} \right)^T \xrightarrow{\left(   
  \adjustbox{max width=0.35\textwidth}{$
 \begin{array}{cccccccc} p_1 & p_2 &  p_3 & p_4 & p_5 & e_1 & e_4 & p_{F_{11}} \end{array}$} 
 \right)^T} R \twoheadrightarrow O_{S \cdot B} \to 0 \, .
\end{align}
Denote this sequence by $F_2 \xrightarrow{M_1} F_1 \twoheadrightarrow O_{S \cdot B} \to 0$. A minimal free resolution is given by
\begin{align}
0 \to F_7 \xrightarrow{M_6} F_6 \xrightarrow{M_5} F_5 \xrightarrow{M_4} F_4 \xrightarrow{M_3} F_3 \xrightarrow{M_2} F_2 \xrightarrow{M_1} F_1 \twoheadrightarrow O_{S \cdot B} \to 0 \, ,
\end{align}
where $\mathrm{rk} ( F_1 ) = \mathrm{rk} ( F_7 ) = 1$,  $\mathrm{rk} ( F_6 ) = 6$, $\mathrm{rk} ( F_2 ) = 7$, $\mathrm{rk} ( F_3 ) = \mathrm{rk} ( F_5 ) = 19$, $\mathrm{rk} ( F_4 ) = 25$.\footnote{The twists of these free modules and the mapping matrices are huge. We therefore omit them here.}

We compute the Euler characteristics of the $\widetilde{F}_i$ by computing their sheaf cohomologies. The latter is performed by use of the Künneth formula. Namely, since $X_5 = P_{39} \times \mathbb{P}_{F_{11}}$, and 
\begin{align}
  H^k\left(X_5,{L}\right)=H^k(P_{39} \times \mathbb{P}_{F_{11}},{L})=\bigoplus_{i+j=k}H^i( P_{39}, {L}) \otimes H^j(  \mathbb{P}_{F_{11}}, {L} )\, .
\end{align} we can easily compute the cohomologies in question from line bundle cohomologies on $P_{39}$ and $\mathbb{P}_{F_{11}}$.
The Euler characteristics of the vector bundles $\widetilde{F}_i$ are 
\begin{align}
  &\chi( \widetilde{F}_1 ) = 1 \, , &\chi( \widetilde{F}_2 ) = -50 \, , &&\chi( \widetilde{F}_3 ) = -82 \, , &&\chi( \widetilde{F}_4 ) = 384 \, ,\\
  &\chi( \widetilde{F}_5 ) = 699 \, , &\chi( \widetilde{F}_6 ) = 266 \, , &&\chi( \widetilde{F}_7 ) = 0 \, .
\end{align}
It follows that
\begin{align}
\begin{split}
S \cdot B = \chi( \widetilde{O}_{S \cdot B} ) &= \chi (\widetilde{F}_1 ) - \chi( \widetilde{F}_2 ) + \chi( \widetilde{F}_3 ) - \chi( \widetilde{F}_4 ) + \chi( \widetilde{F}_5 ) - \chi( \widetilde{F}_6 ) + \chi( \widetilde{F}_7 ) \\
            &= 1 - (-50) + (-82) - 384 + 699 - 266 + 0 = 18 \, . \label{equ:Previous}
\end{split}
\end{align}

\subsubsection{Line bundles from Chow ring of toric ambient space} \label{subsec:LineBundleFromChowX5}

Let us repeat the intersection computation $S \cdot B$ by using
\begin{align}
S \equiv S_{(\one, \two)_{-1/2}}^{(1)} &= \left. V(s_3, p_1) - V(e_1, p_1) - V(s_3, v) \right|_{\hat{Y}_4} \, ,
\end{align}
instead. Similarly, $B = \left. V( e_1, e_4 ) \right|_{\hat{Y}_4}$. We define $S^\prime, T^\prime \in \mathrm{CH}^2( X_5, \mathbb{Z} )$ via
\begin{align}
S^\prime = V( s_3, p_3 ) - V( e_1, p_3 ) - V( s_3, v ), \qquad T^\prime = V( e_1, e_4 ) \, .
\end{align}
Then, it follows $S \cdot_{\widehat{Y}_4} T = S^\prime \cdot_{X_5} T^\prime \cdot_{X_5} V( p_{F_{11}} )$. Explicitly, we find
\begin{align}
  \begin{split}
&V( s_3, p_3 ) \cdot_{X_5} V( e_1, e_4 ) = V( s_3,  s_5 w, e_1, e_4 ) = V( s_3, s_5, e_1, e_4 ) \, , \\
&V( e_1, p_3 ) \cdot_{X_5} V( e_1, e_4 ) =  \emptyset \, , \qquad V( s_3, v ) \cdot_{X_5} V( e_1, e_4 ) = \emptyset \, .
\end{split}
\end{align}
From a primary decomposition, we find $\langle s_3, s_5, e_1, e_4, p_{F_{11}} \rangle = \langle e_1, e_4, s_3, s_5, s_9 \rangle$. Note that $e_1 = e_4 = 0$ fixes all other homogeneous coordinates of $P_{F_{11}}$. Hence
\begin{align}
\pi_\ast \left( S \cdot_{\widehat{Y}_4} T \right) = V \left( s_3, s_5, s_9 \right) \, .
\end{align}
If we consider ${B}_3 = P_{39}$, then it follows from \oref{equ:Previous} that $V( s_3, s_5, s_9 )$ must be a divisor of degree 18 on $C_{(\one, \two)_{-1/2}}$. Indeed, this is true because $\overline{K}_{P_{39}}^3 = 18$. It is not too hard to repeat this computation and find that $\mathcal{A}^\prime$ in \oref{equ:IntegralFlux} gives
\begin{align}
  & D_{(\three, \two)_{1/6}} \left( \mathcal{A}^\prime \right) = 3 \cdot V(\KB,s_3,s_9) \, ,\\
  & D_{(\one, \two)_{-1/2}} \left( \mathcal{A}^\prime \right) =-3 \left[ 5V(s_3,s_5,s_9)-2V(\KB,s_3,P_H) \right] \, , \label{equ:ResultHiggs} \\
  & D_{(\overline{\three}, \one)_{-2/3}} \left( \mathcal{A}^\prime \right) = 3 \cdot V(\KB,s_5,s_9) \, ,\\
  & D_{(\overline{\three}, \one)_{1/3}} \left( \mathcal{A}^\prime \right) = -3 \left[ 5V(s_3,s_6,s_9)-2V(\KB,s_9,P_R) \right] \, ,\\
  & D_{(\one, \one)_{1}} \left( \mathcal{A}^\prime \right) = 3 \cdot V(\KB,s_1,s_5) \, .
\end{align}
In this expression, we are using
\begin{align}
P_H = s_2 s_5^2 + s_1 ( s_1 s_9 - s_5 s_6 ) \, , \qquad P_R = s_3 s_5^2 + s_6 ( s_1 s_6 - s_2 s_5 ) \, .
\end{align}
By considering $\KB^3$-th roots and adding spin bundles on the matter curves, one arrives at the root bundle expressions summarized in \oref{subsec:LineBundlesKbarCubeDifferent}.

\subsubsection{Line bundles from fiber structure} \label{subsec:LineBundleFromFibre}

Finally, let us present a third way to compute the induced line bundles. Even though this approach is equivalent, it provides more intuition than the brute-force intersection computations in $\mathrm{CH}^\ast( X_5 )$. To this end we make use of the genesis of the $G_4$-flux and the fiber structure of $\widehat{Y}_4$, which we outlined in \oref{sec:FibreStructure}.

Let us apply this strategy for the Higgs curve. We first recall that $\mathcal{A}^\prime$ in \oref{equ:IntegralFlux} can be thought of as
\begin{align}
\begin{split}
\mathcal{A}^\prime = \mathcal{A}^\prime_{(\three, \two)_{1/6}} + \mathcal{A}^\prime_{U(1)} = 15 \cdot \mathcal{A}_{(\three, \two)_{1/6}} + 3 \cdot \pi^\ast \left( \KB \right) \cdot \sigma \in CH^2( \widehat{Y}_4, \mathbb{Z} ) \, ,
\end{split}
\end{align}
where (in abuse of notation) $\sigma$ denotes the canonical lift of the 1-form associated to the section $s_1 = V( e_4 )$ via the Shioda map. On general grounds, it now follows that
\begin{align}
   \pi_\ast( S_{\mathbf{R}} \cdot \mathcal{A}^\prime_{U(1)} ) = q_{U(1)} \cdot \left. 3 \KB \right|_{C_{\mathbf{R}}} \, .
\end{align}
For the Higgs curve, we have $q_{U(1)} = -1/2$. Thus,
\begin{align}
\pi_\ast \left( S_{(\mathbf{1},\mathbf{2})_{-1/2}} \cdot \mathcal{A}^\prime_{U(1)} \right) = - \frac{3}{2} \cdot \left. \KB \right|_{C_{(\mathbf{1},\mathbf{2})_{-1/2}}} \, .
\end{align}
Note that (c.f. \oref{equ:IntersectionsOfCurves}) $C_{(\mathbf{1},\mathbf{2})_{-1/2}} \cdot C_{(\mathbf{3},\mathbf{2})_{1/6}} = Y_1 \cup Y_2$. Hence, the intersection number of the Higgs matter surface and $\mathcal{A}^\prime_{(\three, \two)_{1/6}}$ is found in the fiber over $Y_1$ and $Y_2$:
\begin{align}
\left. \mathcal{A}^\prime_{(\three, \two)_{1/6}} \right|_{Y_1} \cdot \left. S_{(\mathbf{1},\mathbf{2})_{-1/2}} \right|_{Y_1} &= \left( 1/2, 2/3, 0, 1, 1/3, 0 \right) \cdot (0, 1,1,1,1,0 ) = -1/2 \, , \\
\left. \mathcal{A}^\prime_{(\three, \two)_{1/6}} \right|_{Y_2} \cdot \left. S_{(\mathbf{1},\mathbf{2})_{-1/2}} \right|_{Y_2} &= \left( 0, 1/2, 2/3, 1, 1/3, 0 \right) \cdot (0, 0,0,0,0,1 ) = +1/2 \, .
\end{align}
This implies
\begin{align}
D_{(\mathbf{1},\mathbf{2})_{-1/2}} \left( \mathcal{A}^\prime_{(\three, \two)_{1/6}} \right) = 15 \cdot \left[ - \frac{1}{2} Y_1 + \frac{1}{2} Y_2 \right] \, .
\end{align}
We now use $Y_1 + Y_2 = \left. \overline{K} \right|_{C_{(\mathbf{1},\mathbf{2})_{-1/2}}}$ (c.f. \oref{equ:IntersectionsOfCurves}) to conclude that
\begin{align}
D_{(\mathbf{1},\mathbf{2})_{-1/2}} \left( \mathcal{A}^\prime_{(\three, \two)_{1/6}} \right) &= 15 \cdot \left[ - \frac{1}{2} Y_1 + \frac{1}{2} Y_2 \right] - \frac{3}{2} \cdot \left. \overline{K} \right|_{C_{(\mathbf{1},\mathbf{2})_{-1/2}}} \\ \nonumber
&= 6 \left. \overline{K} \right|_{C_{(\mathbf{1},\mathbf{2})_{-1/2}}} - 15 Y_1 \ .
\end{align}
Noting that $Y_1 = V( s_3, s_5, s_9)$, and $P_H = s_2 s_5^2 + s_1 ( s_1 s_9 - s_5 s_6 )$, we thus find
\begin{align}
D_{(\mathbf{1},\mathbf{2})_{-1/2}} \left( \mathcal{A}^\prime_{(\three, \two)_{1/6}} \right) = -3 \left[ 5V(s_3,s_5,s_9)-2V(\KB,s_3,P_H) \right] \, .
\end{align}
This is exactly the result that we found in \oref{equ:ResultHiggs}. Similarly, the line bundle expressions found in \oref{subsec:LineBundleFromChowX5} for $C_{(\mathbf{1},\mathbf{1})_1}$, $C_{(\mathbf{\overline{3}},\mathbf{1})_{-2/3}}$, $C_{(\mathbf{\overline{3}},\mathbf{1})_{1/3}}$ can be verified by using this strategy. For the quark-doublet curve, the situation is more involved since the matter surface flux is defined over this very matter curve so that self-intersections are to be taken into account. Equivalently, we can give a quick argument by noting that the divisor in question must be a linear combination of the Yukawa loci on $C_{(\mathbf{3},\mathbf{2})_{1/6}}$. Any of these Yukawa loci $Y_1$, $Y_2$, $Y_3$ admits a pullback description:
\begin{equation}
  \resizebox{0.9\textwidth}{!}{%
$\begin{aligned}
  \mathcal{O}_{C_{(\mathbf{3},\mathbf{2})_{1/6}}}( Y_1 ) \cong \mathcal{O}_{C_{(\mathbf{3},\mathbf{2})_{1/6}}}( Y_3 ) \cong \left. \overline{K}_{{B}_3} \right|_{C_{(\mathbf{3},\mathbf{2})_{1/6}}} \, , \quad \mathcal{O}_{C_{(\mathbf{3},\mathbf{2})_{1/6}}}( Y_2 ) \cong \left. 2 \overline{K}_{{B}_3} \right|_{C_{(\mathbf{3},\mathbf{2})_{1/6}}} \, .
\end{aligned}$%
}
\end{equation}
Therefore, the bundle must be of the form $n \cdot \left. \overline{K}_{{B}_3} \right|_{C_{(\mathbf{3},\mathbf{2})_{1/6}}}$ and the prefactor $n$ is fixed by the chiral index. This gives 
$D_{(\three, \two)_{1/6}} \left( \mathcal{A}^\prime_{(\three, \two)_{1/6}} \right) = 3 \cdot \left. \overline{K}_{{B}_3} \right|_{C_{(\mathbf{3},\mathbf{2})_{1/6}}}$.

\subsection{Root bundle constraints} \label{subsec:LineBundlesKbarCubeDifferent}

By repeating the intersection computations, we obtain the root bundle constraints as functions of $\KB^3$ (c.f. \oref{equ:GeneralRootBundleExpressions}). Since we analyze the case $\KB^3 = 18$ in more detail momentarily, let us list the root bundles for such base spaces explicitly:
\begin{center}
\resizebox{\textwidth}{!}{
  \begin{tabular}{cc|cc|ccc}
  \toprule
  curve & $g$ & ${P}$ & $d$ & \multicolumn{3}{c}{BN-theory} \\
  \midrule
  \multirow{4}{*}{$C_{(\mathbf{3},\mathbf{2})_{1/6}} = V( s_3, s_9 )$} & \multirow{4}{*}{10} & \multirow{4}{*}{${P}_{(\three,\two)_{1/6}}^{\otimes 36} = \left. \overline{K}_{B_3} \right|_{C_{(\three,\two)_{1/6}}}^{\otimes 24}$} & \multirow{4}{*}{12} & $h^0$ & $h^1$ & $\rho$ \\
                    &                                                                                      &    &   & 3 & 0 & 10 \\
                    &                                                                                      &    &   & 4 & 1 & 6 \\
                    &                                                                                      &    &   & 5 & 2 & 0 \\
  \midrule
  & \multirow{6}{*}{82} & \multirow{6}{*}{${P}_{(\mathbf{1},\mathbf{2})_{-1/2}}^{\otimes 36} = \left. \overline{K}_{B_3} \right|_{C_{(\mathbf{1},\mathbf{2})_{-1/2}}}^{\otimes 66}\otimes \mathcal{O}_{C_{(\mathbf{1},\mathbf{2})_{-1/2}}}(-30\cdot Y_1)$} & \multirow{6}{*}{84} & $h^0$ & $h^1$ & $\rho$ \\
  $C_{(\mathbf{1},\mathbf{2})_{-1/2}} =$ &                                               &    &   & 3 & 0 & 82 \\
                    &                                                                    &    &   & 4 & 1 & 78 \\
  $V \left( s_3, s_2 s_5^2 + s_1 ( s_1 s_9 - s_5 s_6 ) \right)$ &                        &    &   & $\vdots$ & $\vdots$ & $\vdots$ \\
                    &                                                                    &    &   & 10 & 7 & 12 \\
  \midrule
  \multirow{4}{*}{$C_{(\overline{\mathbf{3}},\mathbf{1})_{-2/3}} = V( s_5, s_9 )$} & \multirow{4}{*}{10} & \multirow{4}{*}{${P}_{(\overline{\mathbf{3}},\mathbf{1})_{-2/3}}^{\otimes 36} = \left. \overline{K}_{B_3} \right|_{C_{(\overline{\mathbf{3}},\mathbf{1})_{-2/3}}}^{\otimes 24}$} & \multirow{4}{*}{12} & $h^0$ & $h^1$ & $\rho$ \\
                    &                                                                                      &    &   & 3 & 0 & 10 \\
                    &                                                                                      &    &   & 4 & 1 & 6 \\
                    &                                                                                      &    &   & 5 & 2 & 0 \\
  \midrule
  & \multirow{6}{*}{82} & \multirow{6}{*}{${P}_{(\overline{\mathbf{3}},\mathbf{1})_{1/3}}^{\otimes 36} = \left. \overline{K}_{B_3} \right|_{C_{(\overline{\mathbf{3}},\mathbf{1})_{1/3}}}^{\otimes 66} \otimes \mathcal{O}_{C_{(\overline{\mathbf{3}},\mathbf{1})_{1/3}}}(-30\cdot Y_3)$} & \multirow{6}{*}{84} & $h^0$ & $h^1$ & $\rho$ \\
  $C_{(\overline{\mathbf{3}},\mathbf{1})_{1/3}} =$ &                                                   &    &   & 3 & 0 & 82 \\
                &                                                                                      &    &   & 4 & 1 & 78 \\
  $V \left( s_9, s_3 s_5^2 + s_6 ( s_1 s_6 - s_2 s_5 ) \right)$ &                                      &    &   & $\vdots$ & $\vdots$ & $\vdots$ \\
                    &                                                                                  &    &   & 10 & 7 & 12 \\
  \midrule
  \multirow{4}{*}{$C_{(\mathbf{1},\mathbf{1})_{1}} = V( s_1, s_5 )$} & \multirow{4}{*}{10} & \multirow{4}{*}{${P}_{(\mathbf{1},\mathbf{1})_{1}}^{\otimes 36} = \left. \overline{K}_{B_3} \right|_{C_{(\mathbf{1},\mathbf{1})_{1}}}^{\otimes 24}$} & \multirow{4}{*}{12} & $h^0$ & $h^1$ & $\rho$ \\
                    &                                                                                      &    &   & 3 & 0 & 10 \\
                    &                                                                                      &    &   & 4 & 1 & 6 \\
                    &                                                                                      &    &   & 5 & 2 & 0 \\
  \bottomrule
  \end{tabular}}
\end{center}

The parameter $\rho$ from Brill-Noether theory \cite{Brill1874} provides a measure of how likely it is to find a degree $d$ line bundle with certain number of global sections -- the larger $\rho$ is, the more likely such bundles exist. Notably, this parameter does not take the root bundle constraints into account. See \cite{Watari:2016lft,Bies:2020gvf} for an application of Brill-Noether theory to F-theory and further explanations.

For the considered F-theory Standard Model constructions, the toric base spaces must satisfy $\KB^3 \in \{ 6, 10, 18, 30 \}$ \cite{Cvetic:2019gnh}. Therefore, let us list the root bundle constraints for these values of $\KB^3$. Analogous to the above table, the root bundle constraints on $C_{(\overline{\mathbf{3}},\mathbf{1})_{-2/3}}$, $C_{(\overline{\mathbf{3}},\mathbf{1})_{1/3}}$, $C_{(\mathbf{1},\mathbf{1})_{1}}$ follow once the constraints on $C_{(\mathbf{3},\mathbf{2})_{1/6}}$, $C_{(\mathbf{1},\mathbf{2})_{-1/2}}$ are known. For ease of presentation, we will merely list the constraints on $C_{(\mathbf{3},\mathbf{2})_{1/6}}$ and $C_{(\mathbf{1},\mathbf{2})_{-1/2}}$:
\begin{center}
  \resizebox{\textwidth}{!}{
  \begin{tabular}{c|cc|cc|ccc}
  \toprule
  $\overline{K}_{B_3}^3$ & curve & $g$ & ${P}$ & $d$ & \multicolumn{3}{c}{BN-theory} \\
  \midrule
  \multirow{9}{*}{$6$} & \multirow{4}{*}{$C_{(\mathbf{3},\mathbf{2})_{1/6}} = V( s_3, s_9 )$} & \multirow{4}{*}{4} & \multirow{4}{*}{${P}_{(\three,\two)_{1/6}}^{\otimes 12} = \left. \overline{K}_{B_3} \right|_{C_{(\three,\two)_{1/6}}}^{\otimes 12}$} & \multirow{4}{*}{6} & $h^0$ & $h^1$ & $\rho$ \\
  & &                                                                                      &    &   & 3 & 0 & 4 \\
  & &                                                                                      &    &   & 4 & 1 & 0 \\
  & &                                                                                      &    &   & 5 & 2 & -6 \\
  \cmidrule{2-8}
  & & \multirow{6}{*}{28} & \multirow{6}{*}{${P}_{(\mathbf{1},\mathbf{2})_{-1/2}}^{\otimes 12} = \left. \overline{K}_{B_3} \right|_{C_{(\mathbf{1},\mathbf{2})_{-1/2}}}^{\otimes 30}\otimes \mathcal{O}_{C_{(\mathbf{1},\mathbf{2})_{-1/2}}}(-30\cdot Y_1)$} & \multirow{6}{*}{30} & $h^0$ & $h^1$ & $\rho$ \\
  & $C_{(\mathbf{1},\mathbf{2})_{-1/2}} =$ &                                               &    &   & 3 & 0 & 28 \\
  & &                                                                    &    &   & 4 & 1 & 24 \\
  & $V \left( s_3, s_2 s_5^2 + s_1 ( s_1 s_9 - s_5 s_6 ) \right)$ &                        &    &   & $\vdots$ & $\vdots$ & $\vdots$ \\
  & &                                                                    &    &   & 7 & 4 & 0 \\
  \midrule \midrule
  \multirow{9}{*}{$10$} & \multirow{4}{*}{$C_{(\mathbf{3},\mathbf{2})_{1/6}} = V( s_3, s_9 )$} & \multirow{4}{*}{6} & \multirow{4}{*}{${P}_{(\three,\two)_{1/6}}^{\otimes 20} = \left. \overline{K}_{B_3} \right|_{C_{(\three,\two)_{1/6}}}^{\otimes 16}$} & \multirow{4}{*}{8} & $h^0$ & $h^1$ & $\rho$ \\
  & &                                                                                      &    &   & 3 & 0 & 6 \\
  & &                                                                                      &    &   & 4 & 1 & 2 \\
  & &                                                                                      &    &   & 5 & 2 & -4 \\
  \cmidrule{2-8}
  & & \multirow{6}{*}{46} & \multirow{6}{*}{${P}_{(\mathbf{1},\mathbf{2})_{-1/2}}^{\otimes 20} = \left. \overline{K}_{B_3} \right|_{C_{(\mathbf{1},\mathbf{2})_{-1/2}}}^{\otimes 42}\otimes \mathcal{O}_{C_{(\mathbf{1},\mathbf{2})_{-1/2}}}(-30\cdot Y_1)$} & \multirow{6}{*}{48} & $h^0$ & $h^1$ & $\rho$ \\
  & $C_{(\mathbf{1},\mathbf{2})_{-1/2}} =$ &                                               &    &   & 3 & 0 & 46 \\
  & &                                                                    &    &   & 4 & 1 & 42 \\
  & $V \left( s_3, s_2 s_5^2 + s_1 ( s_1 s_9 - s_5 s_6 ) \right)$ &                        &    &   & $\vdots$ & $\vdots$ & $\vdots$ \\
  & &                                                                    &    &   & 8 & 5 & 6 \\
  \midrule \midrule
  \multirow{9}{*}{$18$} & \multirow{4}{*}{$C_{(\mathbf{3},\mathbf{2})_{1/6}} = V( s_3, s_9 )$} & \multirow{4}{*}{10} & \multirow{4}{*}{${P}_{(\three,\two)_{1/6}}^{\otimes 36} = \left. \overline{K}_{B_3} \right|_{C_{(\three,\two)_{1/6}}}^{\otimes 24}$} & \multirow{4}{*}{12} & $h^0$ & $h^1$ & $\rho$ \\
  & &                                                                                      &    &   & 3 & 0 & 10 \\
  & &                                                                                      &    &   & 4 & 1 & 6 \\
  & &                                                                                      &    &   & 5 & 2 & 0 \\
  \cmidrule{2-8}
  & & \multirow{6}{*}{82} & \multirow{6}{*}{${P}_{(\mathbf{1},\mathbf{2})_{-1/2}}^{\otimes 36} = \left. \overline{K}_{B_3} \right|_{C_{(\mathbf{1},\mathbf{2})_{-1/2}}}^{\otimes 66}\otimes \mathcal{O}_{C_{(\mathbf{1},\mathbf{2})_{-1/2}}}(-30\cdot Y_1)$} & \multirow{6}{*}{84} & $h^0$ & $h^1$ & $\rho$ \\
  & $C_{(\mathbf{1},\mathbf{2})_{-1/2}} =$ &                                               &    &   & 3 & 0 & 82 \\
  & &                                                                    &    &   & 4 & 1 & 78 \\
  & $V \left( s_3, s_2 s_5^2 + s_1 ( s_1 s_9 - s_5 s_6 ) \right)$ &                        &    &   & $\vdots$ & $\vdots$ & $\vdots$ \\
  & &                                                                    &    &   & 10 & 7 & 12 \\
  \midrule \midrule
  \multirow{9}{*}{$30$} & \multirow{4}{*}{$C_{(\mathbf{3},\mathbf{2})_{1/6}} = V( s_3, s_9 )$} & \multirow{4}{*}{16} & \multirow{4}{*}{${P}_{(\three,\two)_{1/6}}^{\otimes 60} = \left. \overline{K}_{B_3} \right|_{C_{(\three,\two)_{1/6}}}^{\otimes 36}$} & \multirow{4}{*}{18} & $h^0$ & $h^1$ & $\rho$ \\
  & &                                                                                      &    &   & 3 & 0 & 16 \\
  & &                                                                                      &    &   & 4 & 1 & 12 \\
  & &                                                                                      &    &   & 5 & 2 & 6 \\
  \cmidrule{2-8}
  & & \multirow{6}{*}{136} & \multirow{6}{*}{${P}_{(\mathbf{1},\mathbf{2})_{-1/2}}^{\otimes 60} = \left. \overline{K}_{B_3} \right|_{C_{(\mathbf{1},\mathbf{2})_{-1/2}}}^{\otimes 102}\otimes \mathcal{O}_{C_{(\mathbf{1},\mathbf{2})_{-1/2}}}(-30\cdot Y_1)$} & \multirow{6}{*}{138} & $h^0$ & $h^1$ & $\rho$ \\
  & $C_{(\mathbf{1},\mathbf{2})_{-1/2}} =$ &                                               &    &   & 3 & 0 & 136 \\
  & &                                                                    &    &   & 4 & 1 & 132 \\
  & $V \left( s_3, s_2 s_5^2 + s_1 ( s_1 s_9 - s_5 s_6 ) \right)$ &                        &    &   & $\vdots$ & $\vdots$ & $\vdots$ \\
  & &                                                                    &    &   & 13 & 10 & 6 \\
  \bottomrule
  \end{tabular}}
\end{center}

\subsection{Limit roots in base space \texorpdfstring{$P_{39}$}{P39}} \label{sec:ExampleGeometry}

We consider the smooth, complete toric 3-fold base $P_{39}$, whose Cox ring is $\mathbb{Z}^{8}$-graded
\begin{align}
\begin{tabular}{ccccccccccc}
\toprule
$x_1$ & $x_2$ & $x_3$ & $x_4$ & $x_5$ & $x_6$ & $x_7$ & $x_8$ & $x_9$ & $x_{10}$ & $x_{11}$ \\
\midrule
 1 &  -2 &   1 &   0 &   0 &   0 &   0 &   0 &   0 &   0 &   0 \\
 0 &   1 &   0 &  -2 &   1 &   0 &   0 &   0 &   0 &   0 &   0 \\
 0 &   2 &   0 &  -3 &   0 &   1 &   0 &   0 &   0 &   0 &   0 \\
 0 &  -1 &   0 &   1 &   0 &   0 &   1 &   0 &   0 &   0 &   0 \\
 0 &   1 &   0 &  -1 &   0 &   0 &   0 &   1 &   0 &   0 &   0 \\
 0 &  -1 &   0 &   2 &   0 &   0 &   0 &   0 &   1 &   0 &   0 \\
 0 &   0 &   0 &   1 &   0 &   0 &   0 &   0 &   0 &   1 &   0 \\
 0 &  -1 &   0 &   3 &   0 &   0 &   0 &   0 &   0 &   0 &   1 \\
\bottomrule
\end{tabular} \label{equ:GradingB3}
\end{align}
and whose Stanley-Reisner ideal is given by
\begin{align}
\begin{split}
 I_{\text{SR}} &= \langle
x_8 x_{11},
x_7 x_{11},
x_6 x_{11},
x_5 x_{11},
x_4 x_{11},
x_2 x_{11},
x_9 x_{10},
x_7 x_{10},
x_6 x_{10},
x_5 x_{10}, \\
& \qquad x_4 x_{10},
x_2 x_{10},
x_8 x_9, 
x_6 x_9, 
x_5 x_9, 
x_4 x_9, 
x_2 x_9, 
x_7 x_8, 
x_5 x_8, 
x_4 x_8, 
x_2 x_8, 
x_6 x_7, \\
& \qquad x_5 x_7, 
x_4 x_7, 
x_4 x_6, 
x_2 x_6, 
x_2 x_5, 
x_1 x_3
\rangle \, .
\end{split} \label{equ:SRB3}
\end{align}
$P_{39}$ is a particular triangulation of the 39-th polytope in the Kreuzer-Skarke list of toric threefolds \cite{Kreuzer:1998vb}, hence the name. It follows that $\overline{K}_{P_{39}}^3 = 18$. Furthermore, for $D_i = V( x_i )$, we find non-trivial topological intersection numbers
\begin{align}
\begin{tabular}{cccccc}
\toprule
& $D_{i}$ & $D_3$ & $D_6$ & $D_{11}$ \\
\midrule
$D_i \cdot \overline{K}_{P_{39}}^2$ & 3 &   3 &   6 &   6 \\
\bottomrule
\end{tabular}
\end{align}
The remaining divisors have vanishing topological intersection numbers. Even more, for $D_i \in \{D_4,D_5,D_7,D_9\}$, we find $D_i \cdot V( s_i ) \cdot V( s_j ) = \emptyset$ for any $s_i, s_j \in H^0( P_{39}, \overline{K}_{P_{39}} )$. The divisors $D_2$, $D_8$, $D_{10}$ intersect the generic curve $V( s_i, s_j )$ trivially but admit non-trivial intersections with non-generic curves.

In \oref{subsec:RootBundlesF-theoryMSSMs}, we discussed roots on the quark-doublet curve $C_{(\overline{\mathbf{3}},\mathbf{2})_{1/6}}$. Here, we provide details on the limit roots on  $C_{(\overline{\mathbf{3}},\mathbf{1})_{1/3}} = V \left( s_9, s_3 s_5^2 + s_6 ( s_1 s_6 - s_2 s_5 ) \right)$. We use
\begin{align}
s_1 \to s_6 - s_3 \, , \qquad s_2 \to s_5 - \prod_{i = 1}^{11}{x_i} \, , \qquad s_9 \to \prod_{i = 1}^{11}{x_i} \, ,
\end{align}
and \emph{generic} $s_3$, $s_5$, $s_6$ to deform this curve into
\begin{equation}
\resizebox{0.9\textwidth}{!}{%
$\begin{aligned}
C_{\left(\overline{\mathbf{3}},\mathbf{1}\right)_{1/3}}^\bullet &= V \left( \prod_{i = 1}^{11}{x_i}, s_5 - s_6 \right) \cup V \left( \prod_{i = 1}^{11}{x_i}, s_3 - s_6 \right) \cup V \left( \prod_{i = 1}^{11}{x_i}, s_5 + s_6 \right) \equiv Q_1 \cup Q_2 \cup Q_3 \, .
\end{aligned}$%
}
\end{equation}
It is important to verify that this curve is nodal so that we can apply the limit root techniques outlined in \oref{sec:LimitRoots}. A computationally favorable description is that a point $p$ is a node if and only if the Jacobian matrices vanish identically at $p$ but the Hessian matrix does not \cite{arbarello2011algebraic}. Therefore, it is readily verified that for example, the points $V( x_1, s_5 - s_6, s_3 - s_6 )$ are indeed nodes.

Consequently, we proceed to identify roots $P^\bullet _{(\overline{\mathbf{3}},\mathbf{1})_{1/3}}$ that solve the root bundle \mbox{constraint} in \oref{subsec:LineBundlesKbarCubeDifferent} and admit exactly three sections. For this, it suffices to construct \mbox{solutions} to
\begin{align}
\left( P^\bullet _{(\overline{\mathbf{3}},\mathbf{1})_{1/3}} \right)^{\otimes 6} = \left. \overline{K}_{B_3} \right|_{C^\bullet_{(\overline{\mathbf{3}},\mathbf{1})_{1/3}}}^{\otimes 11}\otimes \mathcal{O}_{C^\bullet_{(\overline{\mathbf{3}},\mathbf{1})_{1/3}}}(-5\cdot Y_3) \, , \quad h^0 \left( C^\bullet_{(\overline{\mathbf{3}},\mathbf{1})_{1/3}}, P^\bullet_{(\overline{\mathbf{3}},\mathbf{1})_{1/3}} \right) = 3 \, , \label{equ:6thRootsOnCRDQ}
\end{align}
where $Y_3 = V( s_3, s_6, s_9 )$. We notice that $Y_3 \cap Q_1 = Y_3 \cap Q_3 = \emptyset$, which implies
\begin{align}
\resizebox{0.9\textwidth}{!}{%
$\begin{aligned}
\left. \left( P_{(\overline{\mathbf{3}},\mathbf{1})_{1/3}}^{\bullet} \right)^{\otimes 6} \right|_{Q_1} &= \left. \overline{K}_{B_3} \right|_{Q_1}^{\otimes 11} \, , \\
\left. \left( P_{(\overline{\mathbf{3}},\mathbf{1})_{1/3}}^{\bullet} \right)^{\otimes 6} \right|_{Q_2} &= \left. \overline{K}_{B_3} \right|_{Q_2}^{\otimes 11} \otimes \mathcal{O}_{Q_2}(-5\cdot Y_3)
= \left. \overline{K}_{B_3} \right|_{Q_2}^{\otimes 11} \otimes \left. \overline{K}_{B_3} \right|_{Q_2}^{\otimes (-5)}
= \left. \overline{K}_{B_3} \right|_{Q_2}^{\otimes 6} \, , \\
\left. \left( P_{(\overline{\mathbf{3}},\mathbf{1})_{1/3}}^{\bullet} \right)^{\otimes 6} \right|_{Q_3} &= \left. \overline{K}_{B_3} \right|_{Q_3}^{\otimes 11} \, .
\end{aligned}$%
}
\end{align}
These observations allow us to draw a weighted graph, which encodes roots $P^\bullet_{(\overline{\mathbf{3}},\mathbf{1})_{1/3}}$ on $C_{(\overline{\mathbf{3}},\mathbf{1})_{1/3}}^\bullet$. This graph is displayed in \oref{fig:ComplicatedCurve}.

We find it important to mention that this graph is non-planar. This is remarkable because all other dual graphs considered in this work are planar. To our knowledge, there does not seem to be any result in the literature which suggests that the dual graph of a nodal curve is necessarily planar. In fact, most of the literature, such as \cite{busonero2006combinatorial} and \cite{arbarello2011algebraic}, only discuss examples of nodal curves with planar dual graphs. Although there are well-known planarity criterion theorems, such as the Kuratowski's theorem \cite{Kuratowski1930}, we resorted to excessive computational checks to verify that $C_{(\overline{\mathbf{3}},\mathbf{1})_{1/3}}^\bullet$ has a non-planar dual graph \oref{fig:ComplicatedCurve}. A more minimalistic example of this sort is the nodal curve whose dual graph is $K_{3,3}$. There are many interesting questions concerning planarity that arise in graph theory, such as criterion theorems \cite{Wagner1937, MacLane1937}, enumeration \cite{2009JAMS...22..309G}, and other variants of planarity \cite{CHMEISS199761, felsner2012geometric}. However, the significance of non-planarity for the geometry of nodal curves is unknown. We hope to return to this interesting question in the future.

Turning back to solving \oref{equ:6thRootsOnCRDQ}, we note that the degrees of the roots encoded by \oref{fig:ComplicatedCurve} are listed in \oref{fig:ComplicatedCurveDegrees}. In particular, the total degree is $d = 84$, as expected for $\chi ( P^\bullet_{(\overline{\mathbf{3}},\mathbf{1})_{1/3}} ) = 3$ on this $g = 82$ curve. Recall that we identify the number of global sections from \oref{cor:CountingSections}, i.e. we add the number of sections on all curve components except the exceptional $\mathbb{P}^1$s, which are colored in blue. Therefore, it suffices to focus on the curves $C_1^{Q_1}$, $C_3^{Q_1}$, $C_1^{Q_2}$, $C_3^{Q_2}$, $C_6^{Q_2}$, $C_1^{Q_3}$ and $C_3^{Q_3}$. Each curve $C_1^{Q_1}$ and $C_1^{Q_3}$ admits 36 roots whereas $C_6^{Q_2}$ only admits a unique root. These roots each have one section.
It follows from \oref{prop:RootsElliptic} that of the roots on $C_3^{Q_1}$, $C_1^{Q_2}$, $C_3^{Q_2}$ and $C_3^{Q_3}$, each curve admits at least 35 roots which have no sections. We have thus found at least $36^2 \cdot 35^4$ solutions to \oref{equ:6thRootsOnCRDQ}. In future works, we wish to investigate which of these root bundles stem from F-theory gauge potentials in $H^4_D( \widehat{Y}_4, \mathbb{Z}(2) )$.

\begin{landscape}
\begin{figure}
\resizebox{\linewidth}{!}{

}
\caption{Weighted diagram of roots $P^\bullet_{(\overline{\mathbf{3}},\mathbf{1})_{1/3}}$ on $C^\bullet_{(\overline{\mathbf{3}},\mathbf{1})_{1/3}}$ which solve \oref{equ:6thRootsOnCRDQ}.}
\label{fig:ComplicatedCurve}
\end{figure}
\end{landscape}

\begin{landscape}
\begin{figure}
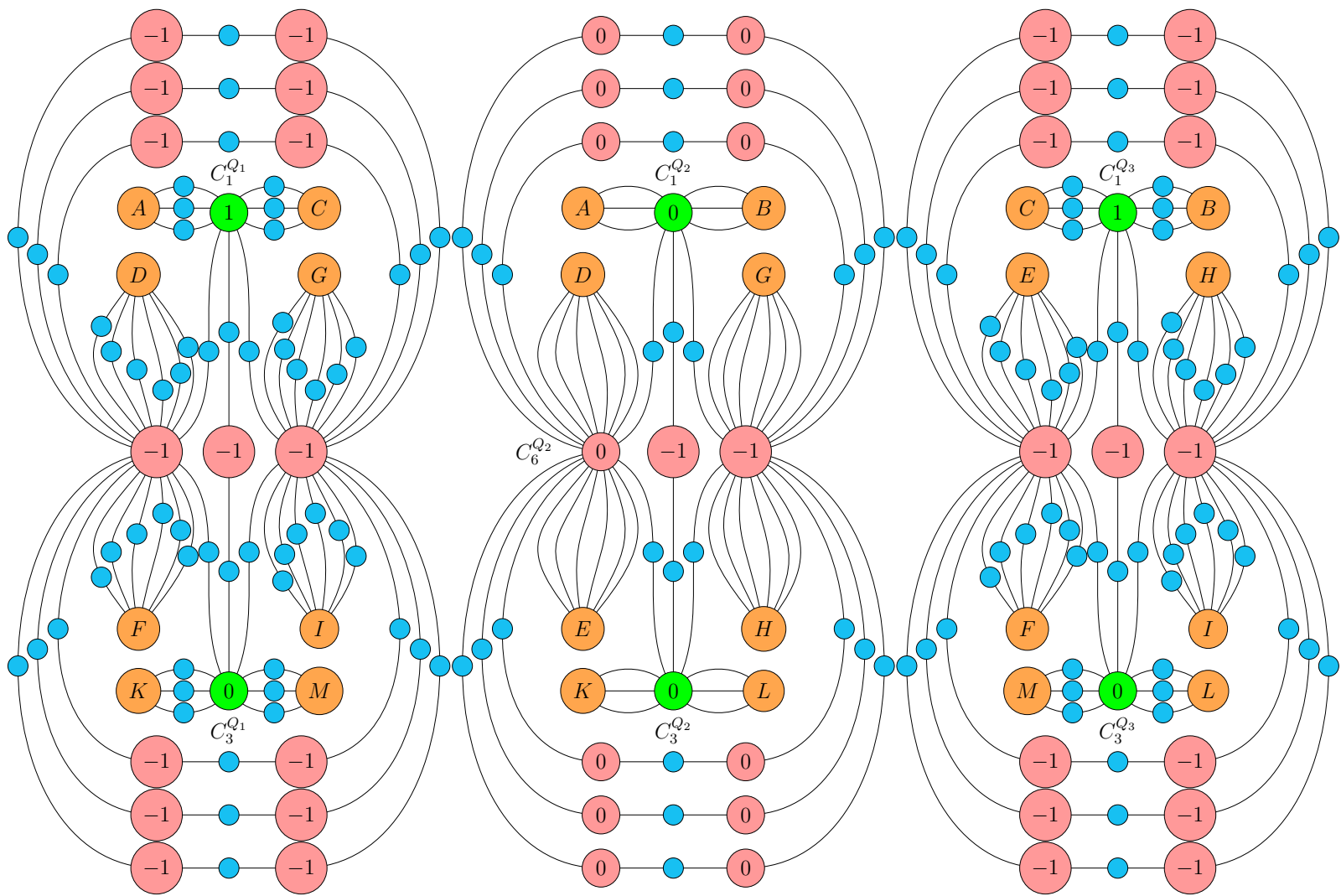

\resizebox{\linewidth}{!}{%
}
\caption{Degrees of roots $P^\bullet_{(\overline{\mathbf{3}},\mathbf{1})_{1/3}}$ on $C^\bullet_{(\overline{\mathbf{3}},\mathbf{1})_{1/3}}$ encoded by \oref{fig:ComplicatedCurve}. Exceptional $\mathbb{P}^1$s are indicated in blue and each carry a line bundle of degree $d = 1$.}
\label{fig:ComplicatedCurveDegrees}
\end{figure}
\end{landscape}

\bibliography{paper}{}
\bibliographystyle{JHEP}

\end{document}